\RequirePackage{fix-cm}
\documentclass[twocolumn,epjc3]{svjour3}  

\smartqed
\journalname{Eur. Phys. J. C}

\RequirePackage{graphicx}
\RequirePackage{amsmath}
\RequirePackage{mathtools}
\RequirePackage{booktabs}
\RequirePackage[tableposition=bottom]{caption}
\RequirePackage{hyperref}
\RequirePackage{makecell}
\RequirePackage{nicefrac}
\RequirePackage{physics}
\RequirePackage{cite}
\RequirePackage[a4paper, total={6.5in, 8.5in}]{geometry}
\RequirePackage{xcolor}
\RequirePackage{amssymb}
\RequirePackage{multirow}
\RequirePackage{hepunits}
\RequirePackage{afterpage}
\RequirePackage[utf8]{inputenc}
\RequirePackage{enumitem}

\newcommand{\xBj}{x_{\mathrm{Bj}}}

\newcounter{comment}

{\refstepcounter{comment}%
\begin{quote}
\small$\blacksquare$ \textbf{\underline{Comment} $\sharp$\thecomment:}}%
{\end{quote}}

{\refstepcounter{comment}%
\begin{quote}
\textcolor{red}{\small$\blacksquare$ \textbf{\underline{Comment} $\sharp$\thecomment:~HM:}}}%
{\end{quote}}

{\refstepcounter{comment}%
\begin{quote}
\textcolor{blue}{\small$\blacksquare$ \textbf{\underline{Comment} $\sharp$\thecomment:~HM:}}}%
{\end{quote}}

{\refstepcounter{comment}%
\begin{quote}
\textcolor{green}{\small$\blacksquare$ \textbf{\underline{Comment} $\sharp$\thecomment:~HM:}}}%
{\end{quote}}

\begin{document}

\title{Border and skewness functions from a leading order fit to DVCS data}

\author{
H.~Moutarde\thanksref{e1,addr1}
\and
P.~Sznajder\thanksref{e2,addr2} 
\and
J.~Wagner\thanksref{e3,addr2} 
}

\thankstext{e1}{e-mail: herve.moutarde@cea.fr}
\thankstext{e2}{e-mail: pawel.sznajder@ncbj.gov.pl}
\thankstext{e3}{e-mail: jakub.wagner@ncbj.gov.pl}


\institute{
IRFU, CEA, Universit\'e Paris-Saclay, F-91191 Gif-sur-Yvette, France \label{addr1}
\and
~National Centre for Nuclear Research (NCBJ), 00-681 Warsaw, Poland \label{addr2}
}

\date{Received: date / Accepted: date}

\maketitle

\sloppy

\begin{abstract}
We propose new parameterizations for the border and skewness functions appearing in the description of 3D nucleon structure in the language of Generalized Parton Distributions (GPDs). These parameterizations are constructed in a way to fulfill the basic properties of GPDs, like their reduction to Parton Density Functions and Elastic Form Factors. They also rely on the power behavior of GPDs in the $x \to 1$ limit and the propounded analyticity property of Mellin moments of GPDs. We evaluate Compton Form Factors (CFFs), the sub-amplitudes of the Deeply Virtual Compton Scattering (DVCS) process, at the leading order and leading twist accuracy. We constrain the restricted number of free parameters of these new parameterizations in a global CFF analysis of almost all existing proton DVCS measurements. The fit is performed within the PARTONS framework, being the modern tool for generic GPD studies. A distinctive feature of this CFF fit is the careful propagation of uncertainties based on the replica method. The fit results genuinely permit nucleon tomography and may give some insight into the distribution of forces acting on partons. 

\keywords{Nucleon Structure \and Generalized Parton Distribution \and GPD \and Skewness Function \and Border Function \and Analyticity \and Power Behavior \and Deeply Virtual Compton Scattering \and DVCS \and PARTONS Framework \and Global Fit \and Nucleon Tomography \and Subtraction Constant}
\PACS{12.38.-t \and 13.60.-r \and 13.60.Fz \and 14.20.-c}

\end{abstract}

\section{Introduction}

Fifty years after the discovery of quarks at SLAC \cite{Bloom:1969kc, Breidenbach:1969kd}, understanding of how partons form a complex object such as the nucleon still remains among the main challenges of nuclear and high energy physics. In the last twenty years we have witnessed a new liveliness in the field of QCD approaches to this problem due to the discovery of Generalized Parton Distributions (GPDs) \cite{Mueller:1998fv, Ji:1996ek, Ji:1996nm, Radyushkin:1996ru, Radyushkin:1997ki}. GPDs draw so much attention because of the wealth of new information they contain. Namely, GPDs allow for the so-called nucleon tomography \cite{Burkardt:2000za, Burkardt:2002hr, Burkardt:2004bv}, which is used to study a spacial distribution of partons in the plane perpendicular to the nucleon motion as a function of parton longitudinal momenta. Before, positions and longitudinal momenta of partons were studied without any connection through other yet less complex non-perturbative QCD objects: Elastic Form Factors (EFFs) and Parton Distribution Functions (PDFs). In addition, GPDs have another unique property, namely they are connected to the QCD energy-momentum tensor of the nucleon. This allows for an evaluation of the contribution of orbital angular momentum of quarks to the nucleon spin through the so-called Ji's sum rule \cite{Ji:1996ek, Ji:1996nm}. This energy-momentum tensor may also help to define ``mechanical properties'' and describe the distribution of forces inside the nucleon \cite{Goeke:2007fp,Polyakov:2018zvc}.

It was recognized from the beginning that Deeply Virtual Compton Scattering (DVCS) is one of the cleanest probes of GPDs. The first measurements of DVCS by HERMES \cite{Airapetian:2001yk} at DESY and by CLAS \cite{Stepanyan:2001sm} at JLab have proved the usability of GPD formalism to interpret existing measurements, and have established a global experimental campaign for GPDs. Indeed, nowadays measurements of exclusive processes are among the main goals of experimental programs carried out worldwide by a new generation of experiments -- those already running, like Hall A and CLAS at JLab upgraded to $12\ \GeV$ and COMPASS-II at CERN, and those foreseen in the future, like Electron Ion Collider (EIC) and Large Hadron Electron Collider (LHeC). Such a vivid experimental status is complemented by a significant progress in the theoretical description of DVCS. In particular, such new developments like NLO \cite{Ji:1997nk, Ji:1998xh, Mankiewicz:1997bk, Belitsky:1999sg, Freund:2001hm, Freund:2001rk, Freund:2001hd, Pire:2011st}, finite-$t$ and mass \cite{Braun:2012hq} corrections are now available. Except DVCS, a variety of other exclusive processes has been described to provide access to GPDs, in particular: Timelike Compton Scattering \cite{Berger:2001xd}, Deeply Virtual Meson Production \cite{Favart:2015umi}, Heavy Vector Meson Production \cite{Ivanov:2004vd}, Double Deeply Virtual Compton Scattering\cite{Guidal:2002kt,Belitsky:2002tf}, two particles \cite{Boussarie:2016qop, Pedrak:2017cpp} and neutrino induced exclusive reactions \cite{Kopeliovich:2012dr, Pire:2017tvv, Pire:2017lfj}. For some of those processes experimental data have been already collected, while other processes are expected to be probed in the future. 

The phenomenology of GPDs is much more involved than that of EFFs and PDFs. It comes from the fact that GPDs are functions of three variables, entering observables in nontrivial convolutions with coefficient functions. In addition, GPDs are separately defined for each possible combination of parton and nucleon helicities, resulting in a plenitude of objects to be constrained at the same time. This fully justifies the need for a global analysis, where a variety of observables coming from experiments covering complementary kinematic ranges is simultaneously analyzed. So far, such analyzes have been done mainly for Compton Form Factors (CFFs), being DVCS sub-amplitudes and the most basic GPD-sensitive quantities as one can unambiguously extract from the experimental data. Recent analyzes include local fits \cite{Dupre:2016mai, Burkert:2018bqq}, where CFFs are independently extracted in each available bin of data, and global fits \cite{Kumericki:2015lhb}, where CFFs parameterizations are constrained in the whole available phase-space. For a review of DVCS phenomenology we direct the reader to Ref. \cite{Kumericki:2016ehc}.

The aim of this analysis is the global extraction of CFFs from the available proton DVCS data obtained by Hall A, CLAS, HERMES and COMPASS experiments. We use the fixed-$t$ dispersion relation technique \cite{Teryaev:2005uj} for the evaluation of CFFs at the Leading Order (LO) and Leading Twist (LT) accuracy. For a given CFF, the dispersion relation together with the analytical regularization techniques requires two components: \emph{i}) the GPD at $\xi = 0$, and \emph{ii}) the skewness ratio at $x=\xi$. Ans\"atze for those two quantities  proposed in our analysis accumulate information encoded in available PDF and EFF parameterizations, and use theory developments like the $x\to 1$ behavior of GPDs \cite{Yuan:2003fs}. They allow to determine a border function \cite{Radyushkin:2011dh, Radyushkin:2012gba}, being a GPD of reduced kinematic dependency $x = \xi$, and the subtraction constant, directly related to the energy-momentum tensor of the nucleon.

Our original approach allows to utilize many basic properties of GPDs at the level of CFFs fits. We analyze PDFs, but also EFF and DVCS data, that is, we combine information coming from (semi-)inclusive, elastic and exclusive measurements. The analysis is characterized by a careful propagation of uncertainties coming from all those sources, which we achieved with the replica method. Obtained results allow for nucleon tomography, while the extracted subtraction constant may give some insight into the distribution of forces acting on partons inside the nucleon. 

This work is done with PARTONS \cite{Berthou:2015oaw} that is the open-source software framework for the phenomenology of GPDs. It serves not only as the main component of the fit machinery, but it is also utilized to handle multithreading computations and MySQL databases to store and retrieve experimental data. PARTONS is also used for the purpose of comparing existing models with the results of this analysis.

This paper is organized as follows. Section \ref{sec:theory} is a brief introduction of GPDs, DVCS and related observables, with details on the evaluation of CFFs given in Sec. \ref{sec:cff}. Ans\"atze for the border and skewness functions are introduced in Sec. \ref{sec:ansatz}. Sections \ref{sec:selection_of_pdfs} and \ref{sec:elasticFF} summarize our analyses of PDFs and EFFs, respectively. DVCS data used in this work are specified in Sec. \ref{sec:dvcs_data}. In Sec. \ref{sec:uncertainties} the propagation of uncertainties is discussed, while the results are given in Sec. \ref{sec:results}. In Sec. \ref{sec:summary} we summarize the content of this paper.

\section{Theoretical framework}
\label{sec:theory}

In this section a brief introduction to the GPD formalism is given. We emphasize the role of quark GPDs, as only those contribute to DVCS at LO. A deep understanding of the basic features of the contributing GPDs is crucial for constructing parameterizations of CFFs. More involved tools, like nucleon tomography, are important for the exploration of the partonic structure of the proton. This section also provides a foundation to DVCS description and illustrates the construction of observables used in our fits. 

For brevity, we suppress in the following the dependence on the factorization and renormalization scales, $\mu_{R}^{2}$ and  $\mu_{F}^{2}$, which in this analysis are identified with the hard scale of the process $Q^{2}$. A detailed preface to the GPD formalism may be found in one of available reviews \cite{Belitsky:2005qn, Diehl:2003ny, Ji:2004gf, Boffi:2007yc}. 

\subsubsection*{Generalized Parton Distributions}

In the following we use the convention for the light cone vectors as in Ref. \cite{Diehl:2003ny}. In the light cone gauge, quark GPDs for a spin-\nicefrac{1}{2} hadron are defined by the following matrix elements:
\begin{flalign}
&F^q (x, \xi, t) = \frac{1}{2} \int \frac{\mathrm{d} z^-}{2 \pi} \, e^{i x P^+ z^-} \times \nonumber \\
&\bra{P+\frac{\Delta}{2}} \bar{q}\left(-\frac{z}{2}\right)\gamma^+ q\left(\frac{z}{2}\right) \ket{P-\frac{\Delta}{2}}\Big|_{\genfrac{}{}{0pt}{}{z^+=0}{z_{\perp}=0}} \;, \\
&\widetilde{F}^q (x, \xi, t) = \frac{1}{2} \int \frac{\mathrm{d} z^-}{2 \pi}\: e^{i x P^+ z^-} \times \nonumber \\
&\bra{P+\frac{\Delta}{2}} \bar{q}\left(-\frac{z}{2}\right)\gamma^+ \gamma_5 q\left(\frac{z}{2}\right) \ket{P-\frac{\Delta}{2}}\Big|_{\genfrac{}{}{0pt}{}{z^+=0}{z_{\perp}=0}} \;. 
\end{flalign}
Here, $x$ is the average longitudinal momentum of the active quark, $\xi = -\Delta^+/(2P^+)$ is the skewness variable and $t = \Delta^{2}$ is the square of four-momentum transfer to the hadron target, with the average hadron momentum $P$ obeying $P^{2} = m^{2} - t/4$, where $m$ is the hadron mass. In this definition the usual convention is used, where the plus-component refers to the projection of any four-vector on a light-like vector $n$. 

With the help of the Dirac spinor bilinears: 
\begin{eqnarray}
h^\mu & = & \bar{u}\left(P + \frac{\Delta}{2}\right) \gamma^\mu u\left(P - \frac{\Delta}{2}\right) \;, \\
e^\mu & = & \frac{i\Delta_\nu}{2 m}\bar{u}\left(P + \frac{\Delta}{2}\right)  \sigma^{\mu\nu} u\left(P - \frac{\Delta}{2}\right) \;, \\
\tilde{h}^\mu & = & \bar{u}\left(P + \frac{\Delta}{2}\right) \gamma^\mu \gamma_5 u\left(P - \frac{\Delta}{2}\right) \;, \\
\tilde{e}^\mu & = & \frac{\Delta^{\mu}}{2 m} \bar{u}\left(P + \frac{\Delta}{2}\right) \gamma_5 u\left(P - \frac{\Delta}{2}\right) \;, 
\end{eqnarray}
which are normalized so that $\bar{u}(p)\gamma^\mu u(p) = 2 p^\mu$, one can decompose $F^{q}$ and $\widetilde{F}^{q}$ into two pairs of chiral-even GPDs:
\begin{flalign}
&F^q(x, \xi, t) = \nonumber \\
&\phantom{xx}\frac{1}{\displaystyle 2 P^+} \big(h^+ H^{q}(x, \xi, t) + e^+ E^{q}(x, \xi, t)\big) \;, \\
&\widetilde{F}^q(x, \xi, t) = \nonumber \\
&\phantom{xx}\frac{1}{\displaystyle 2 P^+} \big(\tilde{h}^+ \widetilde{H}^{q}(x, \xi, t) + \tilde{e}^+ \widetilde{E}^{q}(x, \xi, t)\big) \;, 
\end{flalign}
recognized as two ``unpolarized'' GPDs $H^{q}$ and $E^{q}$, and two ``polarized'' GPDs $\widetilde{H}^{q}$ and $\widetilde{E}^{q}$.

The relation with one-dimensional PDFs and EFFs is essential for the phenomenology of GPDs. In the forward limit of $\xi = t = 0$, when both the hadron and the active quark are untouched, certain GPDs reduce to (one-dimensional) PDFs:
\begin{eqnarray}
H^{q}(x, 0, 0) & \equiv & q(x) \;, \\
\widetilde{H}^{q}(x, 0, 0) & \equiv & \Delta q(x) \;,
\end{eqnarray}
where $q(x)$ and $\Delta q(x)$ are the unpolarized and polarized PDFs, respectively. No similar relations exist for the GPDs $E^{q}$ and $\widetilde{E}^{q}$ that decouple from the forward limit. The relation to EFFs can be obtained by integrating GPDs over the partonic variable $x$:
\begin{align}
\int_{-1}^{\phantom{.}1}\mathrm{d}x H^{q}(x, \xi, t) &\equiv F_{1}^{q}(t) \;, \label{eq:theory:ff1} \\ 
\int_{-1}^{\phantom{.}1}\mathrm{d}x E^{q}(x, \xi, t) &\equiv F_{2}^{q}(t) \;, \label{eq:theory:ff2} \\
\int_{-1}^{\phantom{.}1}\mathrm{d}x \widetilde{H}^{q}(x, \xi, t) &\equiv g_{A}^{q}(t) \;, \label{eq:theory:ff3} \\
\int_{-1}^{\phantom{.}1}\mathrm{d}x \widetilde{E}^{q}(x, \xi, t) &\equiv g_{P}^{q}(t) \;, \label{eq:theory:ff4}
\end{align}
where $F_{1}^{q}(t)$, $F_{2}^{q}(t)$, $g_{A}^{q}(t)$ and $g_{P}^{q}(t)$ are the contribution of the quark flavor $q$ to the Dirac, Pauli, axial and pseudoscalar EFFs, respectively.

The integrals in Eqs. \eqref{eq:theory:ff1}-\eqref{eq:theory:ff4} do not depend on $\xi$ as a consequence of the Lorentz covariance of GPDs. This feature is generally expressed by a non-trivial property of GPDs known as polynomiality. The property states, that any $n$-th Mellin moment of a given GPD is always an even polynomial in $\xi$, of order $n+1$ for the unpolarized GPDs and of order $n$ for the polarized GPDs. In particular:
\begin{flalign}
&\int_{-1}^{\phantom{.}1}\mathrm{d}x~ x^{n} H^{q}(x, \xi, t) = h_{0}^{q, n}(t) + \nonumber \\ 
&\phantom{xx}\xi^{2}h_{2}^{q, n}(t) + \hdots + \textrm{mod}(n,2)\xi^{n+1}h_{n+1}^{q, n}(t) \;, \\
&\int_{-1}^{\phantom{.}1}\mathrm{d}x~ x^{n} \widetilde{H}^{q}(x, \xi, t) = \tilde{h}_{0}^{q, n}(t) + \nonumber \\
&\phantom{xx}\xi^{2}\tilde{h}_{2}^{q, n}(t) + \hdots + \textrm{mod}(n+1,2)\xi^{n}\tilde{h}_{n}^{q, n}(t) \;,
\end{flalign}
where for $n=0$ one has the relations given by Eqs. \eqref{eq:theory:ff1}-\eqref{eq:theory:ff4}. 

The correspondence of GPDs to PDFs and EFFs presages a possibility of studying a spatial distribution of partons inside the nucleon. Indeed, the subfield of hadron structure studies known as nucleon tomography allows one to extract the density of partons carrying a given fraction of the nucleon longitudinal momentum $x$ as a function of the position $\mathbf{b}_{\perp}$ in the plane perpendicular to the nucleon motion. For unpolarized partons inside an unpolarized nucleon this density is expressed by: 
\begin{flalign}
&q(x, \mathbf{b}_{\mathbf{\perp}}) = \nonumber \\ 
&\phantom{xx}\int \frac{\mathrm{d}^{2}\mathbf{\Delta}}{4\pi^{2}} e^{-i \mathbf{b_{\perp}} \cdot \mathbf{\Delta}} H^{q}(x, 0, t=-\mathbf{\Delta}^{2}) \;,
\label{eq:theory:nt_H}
\end{flalign}
where we stress the condition $\xi=0$, meaning no change of the longitudinal momentum of the active parton. This density gets distorted when the nucleon is polarized. This effect is described by adding extra terms related to the GPDs $\widetilde{H}$ and $E$. The longitudinal polarization of partons distributed  in a longitudinally polarized nucleon according to $q(x, \mathbf{b}_{\mathbf{\perp}})$ can be studied with the Fourier transform of GPD $\widetilde{H}$:
\begin{flalign}
&\Delta q(x, \mathbf{b}_{\mathbf{\perp}}) = \nonumber \\
&\phantom{xx}\int \frac{\mathrm{d}^{2}\mathbf{\Delta}}{4\pi^{2}} e^{-i \mathbf{b_{\perp}} \cdot \mathbf{\Delta}} \widetilde{H}^{q}(x, 0, t=-\mathbf{\Delta}^{2}) \;.
\label{eq:theory:nt_Ht}
\end{flalign}
A representation in the impact parameter space is also possible for the GPD $E$:
\begin{flalign}
&e^{q}(x, \mathbf{b}_{\mathbf{\perp}}) = \nonumber \\ 
&\phantom{xx}\int \frac{\mathrm{d}^{2}\mathbf{\Delta}}{4\pi^{2}} e^{-i \mathbf{b_{\perp}} \cdot \mathbf{\Delta}} E^{q}(x, 0, t=-\mathbf{\Delta}^{2}) \;.
\end{flalign}
A probabilistic interpretation of that result is possible if one changes the basis from longitudinal to transverse polarization states of the nucleon \cite{Burkardt:2002hr}. In such a case, $e^{q}(x, \mathbf{b}_{\mathbf{\perp}})$ can be related to a shift of parton density generated in a transversely polarized nucleon. 

As indicated in Refs. \cite{Burkardt:2002hr, Diehl:2004cx}, $\mathbf{b}_{\mathbf{\perp}}$ is the distance between the active parton and the point determined by the positions of individual partons weighted by their momenta, so that $\sum x_{i} \mathbf{b}_{\mathbf{\perp}, i} = 0$, where the sum runs over all partons (the struck parton and all spectators as well). This distance is different than that between the struck parton and the spectator system, which is given by:
\begin{equation}
d_{\perp}(x) = \frac{\left|\mathbf{b}_{\mathbf{\perp}}\right|}{1-x} \;,
\end{equation}
and the parton distribution given as a function of $d_{\perp}$ provides a better estimation of the transverse proton size than $q(x, \mathbf{b}_{\mathbf{\perp}})$. 

Another useful quantity appearing in the context of the nucleon tomography is the normalized second moment of $q(x, \mathbf{b}_{\mathbf{\perp}})$ distribution given by:
\begin{equation}
{\langle b_{\perp}^{2} \rangle}_{q}(x) = 
\frac{
\displaystyle
\int d^{2}\mathbf{b}_{\perp} ~\mathbf{b}_{\perp}^2 q(x, \mathbf{b}_{\perp})
}{
\displaystyle
\int d^{2}\mathbf{b}_{\perp} ~q(x, \mathbf{b}_{\perp})
} \, . 
\label{eq:theory:distance_to_center_unpol}
\end{equation} 
A similar quantity can be also defined for the distribution of longitudinal polarization to check how broad this distribution is and how it corresponds to ${\langle b_{\perp}^{2} \rangle}_{q}(x)$, 
\begin{equation}
{\langle b_{\perp}^{2} \rangle}_{\Delta q}(x) = 
\frac{
\displaystyle
\int d^{2}\mathbf{b}_{\perp} ~\mathbf{b}_{\perp}^2 \Delta q(x, \mathbf{b}_{\perp})
}{
\displaystyle
\int d^{2}\mathbf{b}_{\perp} ~\Delta q(x, \mathbf{b}_{\perp})
} \, .
\label{eq:theory:distance_to_center_pol}
\end{equation}
The need of having the proton size finite requires to keep the mean squared distance between the active parton and the spectator system,
\begin{equation}
\langle d_{\perp}^{2} \rangle_{q}(x) = \frac{{\langle b_{\perp}^{2} \rangle}_{q}(x)}{\left(1-x\right)^{2}} \;,
\label{eq:theory:distance_to_spectator}
\end{equation}
finite as well, also in the limit of $x \to 1$. We will impose it for the valance quarks as an extra constraint on our Ansatz introduced in Sec. \ref{sec:ansatz}. 

To avoid a violation of the positivity of parton densities in the impact parameter space, inequalities studied in a series of papers \cite{Pire:1998nw, Diehl:2000xz, Pobylitsa:2001nt, Pobylitsa:2002gw, Pobylitsa:2002iu, Pobylitsa:2002vi, Pobylitsa:2002vw, Pobylitsa:2002ru} must hold. In particular one has the following inequalities, which are proved to be useful to constrain parameterizations of GPDs: 
\begin{flalign}
&\left| \Delta q(x, \mathbf{b}_{\mathbf{\perp}}) \right| \leq q(x, \mathbf{b}_{\mathbf{\perp}}) \;, \label{eq:theory:ineq1} \\
&\frac{\mathbf{b}_{\mathbf{\perp}}^{2}}{m^{2}} \left( \frac{\partial}{\partial \mathbf{b}_{\mathbf{\perp}}^{2}} e(x, \mathbf{b}_{\mathbf{\perp}}) \right)^{2} \leq \left( q(x, \mathbf{b}_{\mathbf{\perp}}) + \Delta q(x, \mathbf{b}_{\mathbf{\perp}})\right) \times \nonumber \\
&\phantom{xx}\left( q(x, \mathbf{b}_{\mathbf{\perp}}) - \Delta q(x, \mathbf{b}_{\mathbf{\perp}})\right) \;. \label{eq:theory:ineq2}
\end{flalign}

For completeness we also show Ji's sum rule, allowing for the evaluation of total angular momentum carried by partons:
\begin{eqnarray}
\int_{-1}^{\phantom{.}1}\mathrm{d}x~x(H^{q}(x, \xi, 0) + E^{q}(x, \xi, 0)) = 2 J^{q} \;.
\end{eqnarray}
This feature can be used to investigate the nucleon spin decomposition. We note however, that this analysis concentrates on GPDs $H$ and $\widetilde{H}$, and therefore we will not attempt to give any estimation on $J^{q}$. 

\subsubsection*{Deeply Virtual Compton Scattering}

\begin{figure}
\begin{center}
\includegraphics[scale=0.7]{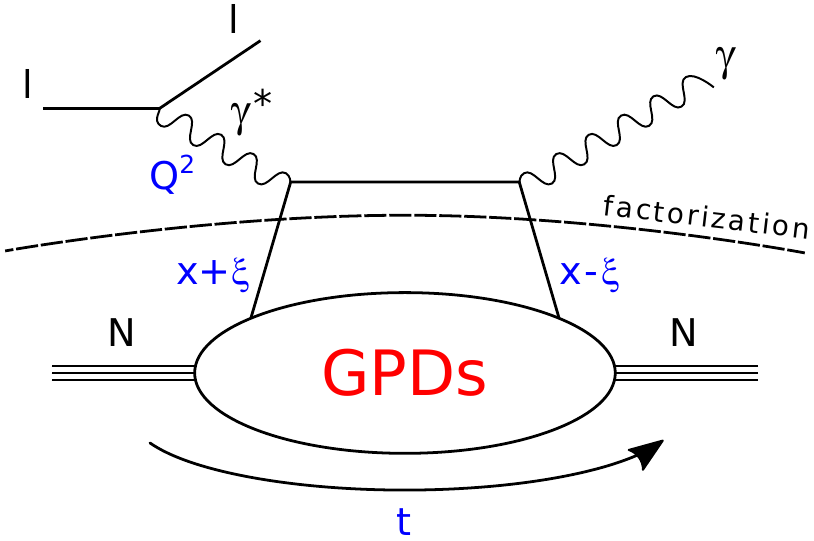}
\caption{Partonic interpretation of the DVCS process.}
\label{fig:theory:partonic-interpretation-dvcs}
\end{center}
\end{figure}

A prominent role in the GPD phenomenology is played by Deeply Virtual Compton Scattering: 
\begin{equation}
l(k) + N(p) \rightarrow l(k') + N(p') + \gamma(q') \;,
\end{equation}
where $l$, $N$ and $\gamma$ denote lepton, nucleon and produced photon, respectively; the four-vectors of these states appear between parenthesis. Under specific kinematic conditions, the factorization theorem allows one to express the DVCS amplitude as a convolution of the hard scattering part, being calculable within the perturbative QCD approach, and GPDs, describing an emission of parton from the nucleon and its subsequent reabsorption, see Fig. \ref{fig:theory:partonic-interpretation-dvcs}. The factorization applies in the Bjorken limit and for $-t/Q^{2} \ll 1$, where $Q^{2}=-(k-k')^2$ is the virtuality of the virtual-photon mediating the exchange of four-momentum between lepton and proton at Born order. 

\begin{figure}[h]
\begin{center}
\includegraphics[scale=0.7]{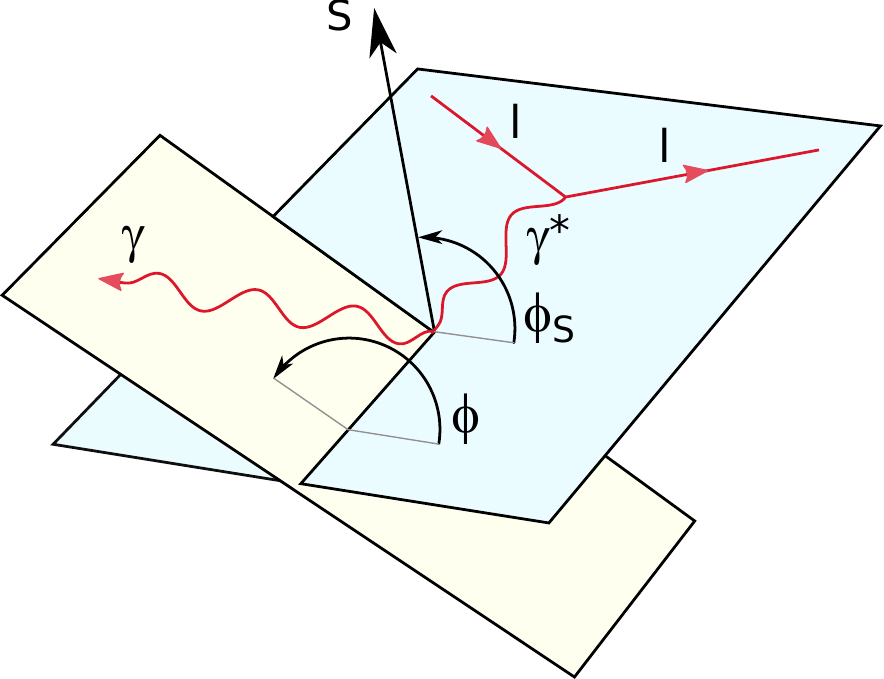}
\end{center}
\caption{Kinematics of DVCS in the target rest frame. The angle between the leptonic plane (spanned by the incoming and outgoing lepton momenta) and the production plane (spanned by the virtual and outgoing photon momenta) is denoted by $\phi$. The angle between the leptonic plane and the nucleon polarization vector is denoted by $\phi_S$.}
\label{fig:theory:angles-dvcs}
\end{figure}

The scattering is described by two angles, see Fig. \ref{fig:theory:angles-dvcs}. These are $\phi$, being the angle between the lepton scattering and production planes, and $\phi_{S}$, being the angle between the lepton scattering plane and the direction of transversely polarized target. 

An electromagnetic process called Bethe-Heitler has the same initial and final states as DVCS. The total amplitude for single photon production, $\mathcal{T}$, is then expressed by a sum of amplitudes for BH and DVCS processes, which leads to:
\begin{equation}
|\mathcal{T}|^{2} =
|\mathcal{T}_{\mathrm{BH}} + \mathcal{T}_{\mathrm{DVCS}}|^{2} = 
|\mathcal{T}_{\mathrm{BH}}|^{2} + |\mathcal{T}_{\mathrm{DVCS}}|^{2} + \mathcal{I} \;,
\end{equation}
where $\mathcal{I}$ denotes the interference term. The cross section for BH is calculable to a high degree of precision and therefore can be easily taken into account in analyses of experimental data. The interference term provides a complementary information to the pure DVCS cross section, and in a certain kinematic domain allows to access GPDs, even if $|\mathcal{T}_{\mathrm{DVCS}}|^{2}$ is small. 

The amplitudes $\mathcal{T}_{\mathrm{DVCS}}$ and $\mathcal{I}$ may be expressed by combinations of CFFs, which are convolutions of GPDs with the hard scattering part of the interaction. CFFs are the most basic quantities that one can unambiguously extract from the experimental data. The way of how CFFs enter the final amplitudes depends on the beam and target helicity states, which provides a welcome experimental filter to distinguish between many possible CFFs and justifies the need of measuring many observables. For brevity we skip the formulas showing how $\mathcal{T}_{\mathrm{DVCS}}$ and $\mathcal{I}$ depend on CFFs. They can be found in Ref. \cite{Belitsky:2012ch}. The evaluation of CFFs is discussed in Sec. \ref{sec:cff}.

\subsubsection*{Observables}

Let us denote a four-fold differential cross section for a single photon production by $d^{4}\sigma_{t}^{b, c}(\xBj, t, Q^{2}, \phi)$, where $t \in \{\leftarrow, \rightarrow\}$ and $b \in \{\leftarrow, \rightarrow\}$ stand for the target and beam helicities, respectively, and $c \in \{+,-\}$ stands for the beam charge. Here, $\xBj = Q^2/(2p \cdot \big(p'-p)\big)$ is the usual Bjorken variable. The cross sections can be used to construct many observables, like cross sections itself, but also differences of cross sections and asymmetries. For instance:
\begin{flalign}
&d^{4} \sigma_{UU}^{-}(\xBj, t, Q^{2}, \phi) = \nicefrac{1}{4} \Big( \nonumber \\
&\phantom{xx}\left( d^{4}\sigma_{\leftarrow}^{\rightarrow, -}(\xBj, t, Q^{2}, \phi) + d^{4}\sigma_{\rightarrow}^{\rightarrow, -}(\xBj, t, Q^{2}, \phi) \right) + \nonumber \\
&\phantom{xx}\left( d^{4}\sigma_{\leftarrow}^{\leftarrow, -}(\xBj, t, Q^{2}, \phi) + d^{4}\sigma_{\rightarrow}^{\leftarrow, -}(\xBj, t, Q^{2}, \phi) \right) \nonumber \\
&\phantom{x}\Big) \;, \\
&\Delta d^{4} \sigma_{LU}^{-}(\xBj, t, Q^{2}, \phi) = \nicefrac{1}{4} \Big( \nonumber \\
&\phantom{xx}\left( d^{4}\sigma_{\leftarrow}^{\rightarrow, -}(\xBj, t, Q^{2}, \phi) + d^{4}\sigma_{\rightarrow}^{\rightarrow, -}(\xBj, t, Q^{2}, \phi) \right) - \nonumber \\
&\phantom{xx}\left( d^{4}\sigma_{\leftarrow}^{\leftarrow, -}(\xBj, t, Q^{2}, \phi) + d^{4}\sigma_{\rightarrow}^{\leftarrow, -}(\xBj, t, Q^{2}, \phi) \right) \nonumber \\
&\phantom{x}\Big) \;, \\
&A_{LU}^{-}(\xBj, t, Q^{2}, \phi) = \frac{\Delta d^{4} \sigma_{LU}^{-}(\xBj, t, Q^{2}, \phi)}{d^{4} \sigma_{UU}^{-}(\xBj, t, Q^{2}, \phi)} \;.
\end{flalign}
Here, the capital letters in the subscripts of observables names denote beam and target polarizations, respectively, with $U$ standing for ``Unpolarized'' and $L$ standing for ``Longitudinally polarized''. We also analyze data for ``Transversely polarized targets'', which are distinguished by the subscript $T$. These data are provided for two moments, $\sin(\phi-\phi_{S})$ and $\cos(\phi-\phi_{S})$, which are distinguished by the corresponding labels in the superscripts, as for instance in $A^{-, \sin(\phi-\phi_{S})}(\xBj, t, Q^{2}, \phi)$. Furthermore there are observables probing only the beam charge dependency (subscript $C$), and those combining cross sections measured with various beam charges to drop either the DVCS or interference contribution (subscripts $\mathrm{I}$ and $\mathrm{DVCS}$, respectively). 

A different group consists of Fourier-like observables related to specific modulations of the $\phi$ angle. For instance:
\begin{flalign}
&A_{C}^{-, \cos 0}(\xBj, t, Q^{2}) = \nonumber \\
&\phantom{xx}\frac{1}{2\pi} \int \mathrm{d}\phi A_{C}^{-}(\xBj, t, Q^{2}, \phi) \;,\\ 
&A_{LU}^{-, \sin \phi}(\xBj, t, Q^{2}) = \nonumber \\
&\phantom{xx}\frac{1}{\pi} \int \mathrm{d}\phi A_{LU}^{-}(\xBj, t, Q^{2}, \phi) \sin \phi \;.
\end{flalign}

Another observable used in this analysis is the slope $b(\xBj, Q^{2})$ of the $t$-distribution of the DVCS cross section integrated over $\phi$. Within the LO formalism one can relate this observable to the transverse extension of partons in the proton. However, this requires the following assumptions, which are expected to hold at small $\xBj$: \emph{i}) dominance of the imaginary part of CFF related to GPD $H$, \emph{ii}) negligible skewness effect at $\xi = x$, \emph{iii}) exponential $t$ dependence of the GPD $H$ at fixed $x$. Since the DVCS $t$-distribution is usually not exactly exponential, in particular because GPDs for valence and sea quarks may have different $t$-dependencies, we evaluate $b(\xBj, Q^{2})$ by probing DVCS $t$-distribution in several equidistant points ranging from $|t| = 0.1~\mathrm{GeV}^{2}$ to $|t| = 0.5~\mathrm{GeV}^{2}$ and perform a linear regression on the logarithmized results. The chosen range of $t$ is typical for the existing measurements of $b(\xBj, Q^{2})$ by COMPASS \cite{Akhunzyanov:2018nut} and HERA experiments \cite{Aktas:2005ty, Chekanov:2008vy, Aaron:2009ac}.

\section{Compton Form Factors}
\label{sec:cff}

\subsubsection*{Imaginary part}

At LO the imaginary part of a given CFF $\mathcal{G} \in \{ \mathcal{H}, \mathcal{E}, \widetilde{\mathcal{H}}, \widetilde{\mathcal{E}} \}$, is proportional to the combination of corresponding GPDs, $G^{q} \in \{ H^{q}, E^{q}, \widetilde{H}^{q}, \widetilde{E}^{q} \}$, probed at $x = \xi$: 
\begin{align}
\mathit{Im} \mathcal{G}(\xi, t) & = \pi G^{(+)}(\xi, \xi, t) \nonumber \\
& = \pi \sum_{q} e_{q}^{2} G^{q (+)}(\xi, \xi, t) \;.
\label{eq:cff:im}
\end{align}
Here, the sum runs over all quark flavors (we remind that at LO gluons do not contribute to the DVCS amplitude), $e_{q}$ is the electric charge of the quark flavor $q$ in units of the positron charge $e$ and $G^{q (+)}$ is the singlet ($C$-even) combination of GPDs: 
\begin{align}
G^{q (+)}(x, \xi, t) = G^{q}(x, \xi, t) - G^{q}(-x, \xi, t)
\label{eq:cff:singlet:Unpol}
\end{align}
for $G \in \{H, E\}$ and:
\begin{align}
G^{q (+)}(x, \xi, t) = G^{q}(x, \xi, t) + G^{q}(-x, \xi, t)
\label{eq:cff:singlet:Pol}
\end{align}
for $G \in \{\widetilde{H}, \widetilde{E}\}$. 

\subsubsection*{Real part}

At LO the real part of a given CFF $\mathcal{G}$ can be evaluated by probing the corresponding GPD $G$ in two ways, that is, by integrating over one of two lines laying in the ($x, \xi$)-plane. This duality is a consequence of the polynomiality property required by the Lorentz invariance of GPDs, see Sec. \ref{sec:theory}. 

The first evaluating method is the "standard" one, where $x$ values of the involved GPDs are probed at fixed $\xi$:
\begin{flalign}
&\mathit{Re} \mathcal{G}(\xi, t) = \nonumber \\
&\phantom{xx}\mathrm{P.V.} \int_{0}^{1} G^{(+)}(x, \xi, t) \left( \frac{1}{\xi - x} \mp \frac{1}{\xi + x}\right) \mathrm{d}x \;. 
\label{eq:cff:standard}
\end{flalign}
Here, the quark propagators, $1/(\xi-x)$ and $1/(\xi+x)$, enter in a combination given by the type of probed GPDs. One has the difference (sum) for the unpolarized (polarized) GPDs.

The second evaluating method is known as fixed-$t$ dispersion relation \cite{Teryaev:2005uj} and it involves the integral probing GPDs at $\xi = x$: 
\begin{flalign}
&\mathit{Re} \mathcal{G}(\xi, t) = C_{G}(t) +\nonumber \\
&\phantom{xx}\mathrm{P.V.} \int_{0}^{1} G^{(+)}(x, x, t) \left( \frac{1}{\xi - x} \mp \frac{1}{\xi + x}\right) \mathrm{d}x \;.
\label{eq:cff:dispersion-relation}
\end{flalign}
Again, the combination of quark propagators depends here on the type of probed GPDs, exactly as for Eq. \eqref{eq:cff:standard}. The additional term in Eq. \eqref{eq:cff:dispersion-relation}, $C_{G}(t)$, is the so-called subtraction constant. It has the same magnitude but the opposite sign for the CFFs $\mathcal{H}$ and $\mathcal{E}$, and it vanishes for the CFFs $\widetilde{\mathcal{H}}$ and $\widetilde{\mathcal{E}}$:  
\begin{gather}
C_{H}(t) = - C_{E}(t) \;,\\
C_{\widetilde{H}}(t) = C_{\widetilde{E}}(t) = 0 \;.
\label{eq:theory:sc-vs-GPD-type}
\end{gather}

After a quick examination of Eqs. \eqref{eq:cff:im} and \eqref{eq:cff:dispersion-relation}, one may notice that the dispersion relation provides a welcome relationship between the real and imaginary parts of the same CFF. As a consequence however, at the LO approximation only GPDs in the limited case of $x = \xi$ and the subtraction constant can be probed.

\subsubsection*{Subtraction constant}

The subtraction constant introduced in Eq. \eqref{eq:cff:dispersion-relation} can be related to $D$-term form factor, $D^{q}(t)$, in the following way:
\begin{equation}
C_{G}^{q}(t) = 2 \int_{-1}^{1} \frac{D^{q}(z, t)}{1-z} \mathrm{d}z \equiv 4 D^{q}(t) \;.
\end{equation}
Here, $z=x/\xi$ and $D^{q}(z, t)$ is the $D$-term \cite{Polyakov:1999gs}. It was originally introduced to restore the polynomiality property in the first models based on double distributions \cite{Radyushkin:1998bz}, but later it has been recognized as an important element of the GPD phenomenology. Because the $D$-term vanishes outside the ERBL region $|x| < |\xi|$, it is not observed in the limit of $\xi=0$, and it can be only studied in the "skewed" case of $\xi \neq 0$.  

By expanding the $D$-term in terms of Gegenbauer polynomials, 
\begin{equation}
D^{q}(z, t) = (1-z^2)\sum_{\substack{i=1 \\ \mathrm{odd}}}^{\infty}d_{i}^{q}(t)C_{i}^{3/2}(z) \;, 
\end{equation}
one can obtain the following series:
\begin{equation}
D^{q}(t) = \sum_{\substack{i=1 \\ \mathrm{odd}}}^{\infty}d_{i}^{q}(t) \;.
\end{equation}
The first term of this expansion, $d_{1}^{q}(t)$, is of a special importance, as it enters the quark part of the QCD energy momentum tensor and it provides an important information on how strong forces are distributed in the nucleon \cite{Polyakov:2002yz}.

The subtraction constant can be evaluated by comparing Eqs. \eqref{eq:cff:standard} and \eqref{eq:cff:dispersion-relation}: 
\begin{flalign}
&C_{G}^{q}(t) = \nonumber \\
&\phantom{xx}\mathrm{P.V.} \int_{0}^{1} G^{q (+)}(x, \xi, t) 
\left(\frac{1}{\xi - x} - \frac{1}{\xi + x}\right) dx\; - \nonumber \\
&\phantom{xx}\mathrm{P.V.} \int_{0}^{1} G^{q (+)}(x, x, t) 
\left(\frac{1}{\xi - x} - \frac{1}{\xi + x}\right) dx\; ,
\end{flalign}
which can be evaluated without principal value prescription, because the singularity at $x=\xi$ is integrable in the expression:
\begin{flalign}
&C_{G}^{q}(t) = \int_{0}^{1} \left( G^{q (+)}(x, \xi, t) - G^{q (+)}(x, x, t)\right)\times\nonumber \\
&\phantom{xx}\left(\frac{1}{\xi - x} - \frac{1}{\xi + x}\right) dx \; .
\end{flalign}
Unfortunately, naively setting $\xi = 0$ in the above formula results in a divergent integral. However, the following moments:
\begin{flalign}
&C_{G,j}^{q}(t) = \nonumber \\
&\phantom{xx}2\int_{0}^{1} \left( G^{q (+)}(x,x, t) - G^{q (+)}(x, 0, t)\right) x^j dx \; ,
\end{flalign}
are well defined for odd positive $j$ and can be analytically continued to $j=-1$, if $ G^{q (+)}(x,x, t) - G^{q (+)}(x, 0, t)$ has a proper analytic behavior, as described in \cite{Polyakov:2008aa}. Such an analytical continuation can be written as:
\begin{flalign}
&C_{G,j}^{q}(t) = \nonumber \\
&\phantom{xx}2\int_{(0)}^{1} \left( G^{q (+)}(x,x, t) - G^{q (+)}(x, 0, t)\right) x^j dx \; , 
\end{flalign}
where we have introduced the analytic regularization technique \cite{Kumericki:2007sa, Kumericki:2008di, Polyakov:2008aa, Radyushkin:2011dh}, given by the following prescription:
\begin{flalign}
&\int_{(0)}^{1}\frac{f(x)}{x^{a+1}} = \nonumber \\ 
&\phantom{xx}\int_{0}^{1}\frac{f(x) - f(0) - xf'(0) - \dots}{x^{a+1}} + \nonumber \\ 
&\phantom{xx}f(0) \int_{(0)}^{1} \frac{dx}{x^{a+1}} + f'(0) \int_{(0)}^{1} \frac{dx}{x^{a}} + \dots = \nonumber \\ 
&\phantom{xx}\int_{0}^{1}\frac{f(x) - f(0) - xf'(0) - \dots}{x^{a+1}} - \nonumber \\
&\phantom{xx}\frac{f(0)}{a} - \frac{f'(0)}{a-1} + \dots 
\label{eq:theory:analytic-regularization}
\end{flalign}
Here, one subtracts as many terms from the Taylor expansion of $f$ around zero as needed to make the integral convergent, and one treats the compensating terms to be convergent as well.

The analytic properties of the Mellin moments of GPDs has been never proved to be a consequence of general principles (neither it has never been proved to contradict general principles) and because of that can be only treated as a model assumption to be \emph{a posteriori} confronted with experimental data \footnote{Analytical properties of the GPDs are the subject of an ongoing discussion, see for instance \cite{Radyushkin:2011dh, Radyushkin:2012gba, Radyushkin:2013hca, Radyushkin:2013bba, Muller:2014wxa, Muller:2015vha}. In particular Sec. 4 of \cite{Radyushkin:2012gba}  illustrates the relation between analytic regularization and the analyticital properties of GPD.}. We will make such an assumption, and calculate the subtraction constant as:
\begin{flalign}
&C_{G}^{q}(t) =C_{G,-1}^{q}(t)
\nonumber \\ 
&\phantom{xx} = 2\int_{(0)}^{1}\frac{ G^{q (+)}(x,x, t) - G^{q (+)}(x, 0, t)}{x} dx \; .
\label{eq:cff:subtraction_constant_at_xi_eq_0}
\end{flalign}
The self-consistency of this approach will lead us to relations \eqref{eq:singregul1} and \eqref{eq:singregul2} among the otherwise non-related parameters of the fitting model. 

\section{Ansatz}
\label{sec:ansatz}

We present in this study a global extraction of CFFs. According to the terminology used within the GPD community, ``global'' refers to constraining parameters of an assumed CFF functional forms from various measurements on a wide kinematic range. On the contrary in local extractions, CFFs are extracted as a set of disconnected values in bins of $\xi$ and $t$ (see for instance Ref. \cite{Dupre:2016mai, Burkert:2018bqq}). We restrict our analysis to the LO approximation and we neglect any contribution coming from higher-twist effects and kinematic (target mass and finite-$t$) corrections. We adopt the description of cross sections in terms of DVCS and BH amplitudes by Guichon and Vanderhaeghen, used for phenomenology for instance in Refs. \cite{Moutarde:2009fg, Kroll:2012sm} and publicly available in the open-source PARTONS framework \cite{Berthou:2015oaw}. We point out, that the limited phase space covered by available data, the precision of those data and the plenitude of involved dependencies force us to keep the parameterizations as simple as possible. Otherwise significant correlations appear between fitted parameters, which somehow obscures the interpretation of obtained results.

The Ansatz introduced in this section is explicitly given for a factorization scale that one may recognize as the reference scale $Q_{0}^{2}$ at which the model is defined. To include the factorization scale dependence in our fit, that is for the comparison with experimental data of $Q^{2} \neq Q_{0}^{2}$, we consider the so-called forward evolution, \emph{i.e.} the one followed by PDFs. The usage of the genuine GPD evolution equations requires the knowledge of GPDs in the full range of $x$ independently on $\xi$, while in this analysis only the GPDs at $x = \xi$ are considered. It was checked however with the GK GPD model \cite{Goloskokov:2005sd, Goloskokov:2007nt, Goloskokov:2009ia}, that the difference between the two evolution schemes is small for $x = \xi$, unless $Q^{2} \gg Q_{0}^{2}$.

\subsubsection*{Decomposition into valence and sea contributions}

In this work we use the decomposition scheme into valence and sea contributions inspired by the double distribution modeling of GPDs \cite{Radyushkin:1998bz, Goeke:2001tz, Belitsky:2001ns}. It gives us: 
\begin{equation}
G^{q}(\xi, \xi, t) = G^{q_{\mathrm{val}}}(\xi, \xi, t) + G^{q_{\mathrm{sea}}}(\xi, \xi, t) \;,
\end{equation} 
for $x = \xi$ and $G \in \{H, E, \widetilde{H}, \widetilde{E}\}$, 
\begin{equation}
G^{q}(-\xi, \xi, t) = - G^{q_{\mathrm{sea}}}(\xi, \xi, t) \;,
\end{equation} 
for $x = -\xi$ and $G \in \{H, E\}$, and
\begin{equation}
G^{q}(-\xi, \xi, t) = G^{q_{\mathrm{sea}}}(\xi, \xi, t) \;,
\end{equation} 
for $x = -\xi$ and $G \in \{\widetilde{H}, \widetilde{E}\}$. Here, $G^{q_{\mathrm{val}}}$ and $G^{q_{\mathrm{sea}}}$ are GPDs for valence and sea quarks, respectively. With this decomposition one can replace Eqs. \eqref{eq:cff:singlet:Unpol} and \eqref{eq:cff:singlet:Pol} by one equivalent expression:
\begin{equation}
G^{q (+)}(\xi, \xi, t) = G^{q_{\mathrm{val}}}(\xi, \xi, t) + 2 G^{q_{\mathrm{sea}}}(\xi, \xi, t) \;.
\end{equation}

\subsubsection*{CFFs $\mathcal{H}$ and $\widetilde{\mathcal{H}}$}

Data used in this analysis are primarily sensitive to the CFFs $\mathcal{H}$ and $\widetilde{\mathcal{H}}$. The LO and LT formalism allows us to evaluate those CFFs with Eq. \eqref{eq:cff:im} for the imaginary part and with Eq. \eqref{eq:cff:dispersion-relation} for the real part. The subtraction constant, which appears in the dispersion relation, is evaluated with Eq. \eqref{eq:cff:subtraction_constant_at_xi_eq_0}, making use of the analytic regularization prescription given by Eq. \eqref{eq:theory:analytic-regularization}. All together, only the GPDs $H^{q}$ and $\widetilde{H}^{q}$ at $\xi=0$ and $\xi=x$ are needed.  

For the GPDs $H^{q}$ and $\widetilde{H}^{q}$ at $\xi=0$ we use an Ansatz that is commonly used in phenomenological analyses of GPDs:
\begin{equation}
G^{q}(x, 0, t) = \mathrm{pdf}_{G}^{q}(x) ~ \exp(f_{G}^{q}(x)t) \; .
\label{eq:ansatz_xi_eq_0}
\end{equation}
Here, $\mathrm{pdf}_{G}^{q}(x)$ is either a parameterization of the unpolarized PDF, $q(x)$, for the GPD $H^{q}$ or a parameterization of the polarized PDF, $\Delta q(x)$, for the GPD $\widetilde{H}^{q}$. The profile function, $f_{G}^{q}(x)$, fixes the interplay between the $x$ and $t$ variables, and it is given by: 
\begin{equation}
f_{G}^{q}(x) = A_{G}^{q}\log(1/x) + B_{G}^{q}(1-x)^{2} + C_{G}^{q}(1-x)x \;,
\label{eq:ansatz_xt_profile}
\end{equation}
where $A_{G}^{q}$, $B_{G}^{q}$ and $C_{G}^{q}$ are free parameters to be constrained by experimental data. This form of $f_{G}^{q}(x)$ allows to modify the classical $A_{G}^{q}\log(1/x)$ term by $B_{G}^{q}(1-x)^{2}$ in the small $x$ region and by $C_{G}^{q}(1-x)x$ in the high $x$ region. The terms proportional to $B_{G}^{q}$ and $C_{G}^{q}$ were found to work best in the analysis of EFF data (see Sec. \ref{sec:elasticFF}), where all combinations of $(1-x)^{i}$ and $(1-x)^{j}x^{k}$ polynomials with $i, j, k = 1, \dots, 5$ were examined. We note that $A_{G}^{q}\log(1/x)$ can not be directly multiplied by a polynomial of $x$, which is imposed by a need of keeping $G^{q}(x, 0, t) / x^{-(\delta + A_{G}^{q}t)}$ analytic at $x=0$. To keep the distance between the active quark and the spectator system finite, see Eq. \eqref{eq:theory:distance_to_spectator}, we require to have $C_{H}^{q_{\mathrm{val}}} = -A_{H}^{q_{\mathrm{val}}}$.

The profile function given by Eq. \eqref{eq:ansatz_xt_profile} is more flexible than that used in the GK model \cite{Goloskokov:2005sd, Goloskokov:2007nt, Goloskokov:2009ia}, $f_{G, \mathrm{GK}}^{q}(x) = A_{G}^{q} + B_{G}^{q}\log(1/x)$, and in the VGG model \cite{Vanderhaeghen:1998uc, Vanderhaeghen:1999xj, Goeke:2001tz, Guidal:2004nd}, $f_{G, \mathrm{VGG}}^{q}(x) = A_{G}^{q}\log(1/x)(1-x)$. In particular, it should be flexible enough to take into account a different interplay between the $x$ and $t$ variables in the valence and sea regions if required by experimental data. We note that $f_{G, \mathrm{DK}}^{q}(x) = A_{G}^{q}\log(1/x)(1-x)^{3} + B_{G}^{q}(1-x)^{3} + C_{G}^{q}(1-x)^{2}x$ used in Refs. \cite{Diehl:2004cx, Diehl:2013xca} to fit EFF data can not be used in this analysis because of the aforementioned issue with the analyticity caused by $A_{G}^{q}\log(1/x)(1-x)^{3}$ term.  

For GPDs $H^{q}$ and $\widetilde{H}^{q}$ at $\xi=x$ we utilize the concept of skewness function:
\begin{equation}
g_{G}^{q}(x, \xi, t) = \frac{G^{q}(x, \xi, t)}{G^{q}(x, 0, t)} \;.
\end{equation}
In our case: 
\begin{equation}
G^{q}(x, x, t) = G^{q}(x, 0, t) ~ g_{G}^{q}(x, x, t) \;,
\label{eq:ansatz_xi_eq_xi}
\end{equation}
where $G^{q}(x, 0, t)$ is given by Eq. \eqref{eq:ansatz_xi_eq_0}. We assume the following form of the skewness function: 
\begin{flalign}
&g_{G}^{q}(x, x, t) \equiv g_{G}^{q}(x, t) = \nonumber \\
&\phantom{x}\frac{a_{G}^{q}}{(1-x^{2})^{2}}
\left( 1 + t(1 - x)(b_{G}^{q} + c_{G}^{q} \log(1 + x)) \right)\;,
\label{eq:skewness_function}
\end{flalign}
where $a_{G}^{q}$ is a free parameter to be constrained by experimental data. Two other parameters, $b_{G}^{q}$ and $c_{G}^{q}$, which govern the $t$-dependence of the skewness function, are fixed in a way to avoid singularities in the evaluation of the subtraction constant. Namely, to use the analytic regularization prescription at fixed $t$ one has:
\begin{equation}
a = \delta + A_{G}^{q}t \;,
\end{equation}
\begin{align} 
f(x) = & \frac{G^{q}(x, x, t) - G^{q}(x, 0, t)}{x^{-a}} = \nonumber \\
&\frac{G^{q}(x, 0, t)\left(g_{G}^{q}(x, t) - 1\right)}{x^{-a}} \;,
\end{align}
where $a$ and $f(x)$ were introduced in Eq. \eqref{eq:theory:analytic-regularization} and $\delta$ describes the behavior of PDFs at $x \to 0$:
\begin{equation}
q(x) \sim x^{-\delta} \;.
\end{equation}
The singularities appear in the two first compensating terms at $a = 0$ and $1-a = 0$, that is for $t \equiv t^{\infty}_{0} = -\delta/A_{G}^{q}$ and $t \equiv t^{\infty}_{1} = (1-\delta)/A_{G}^{q}$, respectively. The problem does not emerge for higher compensating terms as one typically has $0 < \delta < 1$ for valence quarks and $1 < \delta < 2$ for sea quarks. The singularities are regularized by requiring $f(0) = 0$ at $t^{\infty}_{0}$ and $f'(0) = 0$ at $t^{\infty}_{1}$, which is achieved by setting $b_{G}^{q}$ and $c_{G}^{q}$ to:
\begin{flalign}
b_{G}^{q} =& \frac{A_{G}^{q} (a_{G}^{q} - 1)}{a_{G}^{q} \delta} \label{eq:singregul1}\;, \\
c_{G}^{q} =& \frac{\left(a_{G}^{q} - 1\right)}{p_{0} \left(\delta - 1\right) a_{G}^{q} \delta} \big(p_{0}\left(2 B_{G}^{q} - C_{G}^{q}\right) \left(\delta - 1\right) + \nonumber \\ 
&A_{G}^{q} p_{0}\left(\delta - 1 - \alpha\right) + A_{G}^{q} p_{1}\big) \;, \label{eq:singregul2}
\end{flalign}
where $p_{0}$, $p_{1}$ and $\alpha$ are parameters of PDF parameterizations introduced in Sec. \ref{sec:selection_of_pdfs}. 

We stress that our Ansatz for the skewness function is explicitly defined at $x=\xi$ and it can not be generalized to the case of $x\neq\xi$ without a non-trivial modification. The form of the skewness function has been selected because of the following reasons: \emph{i}) for sufficiently small $x$ and $t$, the skewness function coincides with a constant value given by $a_{G}^{q}$. Such a behavior was predicted for HERA kinematics \cite{Frankfurt:1997at} and it was used in one of the first extractions of GPD information \cite{Kumericki:2009uq} from H1 data\cite{Aaron:2007ab}. These data suggest $a_{H}^{q} \approx 1$. \emph{ii}) In the limit of $x \rightarrow 1$ the skewness function is driven by $1/(1-x^{2})^{2}$. This form has been deduced from Ref. \cite{Yuan:2003fs}, where the power behavior of GPDs in the limit of $x \rightarrow 1$ was studied within the pQCD approach. \emph{iii}) The subdominant $t$-dependence in Eq. \eqref{eq:skewness_function} has been inspired by the skewness function evaluated from GK \cite{Goloskokov:2005sd, Goloskokov:2007nt, Goloskokov:2009ia} and VGG \cite{Vanderhaeghen:1998uc, Vanderhaeghen:1999xj, Goeke:2001tz, Guidal:2004nd} GPD models, both being based on the one component double distribution modeling scheme \cite{Radyushkin:1998bz}. In those models the $t$-dependence of the skewness function is dominated by $b_{G}^{q} + c_{G}^{q} \log(1 + x)$ term. In our Ansatz we multiply it by $(1 - x)$ to avoid any $t$-dependence at $x \rightarrow 1$, which is imposed by Ref. \cite{Yuan:2003fs}, where it was shown that GPDs should not depend on $t$ in this limit, regardless the value of $\xi$.

\subsubsection*{CFFs $\mathcal{E}$ and $\widetilde{\mathcal{E}}$}

For $E$ and $\widetilde{E}$ we use a simplified treatment justified by the poor sensitivity of the existing measurements on the corresponding CFFs. Moreover, the forward limit of those GPDs has not been measured, resulting in a need of fixing more parameters than for the GPDs $H$ and $\widetilde{H}$, if an analog modeling was to be adopted.

The modeling of the CFF $\mathcal{E}$ is similar to that of $\mathcal{H}$ and $\widetilde{\mathcal{H}}$, \emph{i.e.} it is based on the dispersion relation with the subtraction constant of the opposite sign as for $\mathcal{H}$. We only consider the valence sector with the forward limit of GPD $E^{q_{\mathrm{val}}}$ taken from Ref. \cite{Diehl:2013xca}: 
\begin{flalign}
&E^{q_{\mathrm{val}}}(x, 0, 0) \equiv e^{q_{\mathrm{val}}}(x) = \nonumber \\
&\phantom{xx}\kappa_{q} N_{q_{\mathrm{val}}} x^{-\alpha_{q_{\mathrm{val}}}} (1 - x)^{\beta_{q_{\mathrm{val}}}} (1 + \gamma_{q_{\mathrm{val}}} \sqrt{x}) \;,
\label{eq:ansatz_xi_eq_0_E}
\end{flalign}
where $\kappa_{u} = 1.67$ and $\kappa_{d} = -2.03$ are the magnetic anomalous moments for up and down quarks, respectively, and where $\beta_{u_{\mathrm{val}}} = 4.65$, $\beta_{d_{\mathrm{val}}} = 5.25$, $\gamma_{u_{\mathrm{val}}} = 4$ and $\gamma_{d_{\mathrm{val}}} = 0$. The parameter $\alpha_{u_{\mathrm{val}}} = \alpha_{d_{\mathrm{val}}} \equiv \alpha$ is fitted to EFF data as described in Sec. \ref{sec:elasticFF}. The normalization parameter:
\begin{align}
N_{q_{\mathrm{val}}}^{-1} = &\Gamma(1 + \beta_{q_{\mathrm{val}}})
\bigg(\frac{\Gamma(1 - \alpha_{q_{\mathrm{val}}})}{\Gamma(2 - \alpha_{q_{\mathrm{val}}} + \beta_{q_{\mathrm{val}}})} + \nonumber \\
&\gamma_{q_{\mathrm{val}}} \frac{\Gamma(1.5 - \alpha_{q_{\mathrm{val}}})}{\Gamma(2.5 - \alpha_{q_{\mathrm{val}}} + \beta_{q_{\mathrm{val}}})}\bigg) \;,
\end{align}
ensures that: 
\begin{equation}
\int_{0}^{1} dx~ e^{q_{\mathrm{val}}}(x) = \kappa_{q} \;.
\end{equation}
At $\xi = 0$  the $t$-dependence is introduced analogously as for the GPDs $H$ and $\widetilde{H}$, \emph{i.e.}:
\begin{equation}
E^{q_{\mathrm{val}}}(x, 0, t) = e^{q_{\mathrm{val}}}(x) \exp(f_{E}^{q_{\mathrm{val}}}(x)t) \;,
\label{eq:ansatz_xi_eq_xi_E}
\end{equation}
with $f_{E}^{q_{\mathrm{val}}}(x)$ given by Eq. \eqref{eq:ansatz_xt_profile}, where:  $A_{E}^{q_{\mathrm{val}}}$, $B_{E}^{q_{\mathrm{val}}}$, $C_{E}^{q_{\mathrm{val}}}$ are free parameters being fitted to EFF data as described in Sec. \ref{sec:elasticFF}. It is assumed that $A_{E}^{q_{\mathrm{val}}} = A_{H}^{q_{\mathrm{val}}}$.
 
The skewness function used to evaluate $E^{q_{\mathrm{val}}}(x, x, t)$ from $E^{q_{\mathrm{val}}}(x, 0, t)$ is of the following form: 
\begin{align}
g_{E}^{q_{\mathrm{val}}}(x, x, t) \equiv g_{E}^{q_{\mathrm{val}}}(x, t) = \frac{a_{E}^{q_{\mathrm{val}}}}{(1-x^{2})^{3}} \frac{f(x)}{f(0)} \;,
\label{eq:skewness_function_E}
\end{align}
which again is deduced from Ref. \cite{Yuan:2003fs}. Here, $a_{E}^{u_{\mathrm{val}}} = a_{E}^{d_{\mathrm{val}}} \equiv a_{E}^{q_{\mathrm{val}}} = 1$ and any $t$-dependence is neglected. An additional $\xi$-dependence coming from twist-three and twist-four distribution amplitudes of the nucleon is denoted by $f(x)$. In the present analysis, where only the leading twist is considered, it is assumed that $f(x)/f(0) = 1$.

For the GPD $\widetilde{E}$ the shape of the corresponding form factor is fixed by the GK model \cite{Goloskokov:2005sd, Goloskokov:2007nt, Goloskokov:2009ia}. Only the normalization parameter, $N_{\widetilde{E}}$, is fitted to the experimental data, 
\begin{align}
\widetilde{\mathcal{E}}(\xi, t) = N_{\widetilde{E}} \widetilde{\mathcal{E}}_{\mathrm{GK}}(\xi, t) \;.
\end{align}

\subsubsection*{Inequalities}

We impose two extra constraints implied by the positivity of parton densities in the impact parameter space. Namely, for the profile functions we require to have:
\begin{align}
f_{H}^{q}(x) &> 0 \label{eq:ansatz_summary_ineq_1} \;, \\
f_{H}^{q}(x) &\geq f_{\widetilde{H}}^{q}(x) \label{eq:ansatz_summary_ineq_2} \;,
\end{align}
being imposed by Eq. \eqref{eq:theory:ineq1}. The requirements are implemented in a way to penalize combinations of fitted parameters that do not hold the inequalities. It is achieved by introducing a penalty to the $\chi^{2}$ function, which value is proportional to the maximum violation of a given inequality. Such an implementation allows us to keep the $\chi^{2}$ function smooth. We report that the inequalities given by Eqs. \eqref{eq:ansatz_summary_ineq_1} and \eqref{eq:ansatz_summary_ineq_2} have proved to be important to constrain parameterizations for the sea contribution to GPD $H$ and for the valence contribution to GPD $\widetilde{H}$. It is due to a low sensitivity to those contributions and a limited phase space covered by the available data.

The inequality \eqref{eq:theory:ineq2} is not checked during the minimization. Namely, by following Ref. \cite{Diehl:2004cx} one can obtain:
\begin{flalign}
&\frac{e^{q}(x)^2}{16 m^2 f_{E}^{q}(x)^4} \leq \frac{1}{\mathbf{b}_{\mathbf{\perp}}^2}\bigg[ \nonumber \\
&\phantom{xx}\frac{q(x)^2}{f_{H}^{q}(x)^2} \exp(\frac{\mathbf{b}_{\mathbf{\perp}}^2}{2f_{E}^{q}(x)}-\frac{\mathbf{b}_{\mathbf{\perp}}^2}{2f_{H}^{q}(x)})-  \nonumber \\
&\phantom{xx}\frac{\Delta q(x)^2}{f_{\widetilde H}^{q}(x)^2} \exp(\frac{\mathbf{b}_{\mathbf{\perp}}^2}{2f_{E}^{q}(x)}-\frac{\mathbf{b}_{\mathbf{\perp}}^2}{2f_{\widetilde H}^{q}(x)})\bigg] 
\label{eq:ansatz_summary_ineq_3}
\;.
\end{flalign}
where for $f_{\widetilde H}^{q}(x)=f_{H}^{q}(x)$ (condition not imposed in this analysis) one can minimize right hand side over $\mathbf{b}_{\mathbf{\perp}}^2$, and get the strongest inequality:
\begin{align}
&\frac{e^{q}(x)^2}{8 m^2} \leq
\exp(1) \left(\frac{f_{E}^{q}(x)}{f_{H}^{q}(x)}\right)^3 \big(f_{H}^{q}(x)-f_{E}^{q}(x)\big) \times \nonumber \\
&\phantom{xx}
	\left[q(x)^2-\Delta q(x)^2\right]
\; ,
\label{eq:ansatz_summary_ineq_4}
\end{align} 
which is used for instance in Ref. \cite{Diehl:2013xca}. To effectively use the inequality \eqref{eq:ansatz_summary_ineq_3} in the minimization, one should simultaneously fit all free parameters of $f_{H}^{q}(x)$, $f_{\widetilde H}^{q}(x)$, $f_{E}^{q}(x)$ and $e(x)$, which presently is not the case, as we fit EFF and DVCS data separately. In addition, for the GPDs $\widetilde{H}$ and $E$ we consider only the valence contribution, while the sea sector may have a significant impact on \eqref{eq:ansatz_summary_ineq_3} in the range of small $\xBj$. The validity of \eqref{eq:ansatz_summary_ineq_3} has however been checked \emph{a posteriori}, that is after the $\chi^2$-minimization. The result of this test shows that the inequality is violated for most of the replicas for very small $\mathbf{b}_{\mathbf{\perp}}^2$. To correct this one should change the simplified Ansatz used for GPD $E$ and preferably fit its free parameters simultaneously with those for other GPDs, which is beyond the scope of this analysis.  

\subsubsection*{Approximations and summary}

For the GPD $H$ and valence quarks, all parameters of the profile function given by Eq. \eqref{eq:ansatz_xt_profile} are fitted to EFF data, where we fixed $C_{H}^{u_{\mathrm{val}}} = -A_{H}^{u_{\mathrm{val}}}$ and $C_{H}^{d_{\mathrm{val}}} = -A_{H}^{d_{\mathrm{val}}}$. The unconstrained parameter in the skewness function given by Eq. \eqref{eq:skewness_function} is fitted to DVCS data, where it is assumed that $a_{H}^{u_{\mathrm{val}}} = a_{H}^{d_{\mathrm{val}}} \equiv a_{H}^{q_{\mathrm{val}}}$. For sea quarks all parameters for both profile and skewness functions are fitted to DVCS data. Due to a limited sensitivity to the sea sector, we assume the symmetry with respect to the change of quark flavor, \emph{i.e.} one has $a_{H}^{u_{\mathrm{sea}}} = a_{H}^{d_{\mathrm{sea}}} = a_{H}^{s} \equiv a_{H}^{q_{\mathrm{sea}}}$, $A_{H}^{u_{\mathrm{sea}}} = A_{H}^{d_{\mathrm{sea}}} = A_{H}^{s} \equiv A_{H}^{q_{\mathrm{sea}}}$, \emph{etc}. Sea components are not yet fully symmetric in our fit, as we do not impose the flavor symmetry in PDFs, see Sec. \ref{sec:selection_of_pdfs}. Due to the lack of precision of axial EFF data, for the GPD $\widetilde{H}$ all parameters for both profile and skewness functions are fitted to DVCS data. As the contribution coming from $\Delta q_{\mathrm{sea}}$ is subdominant, we neglect it entirely. For valence quarks, similarly to the GPD $H$, we allow the profile function to be different for $u_{\mathrm{val}}$ and $d_{\mathrm{val}}$, while for the skewness function one has $a_{\widetilde{H}}^{u_{\mathrm{val}}} = a_{\widetilde{H}}^{d_{\mathrm{val}}} \equiv a_{\widetilde{H}}^{q_{\mathrm{val}}}$. For $E$ and $\widetilde{E}$ only the valence quarks are considered. For the GPD $E$ all free parameters for both profile function and forward limit given by Eq. \eqref{eq:ansatz_xi_eq_0_E} are fitted to EFF data. For the GPD $\widetilde{E}$ the normalization parameter $N_{\widetilde{E}}$ is fitted to DVCS data. In total, we fit $9$ parameters to EFF data (see Tab. \ref{tab:elasticFF_results}) and $13$ parameters are constrained by DVCS data (see Tab. \ref{tab:fitted_param}).

\section{Analysis of PDFs}
\label{sec:selection_of_pdfs}

The analytic regularization prescription introduced in Eq. \eqref{eq:theory:analytic-regularization} requires the function $q(x)/x^{-\delta}$ to be non-zero and analytic at $x = 0$. However, the numeric evaluation of such a function and its derivatives near $x = 0$ is difficult for typical PDF sets, like those published by NNPDF \cite{Ball:2014uwa} and CTEQ \cite{Nadolsky:2008zw} groups, because of several numerical issues. Namely, interpolations in PDF grids and extrapolations outside those grids for small $x$, and a delicate cancellation of the numerator and denominator of $q(x)/x^{-\delta}$ make the evaluation numerically unstable. The problem can be avoided with functional parameterizations of PDFs, which can be used for a straightforward evaluation of $q(x)/x^{-\delta}$ and its derivatives. In addition, such parameterizations allow for a significant reduction of computation time and a precise determination of $\delta$. 

In this analysis the parameterization used for both unpolarized and polarized PDFs reads:  
\begin{flalign}
&\mathrm{pdf}_{G}(x, Q^{2}) = x^{-g(\delta_{p}, \delta_{q}, Q^{2})} \times \nonumber \\ 
&\phantom{xx}(1-x)^{\alpha} \sum_{i=0}^{4} g(p_{i}, q_{i}, Q^{2}) x^{i} \;,
\label{eq:pdf_parameterization}
\end{flalign}
where:
\begin{equation}
g(p, q, Q^{2}) = p + q\log\frac{Q^{2}}{Q_{0}^{2}} \;,
\end{equation}
describes the evolution in the renormalization scale, recognized as $Q^{2}$, where $Q_{0}^{2} = 2~\mathrm{GeV}^{2}$ can be identified as the initial scale. The parameterization contains thirteen parameters,
\begin{equation}
\delta_{p}, \delta_{q}, \alpha, p_{i}, q_{i},~~~\mathrm{where}~i=0, 1, \dots, 4 \;,
\end{equation}
constrained for each quark flavor in a fit to "NNPDF30\_lo\_as\_0118\_nf\_3" set \cite{Ball:2014uwa} for unpolarized PDFs and to "NNPDFpol11\_100" set \cite{Nocera:2014gqa} for polarized PDFs. These fits are performed for a grid of nearly $10000$ points equidistantly distributed in the ranges of $10^{-4} < x < 0.9$ and $1~\mathrm{GeV}^{2} < Q^{2} < 20~\mathrm{GeV}^{2}$. We have selected the sets by NNPDF group as: \emph{i}) they are provided for both $q(x)$ and $\Delta q(x)$, \emph{ii}) for $q(x)$ they are provided at LO and three active flavors, \emph{iii}) they provide reliable estimation of uncertainties extracted through the neural network approach and the replica method. The PDF sets are handled with LHAPDF interface \cite{Buckley:2014ana}.

\begin{figure*}[!ht]
\begin{center}
\includegraphics[width=0.45\textwidth]{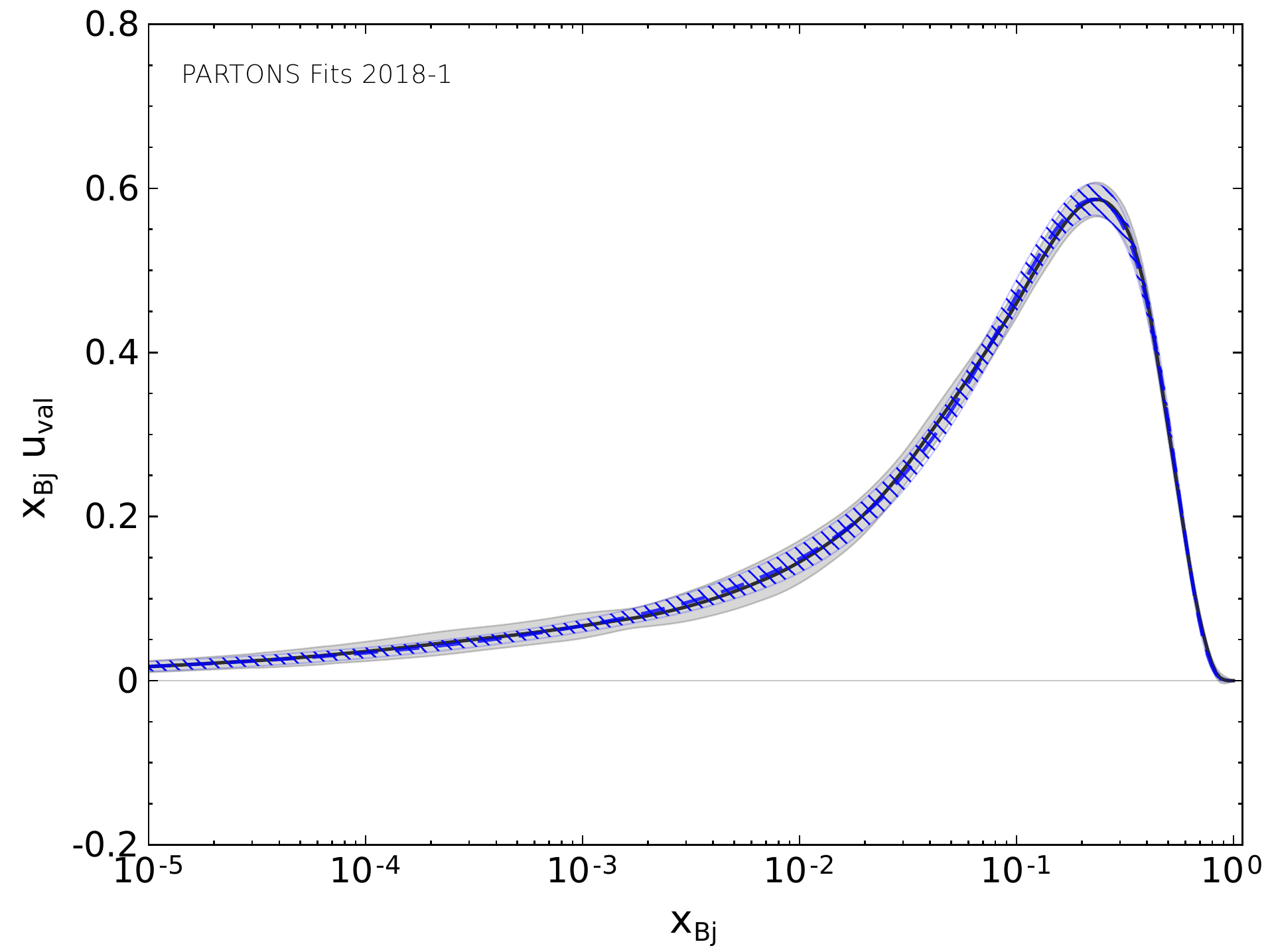}
\includegraphics[width=0.45\textwidth]{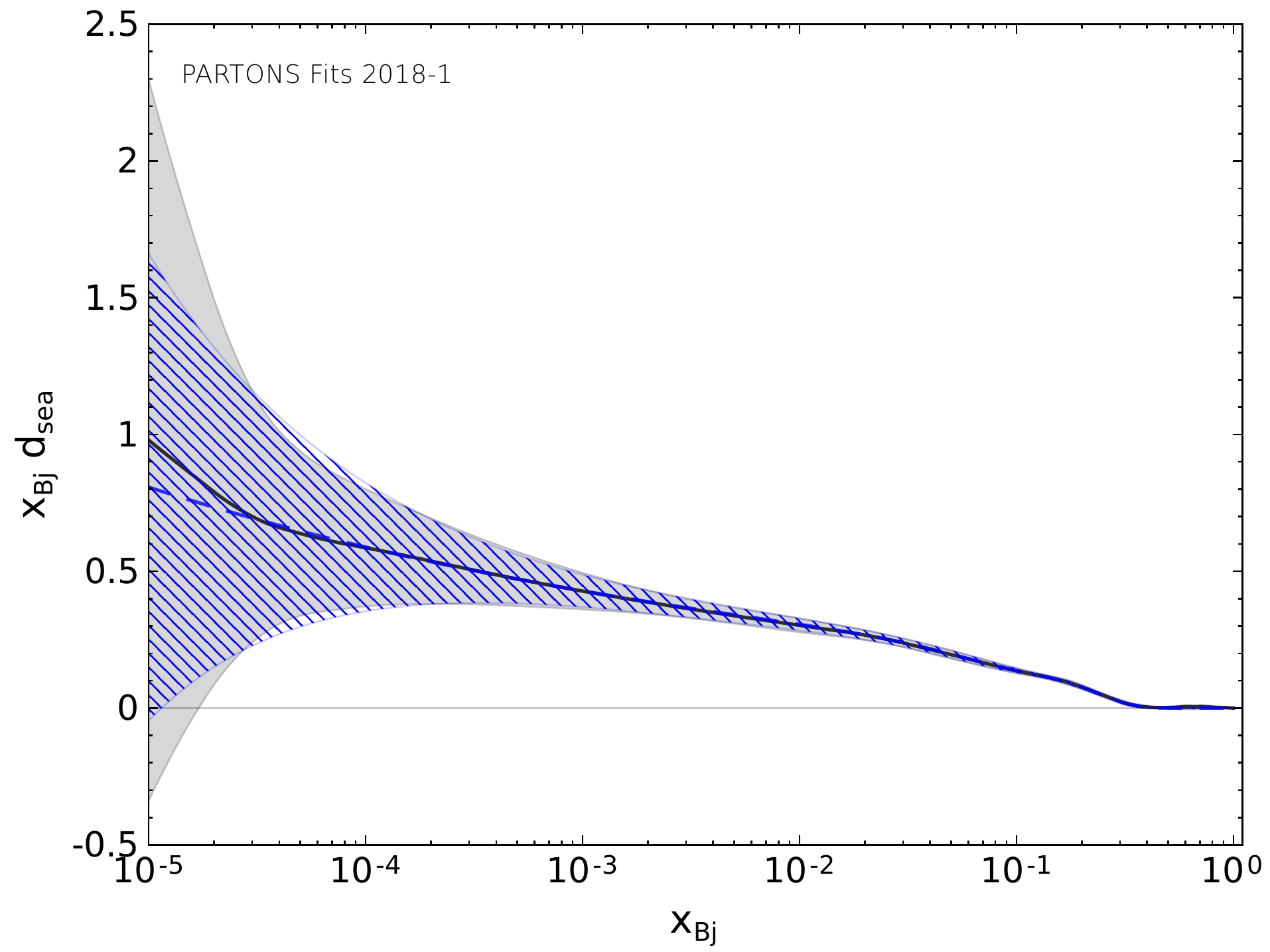}
\includegraphics[width=0.45\textwidth]{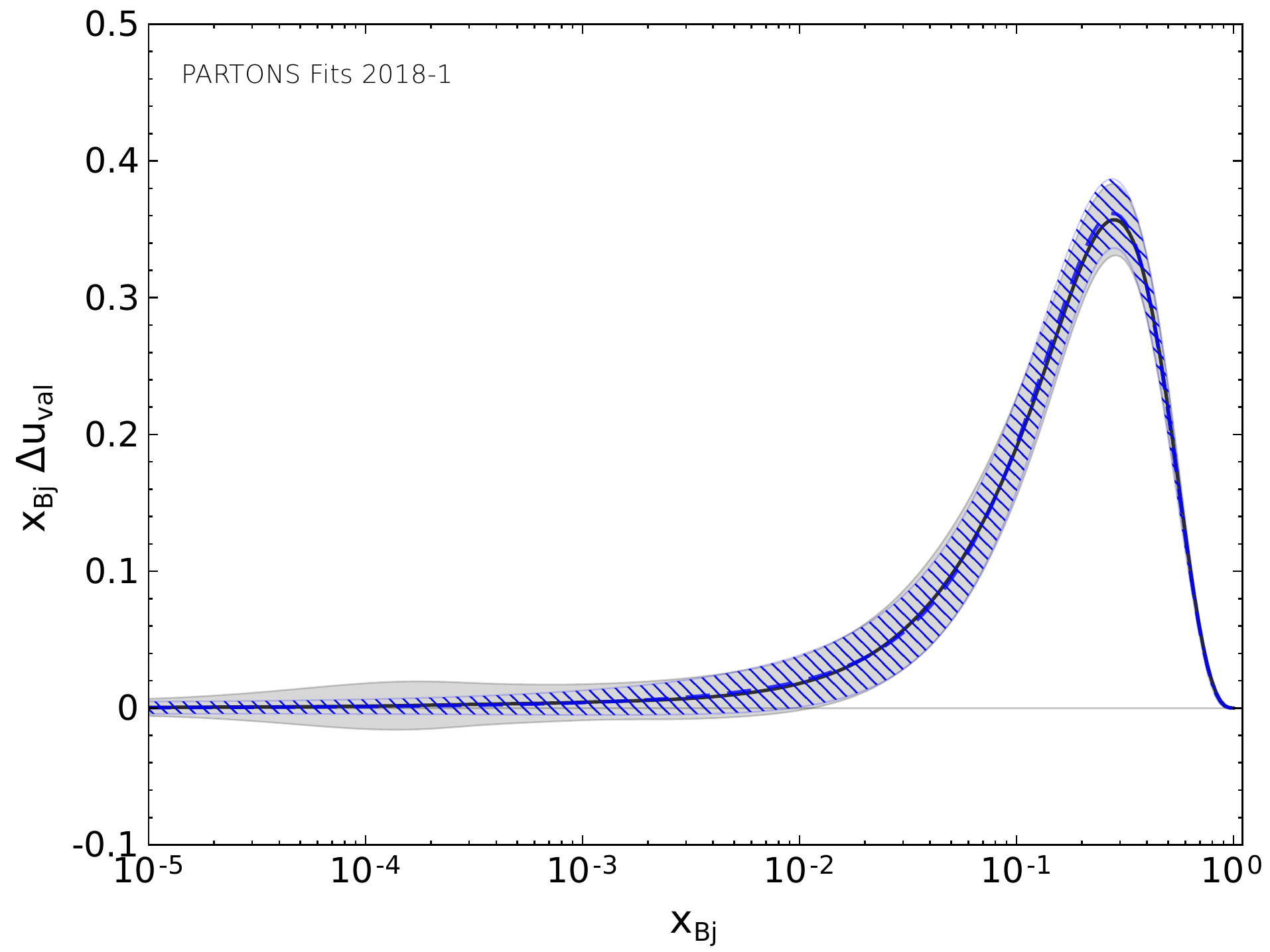}
\includegraphics[width=0.45\textwidth]{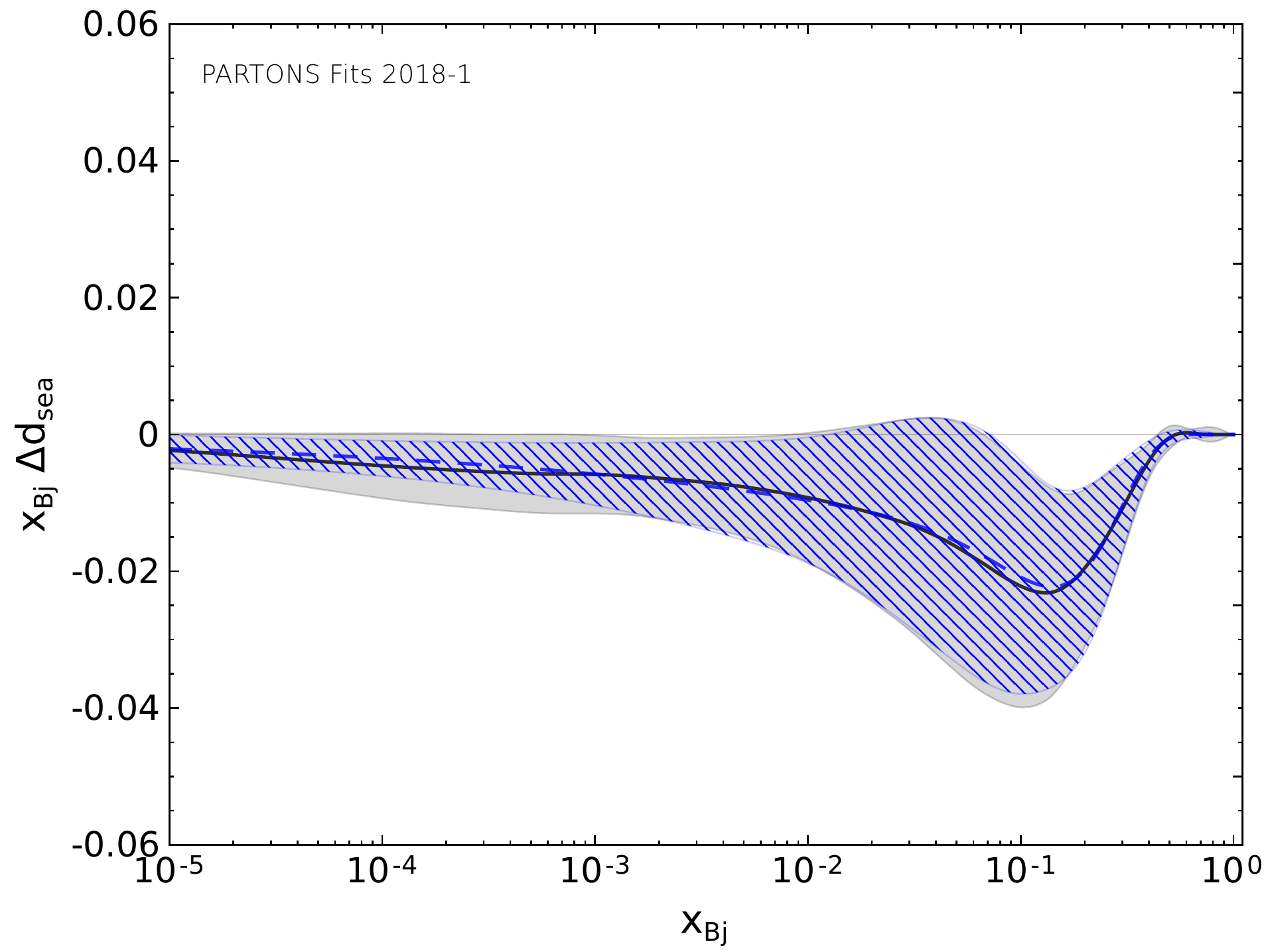}
\caption{Comparison between PDF sets by NNPDF group \cite{Ball:2014uwa, Nocera:2014gqa} and parameterizations based on Eq. \eqref{eq:pdf_parameterization}. The upper (lower) row is for unpolarized (polarized) PDFs. For a given row, the left plot is for $u_{\mathrm{val}}$ quarks, while the right one is for $d_{\mathrm{sea}}$ quarks. For a given figure, the black solid curve with the grey band representing $68\%$ confidence level is for PDFs by NNPDF group, while the blue dashed curve with the hatched band is for our fit. The curves are evaluated at $Q^{2} = 2~\mathrm{GeV}^{2}$.}
\label{fig:pdf_comparison}
\end{center}
\end{figure*}

A consistent fit to all replicas of used NNPDF sets allows us to reproduce the original uncertainties of PDFs. Figure \ref{fig:pdf_comparison} demonstrates the agreement between the original sets and our fits. The comparison is satisfactory and in particular the central values of fitted PDFs stay within the original uncertainty bands.

\section{Analysis of Dirac and Pauli Form Factors}
\label{sec:elasticFF}

Equations \eqref{eq:theory:ff1}-\eqref{eq:theory:ff4} link GPDs to EFFs, thus they can be used to constrain GPDs from elastic data. In this analysis we use this feature to constrain the interplay between the $x$ and $t$ variables for the GPDs $H$ and $E$, however only for the valence quarks. More precisely, we fix the parameters of $f_{H/E}^{q_{\mathrm{val}}}(x)$ defined in Eq. \eqref{eq:ansatz_xt_profile} by using the Ansatz for $H^{q_{\mathrm{val}}}(x, 0, t)$ given in Eq. \eqref{eq:ansatz_xi_eq_0} and that for $E^{q_{\mathrm{val}}}(x, 0, t)$ given in Eq. \eqref{eq:ansatz_xi_eq_xi_E}. We achieve this by studying proton, $F_{i}^{p}(t)$, and neutron, $F_{i}^{n}(t)$, Dirac and Pauli form factors: 
\begin{align}
F_{i}^{p} &= e_{u} F_{i}^{u} + e_{d} F_{i}^{d} \;, \nonumber \\
F_{i}^{n} &= e_{u} F_{i}^{d} + e_{d} F_{i}^{u},~~~\mathrm{where}~i=1, 2 \;,
\end{align}
related to Sachs form factors as follows:
\begin{align}
G_{M}^{i} &= F_{1}^{i} + F_{2}^{i} \;,  \nonumber \\
G_{E}^{i} &= F_{1}^{i} + \frac{t}{4m^{2}}F_{2}^{i},~~~\mathrm{where}~i=p, n \;.
\end{align}
Sachs form factors can be used to define many observables, in particular:
\begin{itemize}[label=\small$\bullet$]
\item the magnetic form factor normalized to both magnetic moment, $\mu_{p}$ or $\mu_{n}$, and dipole form factor:
\begin{equation}
G_{M, N}^{i}(t) = \frac{G_{M}^{i}(t)}{\mu_{i}G_{D}(t)} \;,
\end{equation}
where $i=p, n$ and
\begin{equation}
G_{D}(t) = \frac{1}{\left( 1 - t/M_D^2 \right)^{2}} \;,
\end{equation}
with $M_D^2 = 0.71~\mathrm{GeV}^2$.
\item the ratio of electric and normalized magnetic form factors
\begin{equation}
R^{i}(t) = \frac{\mu_{i}G_{E}^{i}(t)}{G_{M}^{i}(t)} \;.
\end{equation}
where $i=p, n$. 
\item the squared charge radius of neutron:
\begin{equation}
r_{nE}^{2} = 6 \frac{dG_{E}^{n}(t)}{dt}\bigg\rvert_{t = 0} \;.
\end{equation}
\end{itemize}

Experimental data for those observables and for $G_{E}^{n}(t)$ come from the sources summarized in Table \ref{tab:elasticFF_usedData}. For the selection of observables and related data we follow Ref. \cite{Diehl:2013xca}, which deals with the subject in great detail. In total, the used sample of data consists of $178$ data points covering the range of $0.017~\mathrm{GeV}^{2} \leq |t| \leq 31.2~\mathrm{GeV}^{2}$. Our fit to those data ends  for the central PDF replica with $\chi^{2}/\mathrm{ndf} = 129.6/(178-9) \approx 0.77$. The fit is repeated for all other PDF replicas and for $100$ instances of replicated data to propagate the uncertainties of EFF data to the analysis of CFFs. A single instance of replicated data is generated with the following prescription:
\begin{equation}
v_{i} \pm {\Delta}_{i}^{\mathrm{tot}} \xrightarrow[]{\mathrm{replica}~j} \mathrm{rnd}_{j}(v_{i}, {\Delta}_{i}^{\mathrm{tot}}) \pm {\Delta}_{i}^{\mathrm{tot}} \;,
\label{eq:elasticFF_replication}
\end{equation}
where $v_{i}$ is the measured value associated to the experimental point $i$. The total uncertainty, which is also used to evaluate the $\chi^{2}$ value in the fit to EFF data, reads: 
\begin{equation}
{\Delta}_{i}^{\mathrm{tot}} = \sqrt{\left({\Delta}_{i}^{\mathrm{stat}}\right)^{2} + \left({\Delta}_{i}^{\mathrm{sys}}\right)^{2}} \;,
\label{eq:elasticFF:total_err}
\end{equation}
where ${\Delta}_{i}^{\mathrm{stat}}$ and ${\Delta}_{i}^{\mathrm{sys}}$ are  statistical and systematic uncertainties, respectively, both linked to the point $i$. The generator of random numbers following a specified normal distribution, $f(x | \mu, \sigma)$, is denoted by $\mathrm{rnd}_{j}(\mu, \sigma)$, where $j$ is both the identifier of a given replica and a unique random seed.  

\begin{table}[!ht]
\centering
\caption{EFF data used in this analysis. For $R^{n}$ we use data coming from Ref. \cite{Diehl:2013xca}, which are evaluated from those specified in this table.}
\label{tab:elasticFF_usedData}
\begin{tabular}{@{}ccc@{}}
\toprule
Observable			& Reference					& No. of points \\ \midrule
$G_{M, N}^{p}$ 	& \cite{Arrington:2007ux}	& $54$ \\    
$R^{p}$           & \cite{Arrington:2007ux, Milbrath:1997de, Pospischil:2001pp, Gayou:2001qt, Gayou:2001qd, Punjabi:2005wq, MacLachlan:2006vw, Puckett:2010ac, Paolone:2010qc, Ron:2011rd, Zhan:2011ji} & $54$ \\
$G_{M, N}^{n}$    & \cite{Anklin:1998ae, Kubon:2001rj, Anklin:1994ae, Lachniet:2008qf, Anderson:2006jp} & $36$ \\   
$R^{n}$           & \cite{Herberg:1999ud, Glazier:2004ny, Plaster:2005cx, Passchier:1999cj, Zhu:2001md, Warren:2003ma, Geis:2008aa, Bermuth:2003qh, rohe_pc, Riordan:2010id} & $21$ \\    
$G_{E}^{n}$       & \cite{Schiavilla:2001qe} & $12$ \\   
$r_{nE}^{2}$      & \cite{Beringer:1900zz} & $1$  \\ \bottomrule
\end{tabular}
\end{table}
 
\begin{figure*}[!ht]
\begin{center}
\includegraphics[width=0.95\textwidth]{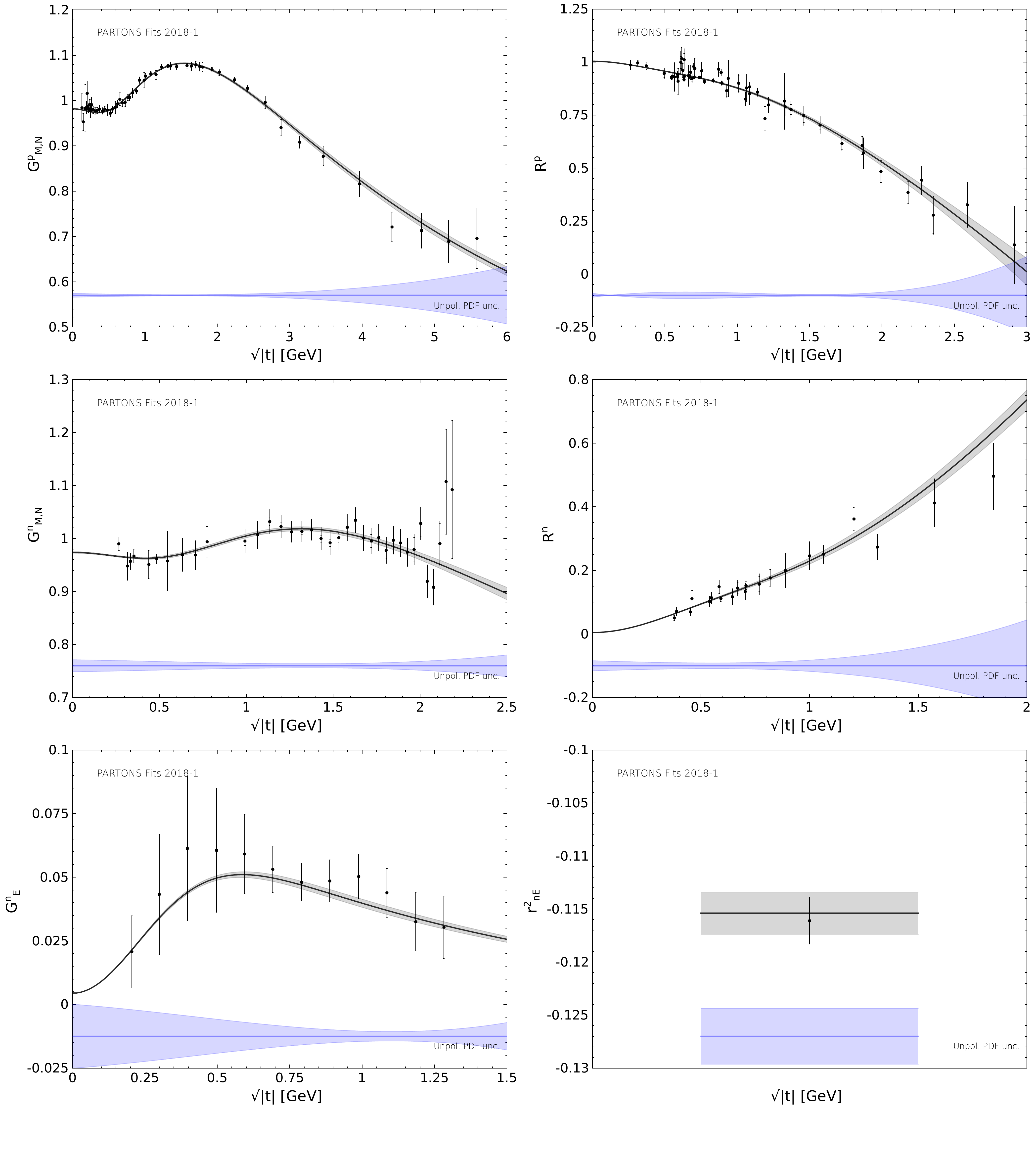}
\caption{Elastic Form Factor data listed in Table \ref{tab:elasticFF_usedData} and fitted according to the text. The gray bands indicate $68 \%$ confidence level for uncertainties coming from EFF data. The corresponding bands for unpolarized PDFs are indicated by the labels. For data points provided with systematic uncertainties, the inner bars represent statistical uncertainties, while the outer ones are for the quadratic sum of statistical and systematic uncertainties. Otherwise, statistical uncertainties are shown.}
\label{fig:elasticFF_all}
\end{center}
\end{figure*}

\begin{figure*}[!ht]
\begin{center}
\includegraphics[width=0.45\textwidth]{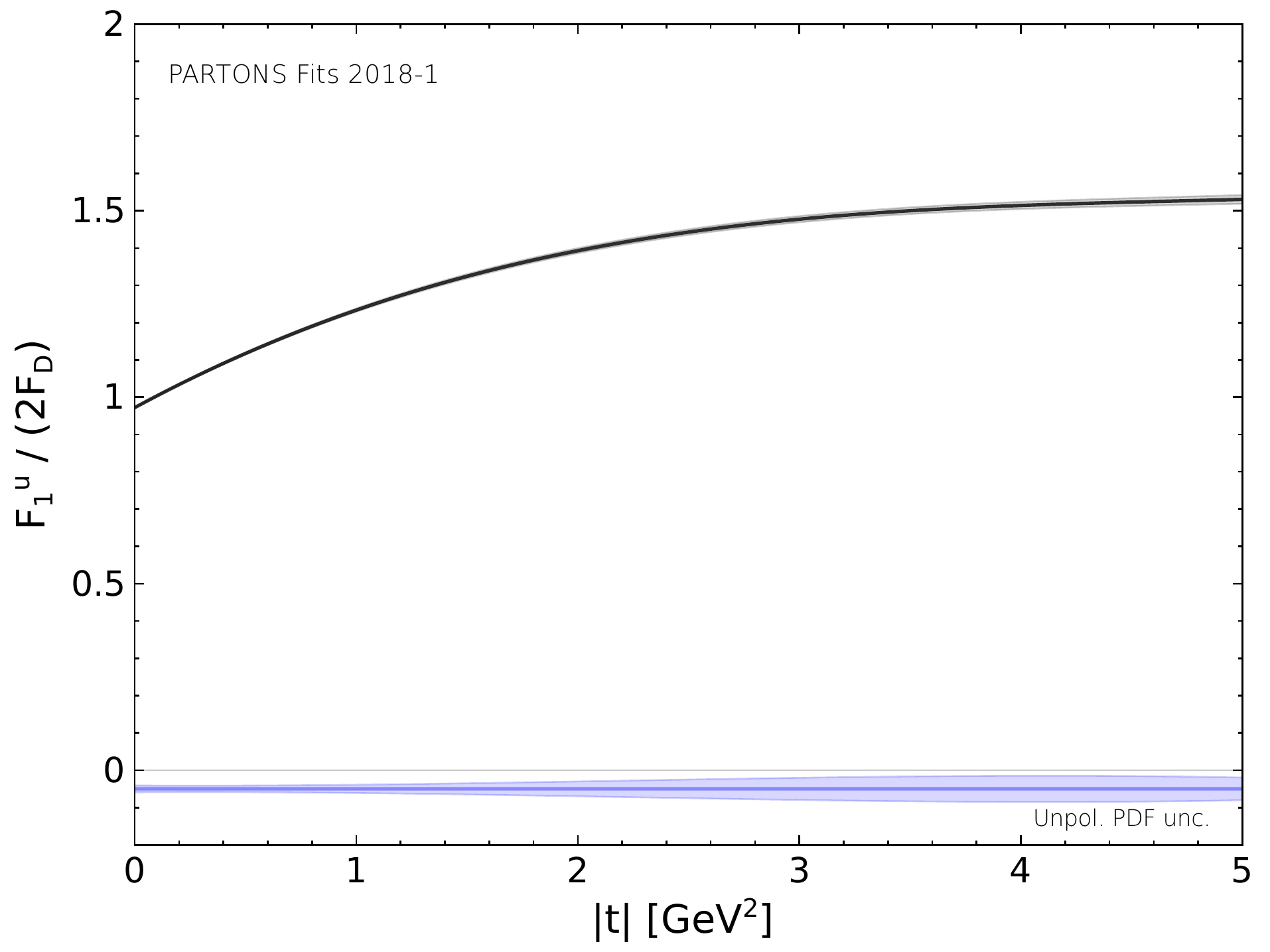}
\includegraphics[width=0.45\textwidth]{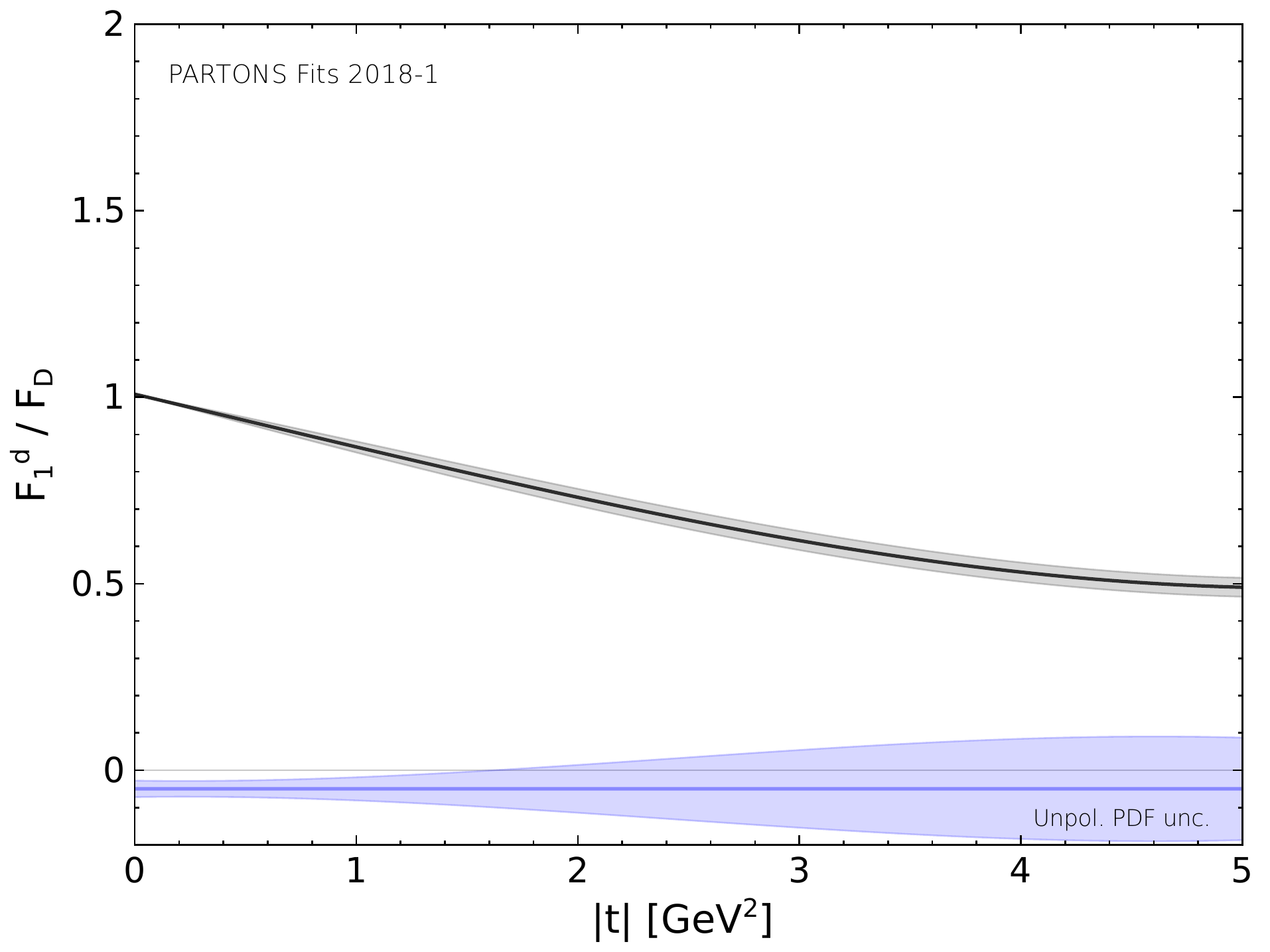}
\includegraphics[width=0.45\textwidth]{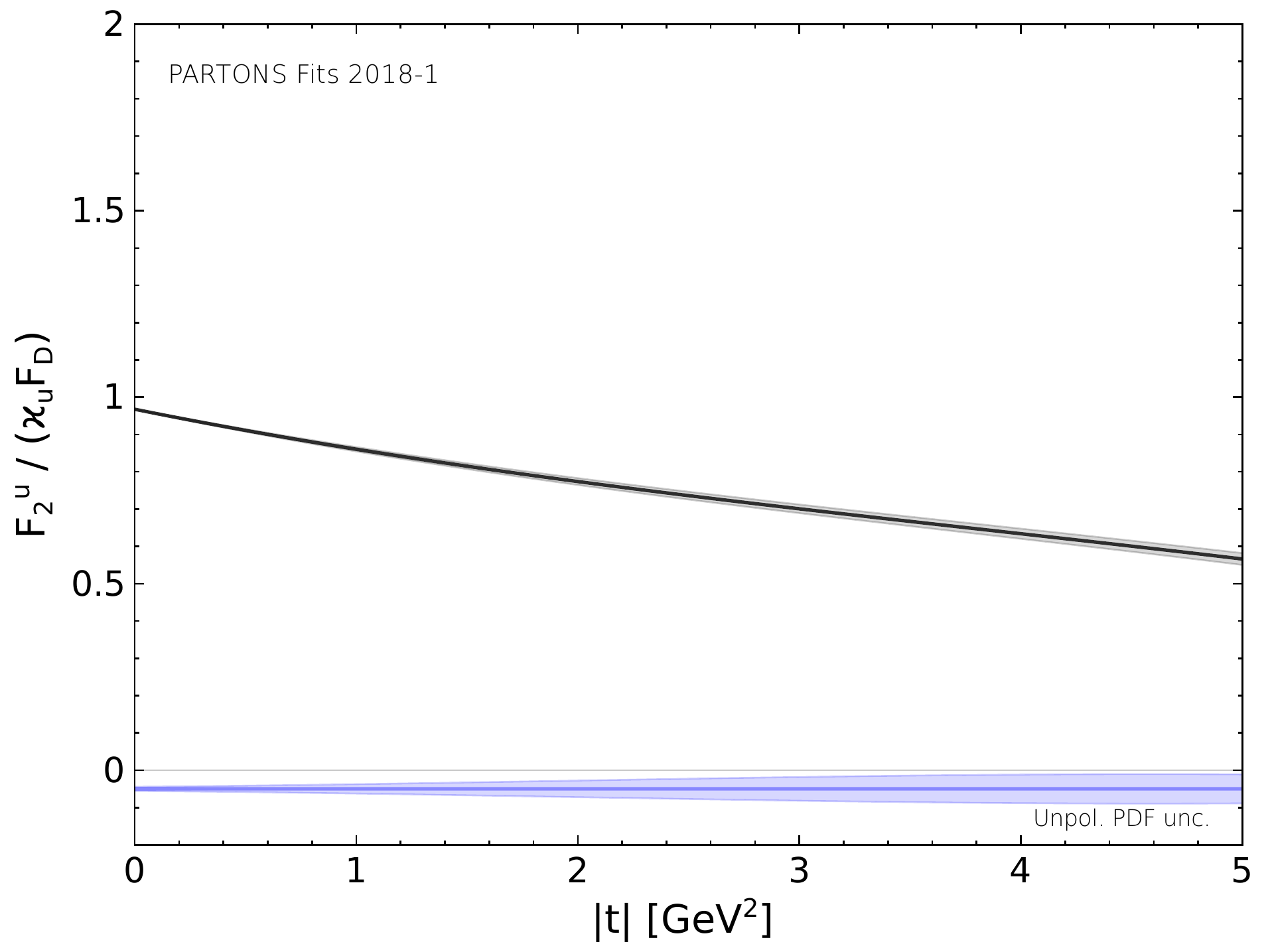}
\includegraphics[width=0.45\textwidth]{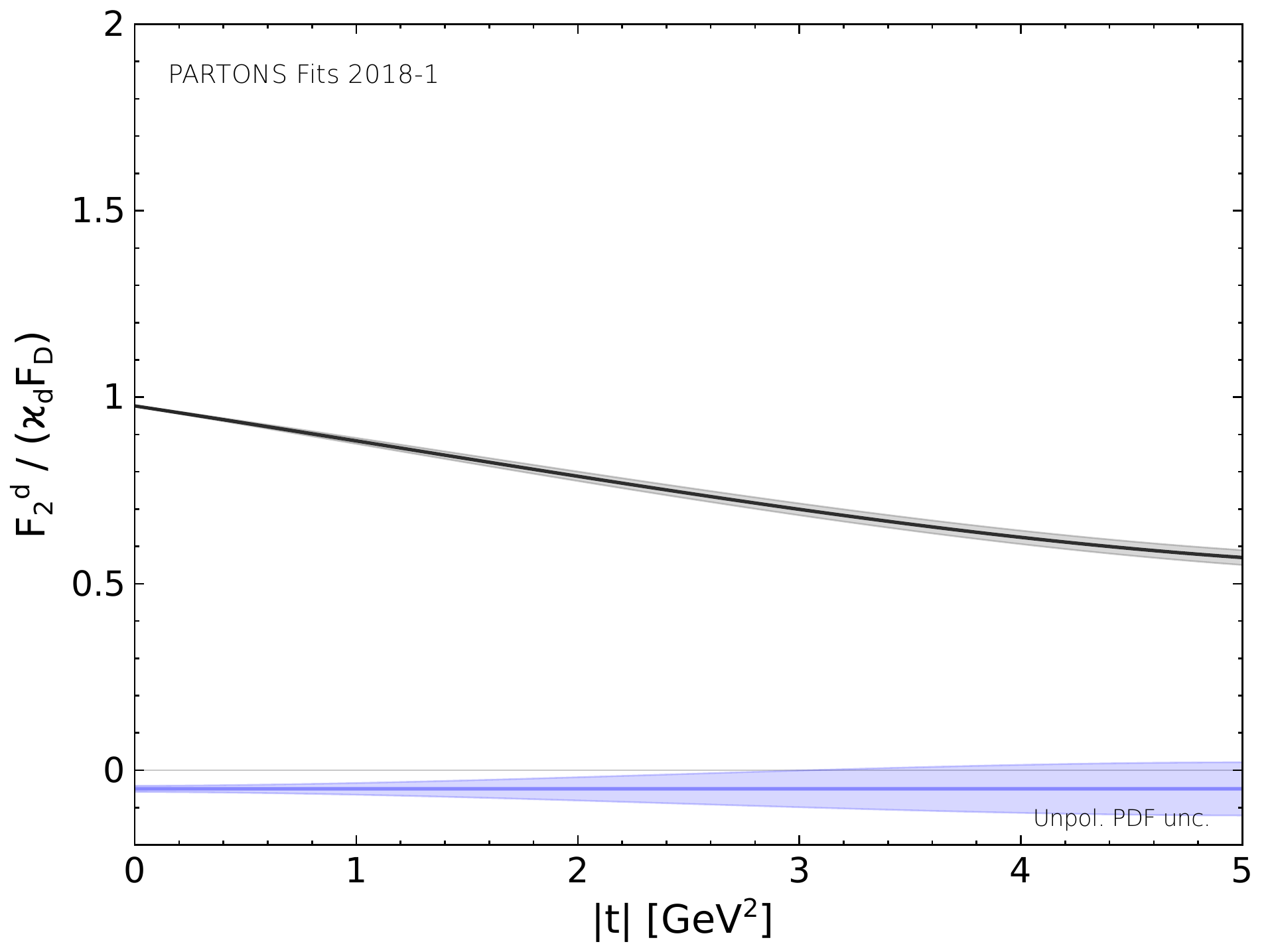}
\caption{Parton Dirac and Pauli form factors obtained in this analysis. For further description see Fig. \ref{fig:elasticFF_all}.}
\label{fig:ff_comparison}
\end{center}
\end{figure*}

The final result consists of $201$ replicas, where each replica represents a possible realization of fitted EFFs. Such a set of replicas can be used to estimate mean values and uncertainties of the fitted parameters, which we summarize in Table \ref{tab:elasticFF_results}. Experimental data are superimposed with the results of our fit in Fig. \ref{fig:elasticFF_all}, while distributions of quark EFFs are shown in Fig. \ref{fig:ff_comparison}. 

\begin{table}[!ht]
\centering
\caption{Values of parameters fitted to EFF data together with estimated uncertainties coming from those data and used PDF parameterizations.}
\label{tab:elasticFF_results}
\begin{tabular}{@{}cccc@{}}
\toprule
Parameter & Mean & Data unc. & Unpol. PDF unc. \\ \midrule
$A_{H/E}^{u_{\mathrm{val}}}$		& $\phantom{-}0.99$ &    $0.01$ &  $0.08$ \\
$B_{H}^{u_{\mathrm{val}}}$    		& 			$-0.50$ &    $0.02$ &  $0.14$ \\
$A_{H/E}^{d_{\mathrm{val}}}$    	& $\phantom{-}0.70$ &    $0.02$ &  $0.08$ \\
$B_{H}^{d_{\mathrm{val}}}$    		& $\phantom{-}0.47$ &    $0.07$ &  $0.24$ \\
$\alpha$    			        	& $\phantom{-}0.69$ &    $0.01$ &  $0.03$ \\ 
$B_{E}^{u_{\mathrm{val}}}$    		& 			$-0.69$ &    $0.04$ &  $0.18$ \\
$C_{E}^{u_{\mathrm{val}}}$    		& 			$-0.92$ &    $0.04$ &  $0.09$ \\
$B_{E}^{d_{\mathrm{val}}}$    		& 			$-0.54$ &    $0.06$ &  $0.20$ \\
$C_{E}^{d_{\mathrm{val}}}$    		& 			$-0.73$ &    $0.06$ &  $0.22$ \\
\bottomrule
\end{tabular}
\end{table}

\section{DVCS data sets}
\label{sec:dvcs_data}

Table \ref{tab:used_data} summarizes DVCS data used in this analysis. Currently, only proton data are used, while those sparse ones for neutron targets are foreseen to be included in the future. We note, that recent Hall A data \cite{Defurne:2015kxq, Defurne:2017paw} published for unpolarized cross sections, $d^{4}\sigma_{UU}^{-}$, are not used in the present analysis since it was not possible to correctly describe them with our fitting Ansatz, and their inclusion was causing a bias on the final extraction of GPD information. This specific question will be addressed in a future study. However we keep the recent Hall A data for differences of cross sections, $\Delta d^{4}\sigma_{LU}^{-}$. In addition, we skip all DVCS data published by HERA experiments since they should be dominated by gluons, and would probably lead to misleading conclusions in our analysis which is adapted to the quark sector. We will back to these problems in Sec. \ref{sec:results}.   

We apply two kinematic cuts on experimental data:
\begin{flalign}
&Q^{2} > 1.5~\mathrm{GeV}^{2} \label{eq:dvcs_data:cut_1} \;, \\
&-t/Q^{2} < 0.25 \label{eq:dvcs_data:cut_2} \;.
\end{flalign} 
The purpose of those cuts is to restrict the phase-space covered by experimental data to the deeply virtual region where one can rely on the factorization between GPDs and the hard scattering kernel. The values of those cuts have been selected with the correspondence to KM analysis \cite{Kumericki:2015lhb}, which may help to compare results of both analyses in the future.

\begin{table*}[!ht]
\centering
\caption{DVCS data used in this analysis.}
\label{tab:used_data}
\begin{tabular}{ccccclcc}
\toprule
No. & Collab. 	& Year & Ref. & \multicolumn{2}{c}{Observable} & \makecell{Kinematic \\ dependence} & \makecell{No. of points \\ used /\ all} \\ \midrule
1   & HERMES 	& 2001 & \cite{Airapetian:2001yk} & $A_{LU}^{+}$ & & $\phi$ & 10 /\ 10 \\
2   &  		& 2006 & \cite{Airapetian:2006zr} & $A_{C}^{\cos i \phi}$ & $i = 1$ & $t$ & 4 /\ 4 \\ 
3   &  		& 2008 & \cite{Airapetian:2008aa} & $A_{C}^{\cos i \phi}$ & $i = 0, 1$ & $\xBj$ & 18 /\ 24 \\ 
    &  		&      & 			  & $A_{UT, \mathrm{DVCS}}^{\sin(\phi-\phi_{S})\cos i \phi}$ & $i = 0$ & & \\
    &  		&      & 			  & $A_{UT, \mathrm{I}}^{\sin(\phi-\phi_{S})\cos i \phi}$ & $i = 0, 1$ & & \\
    &  		&      & 			  & $A_{UT, \mathrm{I}}^{\cos(\phi-\phi_{S})\sin i \phi}$ & $i = 1$ & & \\                    
4   &  		& 2009 & \cite{Airapetian:2009aa} & $A_{LU, \mathrm{I}}^{\sin i \phi}$ & $i = 1, 2$ & $\xBj$ & 35 /\ 42 \\
    &  		&      & 			  & $A_{LU, \mathrm{DVCS}}^{\sin i \phi}$ & $i = 1$ & & \\     
    &  		&      & 			  & $A_{C}^{\cos i \phi}$ & $i = 0, 1, 2, 3$ & & \\   
5   &  		& 2010 & \cite{Airapetian:2010ab} & $A_{UL}^{+, \sin i \phi}$ & $i = 1, 2, 3$ & $\xBj$ & 18 /\ 24 \\
    &  		&      & 			  & $A_{LL}^{+, \cos i\phi}$ & $i = 0, 1, 2$ & & \\ 
6   &  		& 2011 & \cite{Airapetian:2011uq} & $A_{LT, \mathrm{DVCS}}^{\cos(\phi-\phi_{S})\cos i \phi}$ & $i = 0, 1$ & $\xBj$ & 24 /\ 32 \\   
    &  		&      & 			  & $A_{LT, \mathrm{DVCS}}^{\sin(\phi-\phi_{S})\sin i \phi}$ & $i = 1$ & & \\
    &  		&      & 			  & $A_{LT, \mathrm{I}}^{\cos(\phi-\phi_{S})\cos i \phi}$ & $i = 0, 1, 2$ & & \\
    &  		&      & 			  & $A_{LT, \mathrm{I}}^{\sin(\phi-\phi_{S})\sin i \phi}$ & $i = 1, 2$ & & \\
7   &  		& 2012 & \cite{Airapetian:2012mq} & $A_{LU, \mathrm{I}}^{\sin i \phi}$ & $i = 1, 2$ & $\xBj$ & 35 /\ 42 \\
    &  		&      & 			  & $A_{LU, \mathrm{DVCS}}^{\sin i \phi}$ & $i = 1$ & & \\ 
    &  		&      & 			  & $A_{C}^{\cos i \phi}$ & $i = 0, 1, 2, 3$ & & \\
8   & CLAS 	& 2001 & \cite{Stepanyan:2001sm}  & $A_{LU}^{-, \sin i \phi}$ & $i = 1, 2$ & --- & 0 /\ 2 \\
9   & 	 	& 2006 & \cite{Chen:2006na} 	  & $A_{UL}^{-, \sin i \phi}$ & $i = 1, 2$ & --- & 2 /\ 2 \\
10  & 	 	& 2008 & \cite{Girod:2007aa} 	  & $A_{LU}^{-}$ & & $\phi$ & 283 /\ 737\\
11  & 	 	& 2009 & \cite{Gavalian:2008aa}   & $A_{LU}^{-}$ & & $\phi$ & 22 /\ 33 \\
12  & 	 	& 2015 & \cite{Pisano:2015iqa}    & $A_{LU}^{-}$, $A_{UL}^{-}$, $A_{LL}^{-}$ & & $\phi$ & 311 /\ 497 \\
13  & 		& 2015 & \cite{Jo:2015ema}	  & $d^{4}\sigma_{UU}^{-}$ & & $\phi$ & 1333 /\ 1933 \\
14  & Hall A 	& 2015 & \cite{Defurne:2015kxq}   & $\Delta d^{4}\sigma_{LU}^{-}$ & & $\phi$ & 228 /\ 228 \\
15  & 		& 2017 & \cite{Defurne:2017paw}   & $\Delta d^{4}\sigma_{LU}^{-}$ & & $\phi$ & 276 /\ 358 \\ 
16  & COMPASS 	& 2018 & \cite{Akhunzyanov:2018nut} & $b$ & & --- & 1 /\ 1\\
    & 	 	&      & & & & & \\
    & 	 	&      & & & & SUM: & 2600 /\ 3970\\ \bottomrule
\end{tabular}
\end{table*}

\section{Uncertainties}
\label{sec:uncertainties}

In this analysis all the uncertainties are evaluated with the replica method, which for a sufficiently large set of replicas allows one to accurately reproduce the probability distribution of a given problem. We distinguish four types of uncertainties on the extracted parameterizations of CFFs. They origin from: \emph{i)} DVCS data, \emph{ii}) unpolarized PDFs, \emph{iii}) polarized PDFs) and \emph{iv}) EFFs. Each type of uncertainty is estimated independently, as described in the following. 

The uncertainties coming from DVCS data are estimated with a set of $100$ instances of replicated data. Similarly to the analysis of EFFs, see Sec. \ref{sec:elasticFF}, a single instance of replicated data is generated with the following prescription:
\begin{flalign}
&v_{i} \pm {\Delta}_{i}^{\mathrm{tot}} \xrightarrow[]{\mathrm{replica}~j} \nonumber \\
&\phantom{xx}\left(\mathrm{rnd}_{j}(v_{i}, {\Delta}_{i}^{\mathrm{tot}}) \pm {\Delta}_{i}^{\mathrm{tot}} \right) \times \mathrm{rnd}_{j}(1, {\Delta}_{i}^{\mathrm{norm}}) \;.
\label{eq:replication}
\end{flalign}
Here, $v_{i}$ is the measured value associated to the experimental point $i$, which comes with statistical, ${\Delta}_{i}^{\mathrm{stat}}$, systematic, ${\Delta}_{i}^{\mathrm{sys}}$, and normalization, ${\Delta}_{i}^{\mathrm{norm}}$, uncertainties. The latter one appears whenever the observable is sensitive to either beam or target polarization, or to both of them. In such cases the polarization describes the analyzing power for the extraction of those observable and the normalization uncertainties are related to the measurement or other determination of the involved polarizations. The total uncertainty, which is also used in the fit of CFFs to evaluate the $\chi^{2}$ value, is evaluated according to Eq. \eqref{eq:elasticFF:total_err}. 

The uncertainties coming from unpolarized and polarized PDFs are estimated by propagating our parameterizations of replicas by NNPDF group, see Sec. \ref{sec:selection_of_pdfs}. Namely, we repeat our fit to DVCS data separately for each PDF replica, that is $100$ times for unpolarized PDFs and $100$ times for polarized PDFs. A similar method is used to evaluate the uncertainties coming from EFF parameterizations. We repeat our fit to DVCS data for each replica obtained in the analysis of EFF data, see Sec. \ref{sec:elasticFF}.

As a result of this analysis we obtain a set of $401$ replicas, where each of them represents a possible realization of CFF parameterizations. For a given kinematic point the mean and uncertainties can be then estimated by taking the mean and the standard deviation of values evaluated from those replicas. 

\section{Results}
\label{sec:results}

\subsubsection*{Performance}

For the central PDF and EFF replicas the minimum value of the $\chi^{2}$ function returned by the minimization routine (Minuit \cite{James:1975dr} ported to ROOT \cite{Brun:1997pa}) is $2346.3$ for $2600$ experimental points and $13$ free parameters, which gives the reduced value equal to $2346.3/(2600-13) \approx 0.91$. The values of $\chi^{2}$ function per experimental data set are summarized in Table \ref{tab:chi2_per_dataset}. In general the agreement between our fit and experimental data is quite good. The only exception is for the COMPASS point, on which we will comment in the following. 

In this analysis a set of $401$ replicas is obtained that can be used to estimate mean values and uncertainties of the fitted parameters. We summarize this in Table \ref{tab:fitted_param}, where we also indicate ranges in which the minimization routine was allowed to vary the fitted parameters. As one can see from this table, the parameter $B_{H}^{q_{\mathrm{sea}}}$ is found to be at the lower limit of the allowed range. Without this limit the fit would end with a much smaller value of $B_{H}^{q_{\mathrm{sea}}}$ compensated by a higher value of $A_{H}^{q_{\mathrm{sea}}}$, which we consider to be unphysical. The problem with $B_{H}^{q_{\mathrm{sea}}}$ is a consequence of the sparse data covering low-$\xBj$ region, but it may be also a sign of the breaking of the LO description in the range dominated by sea quarks and gluons. 

In addition to the fitted parameters, in Table \ref{tab:fitted_param} we also show typical values of $b_{G}^{q}$ and $c_{G}^{q}$ evaluated from Eqs. \eqref{eq:singregul1} and \eqref{eq:singregul2}, respectively. These values indicate that the $t$-dependence in the skewness function is subdominant, which we consider to be an expected feature. 

\begin{figure*}[!ht]
\begin{center}
\includegraphics[width=0.49\textwidth]{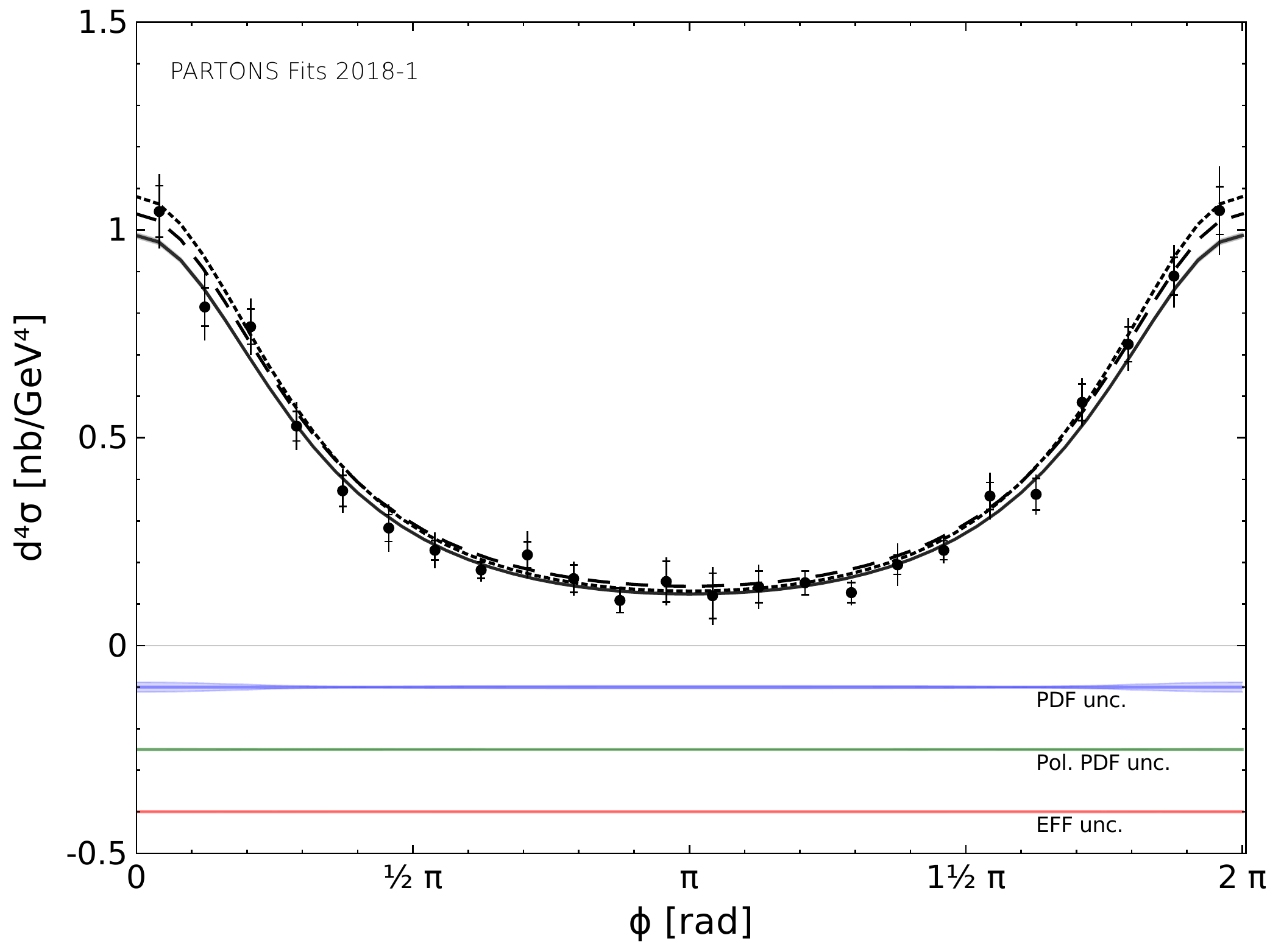}
\includegraphics[width=0.49\textwidth]{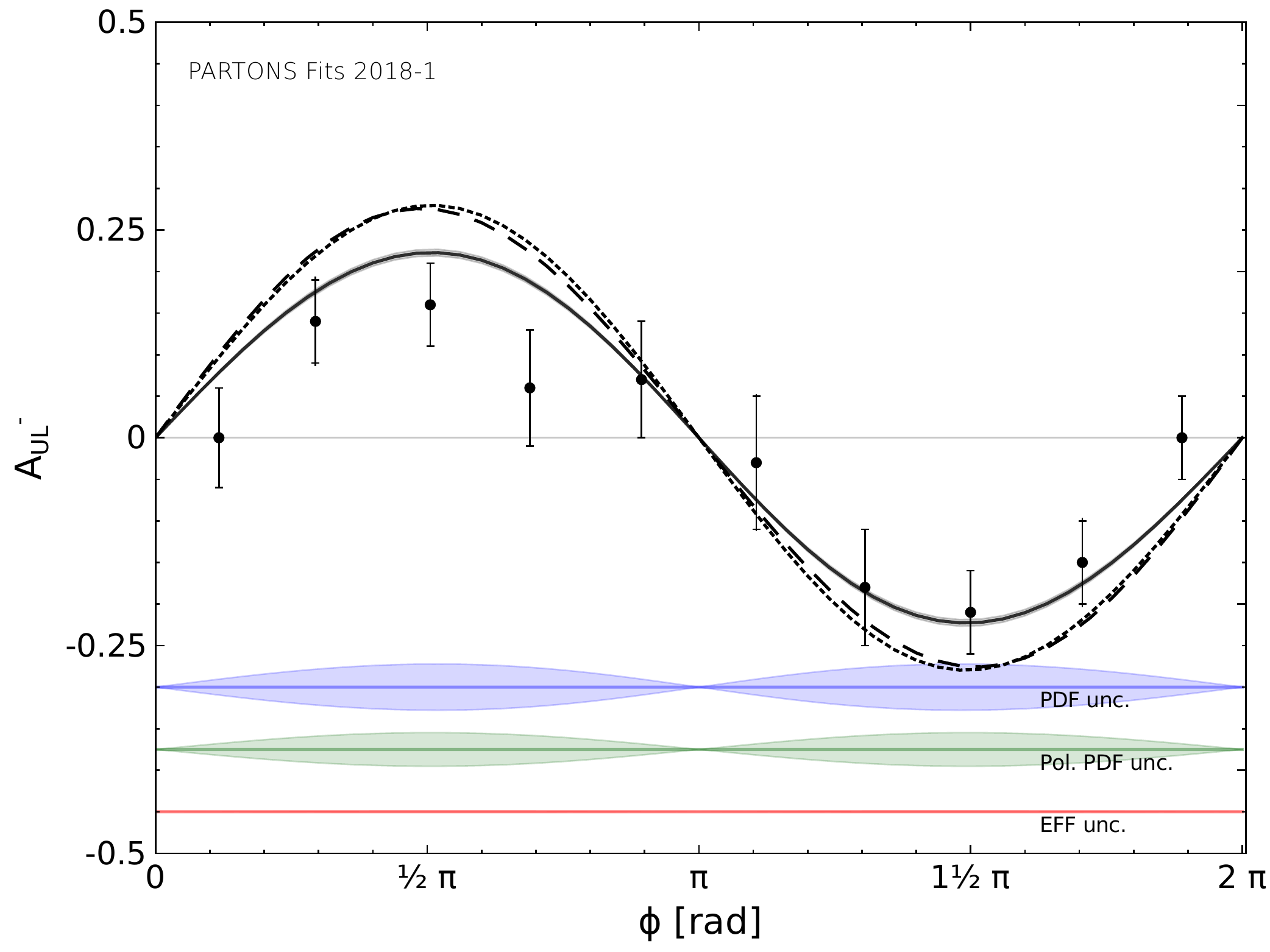}
\caption{Comparison between the results of this analysis, some selected GPD models and experimental data published by CLAS in Refs. \cite{Jo:2015ema, Pisano:2015iqa} for $d^{4}\sigma_{UU}^{-}$ at $\xBj = 0.244$, $t = -0.15~\mathrm{GeV}^2$ and $Q^{2} = 1.79~\mathrm{GeV}^2$ (left) and for $A_{UL}^{-}$ at $\xBj = 0.2569$, $t = -0.23~\mathrm{GeV}^2$, $Q^{2} = 2.019~\mathrm{GeV}^2$ (right). The solid curves and the gray bands surrounding those curves correspond to the results of this analysis and 68\% confidence levels for the uncertainties coming from DVCS data, respectively. The magnitudes of the additional uncertainties to the plotted observables and coming from unpolarized PDFs, polarized PDFs and EFFs, can be separately estimated from the bands located below the curves and labeled with ``PDF unc.'', ``Pol. PDF unc.'' and ``EFF unc.'', respectively. Each of those four uncertainties is evaluated in the analysis of the corresponding set of replicas. The inner bars on data points are for statistical uncertainties, while those outer ones are for the quadratic sum of statistical and systematic uncertainties. The dotted curve is for the GK GPD model \cite{Goloskokov:2005sd, Goloskokov:2007nt, Goloskokov:2009ia}, while the dashed one is for VGG \cite{Vanderhaeghen:1998uc, Vanderhaeghen:1999xj, Goeke:2001tz, Guidal:2004nd}. The curves are evaluated at the kinematics of experimental data.}
\label{fig:results:clas}
\end{center}
\end{figure*}

\begin{figure*}[!ht]
\begin{center}
\includegraphics[width=0.49\textwidth]{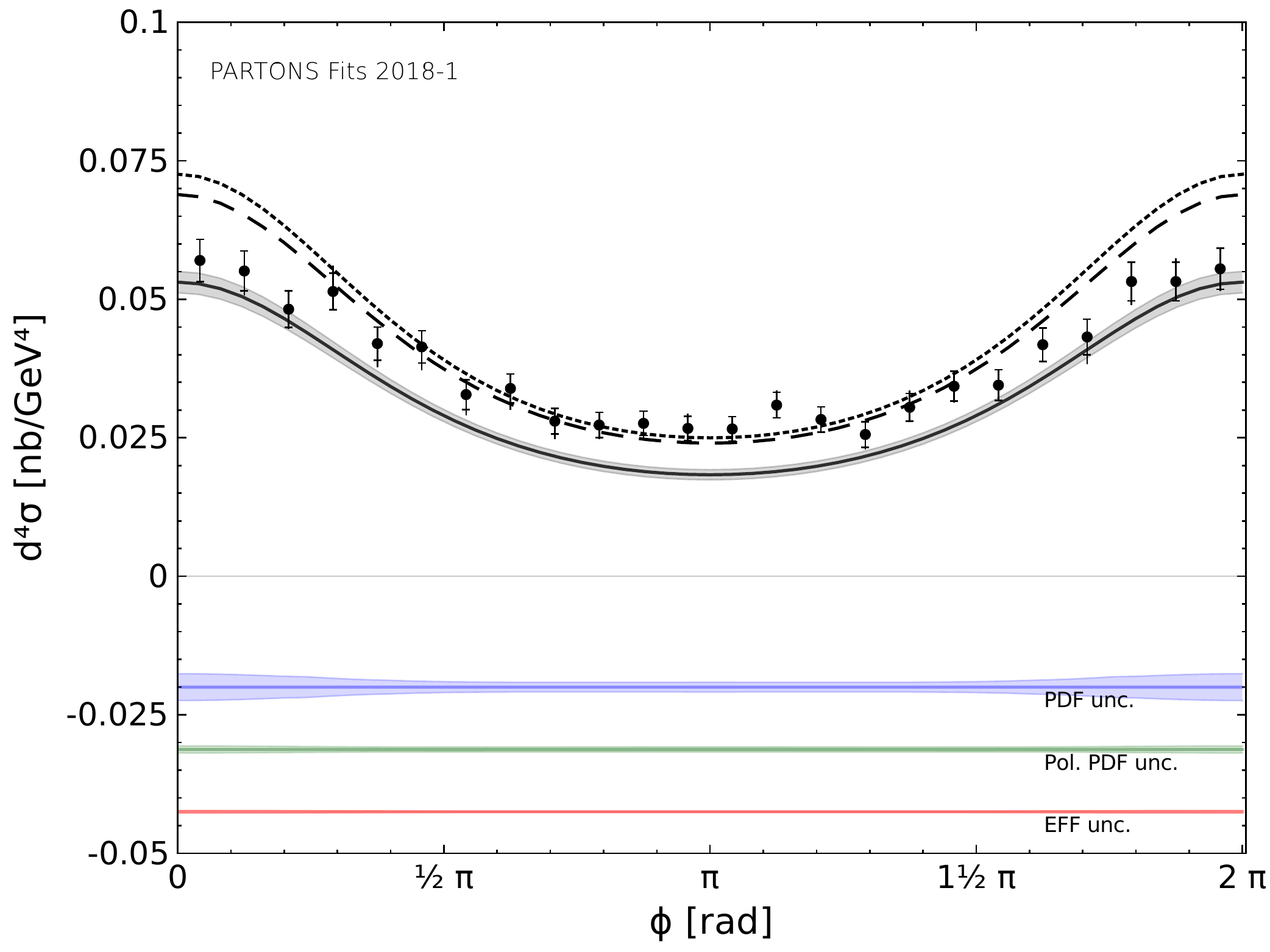}
\includegraphics[width=0.49\textwidth]{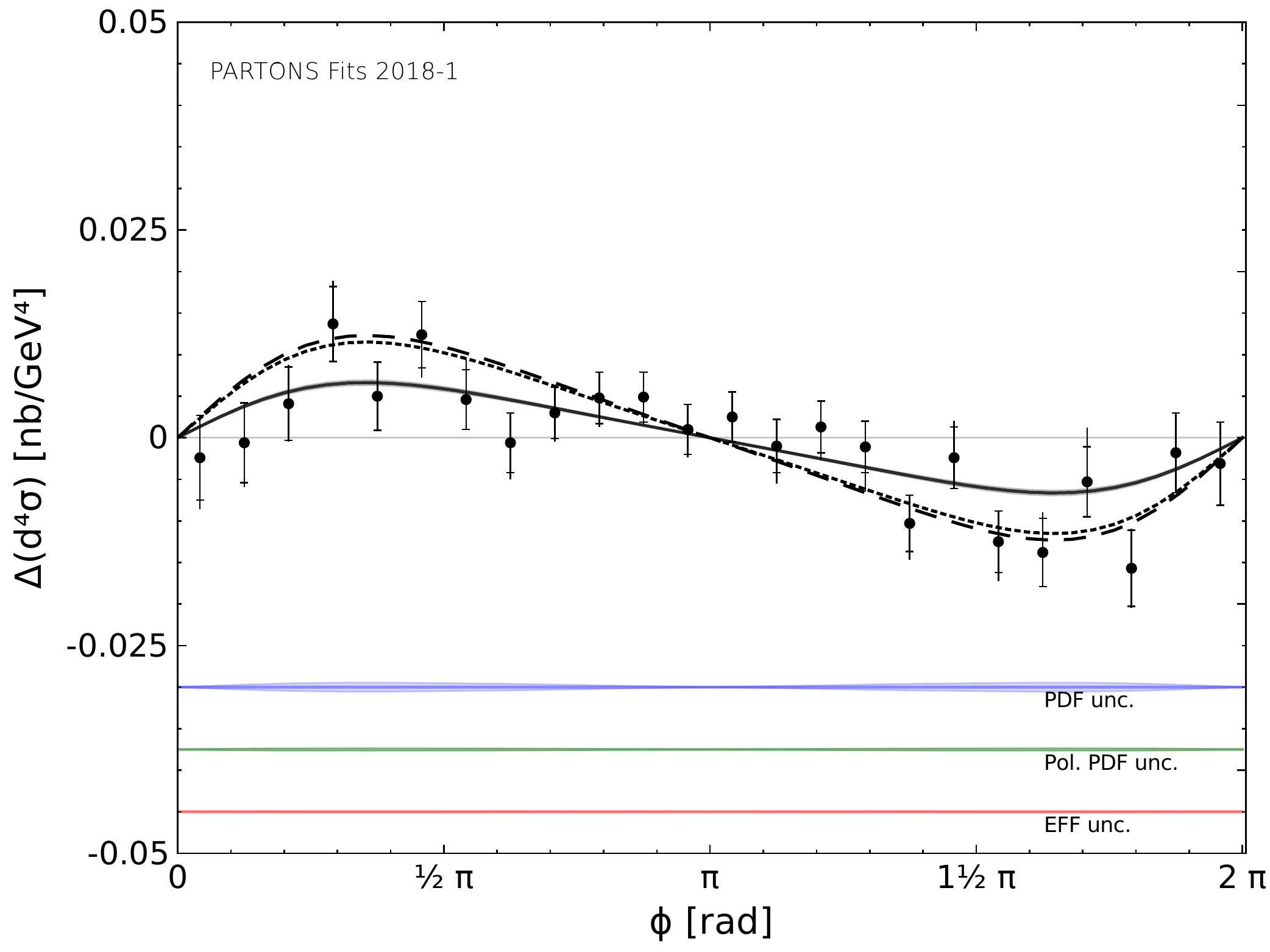}
\caption{Comparison between the results of this analysis, some selected GPD models and experimental data published by Hall A in Ref. \cite{Defurne:2015kxq} for $d^{4}\sigma_{UU}^{-}$ (left) and $\Delta d^{4}\sigma_{LU}^{-}$ (right) at $\xBj = 0.392$, $t = -0.233~\mathrm{GeV}^2$ and $Q^{2} = 2.054~\mathrm{GeV}^2$. For further description see Fig. \ref{fig:results:clas}.}
\label{fig:results:halla}
\end{center}
\end{figure*}

\begin{figure*}[!ht]
\begin{center}
\includegraphics[width=0.49\textwidth]{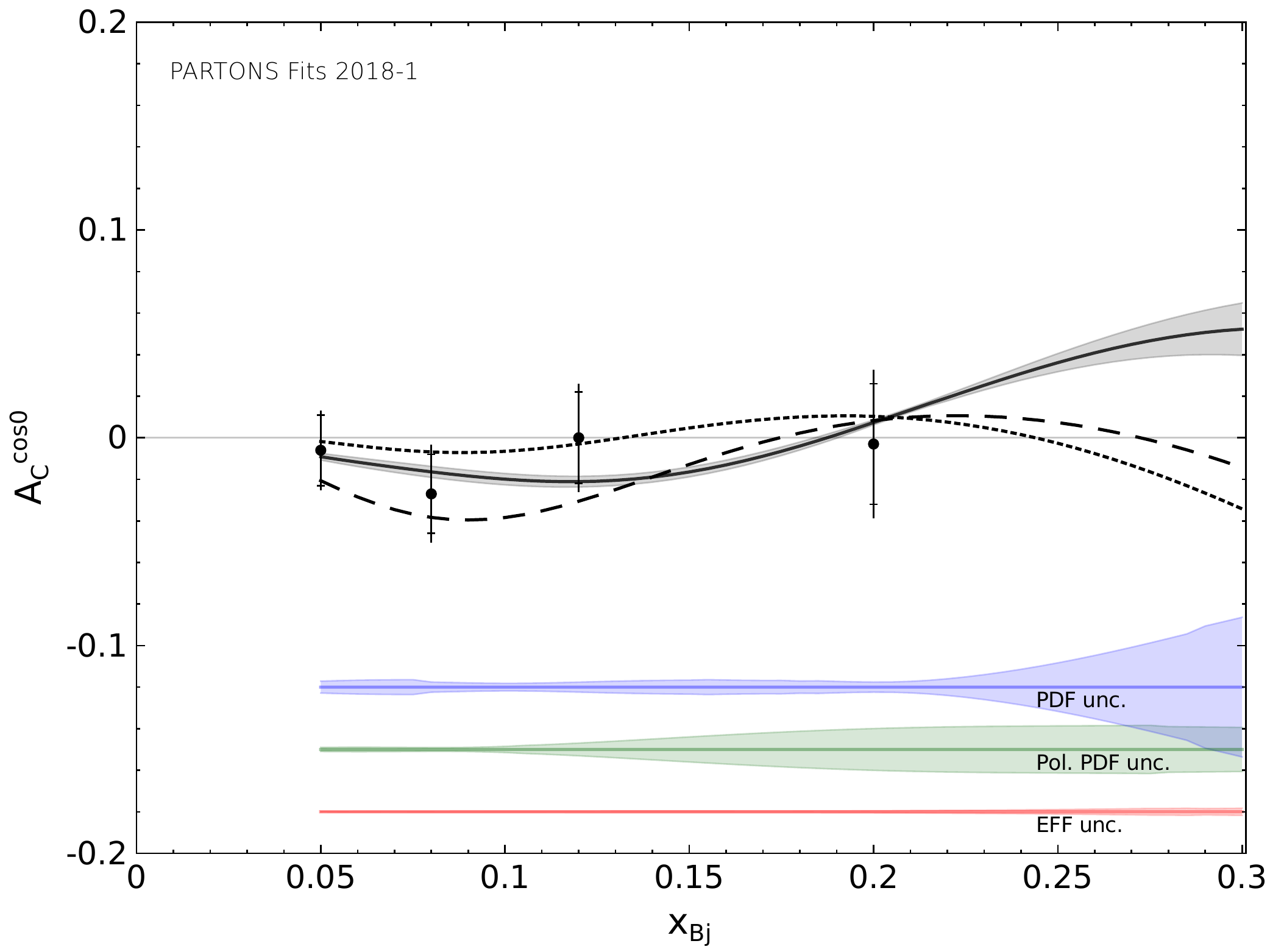}
\includegraphics[width=0.49\textwidth]{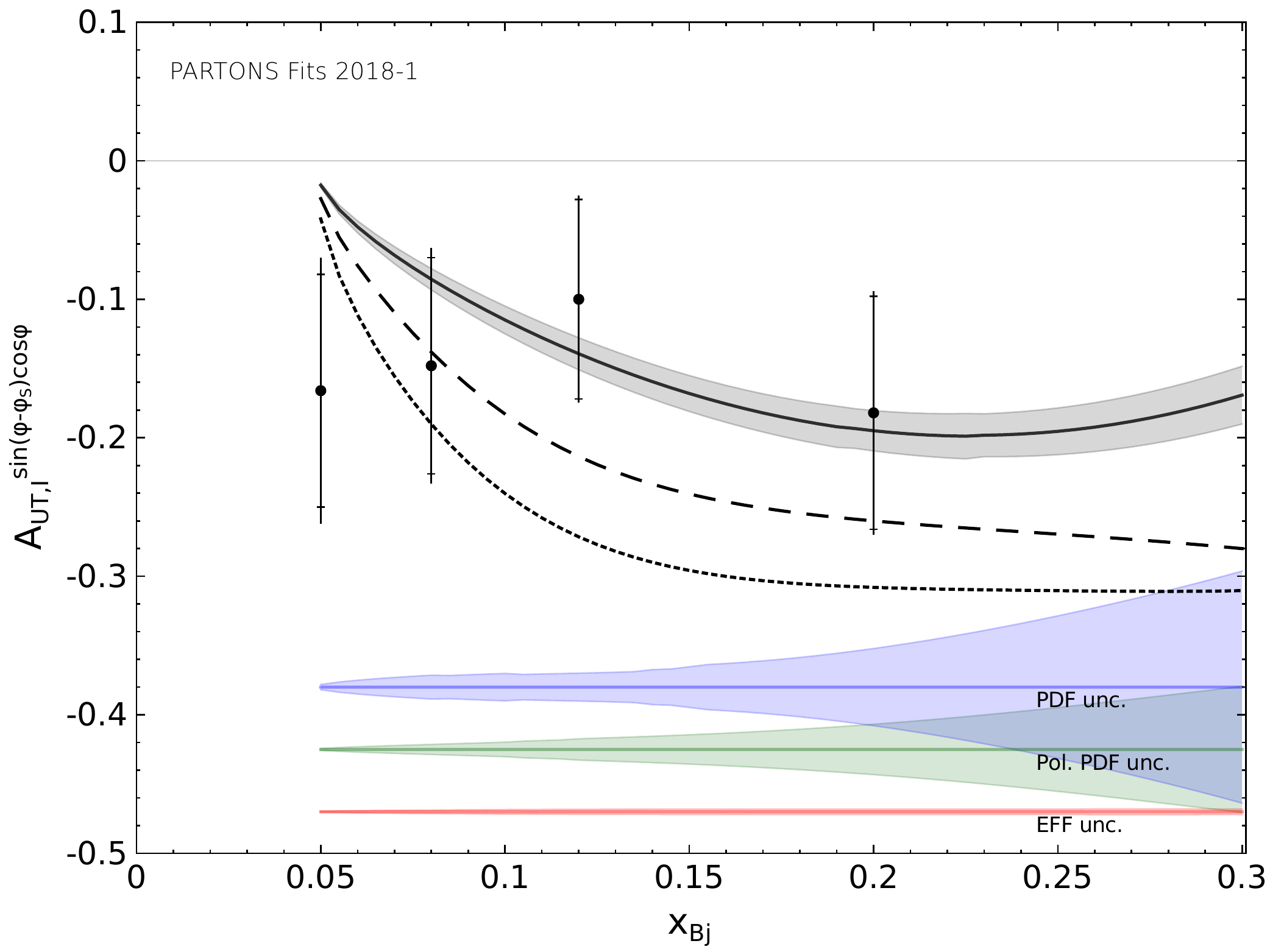}
\caption{Comparison between the results of this analysis, some selected GPD models and experimental data published by HERMES in Ref. \cite{Airapetian:2008aa} for $A_{C}^{\cos 0 \phi}$ (left) and $A_{UT, \mathrm{I}}^{\sin(\phi-\phi_{S})\cos \phi}$ (right) at $t = -0.12~\mathrm{GeV}^2$ and $Q^{2} = 2.5~\mathrm{GeV}^2$. For further description see Fig. \ref{fig:results:clas}.}
\label{fig:results:hermess}
\end{center}
\end{figure*}

\begin{figure*}[!ht]
\begin{center}
\includegraphics[width=0.49\textwidth]{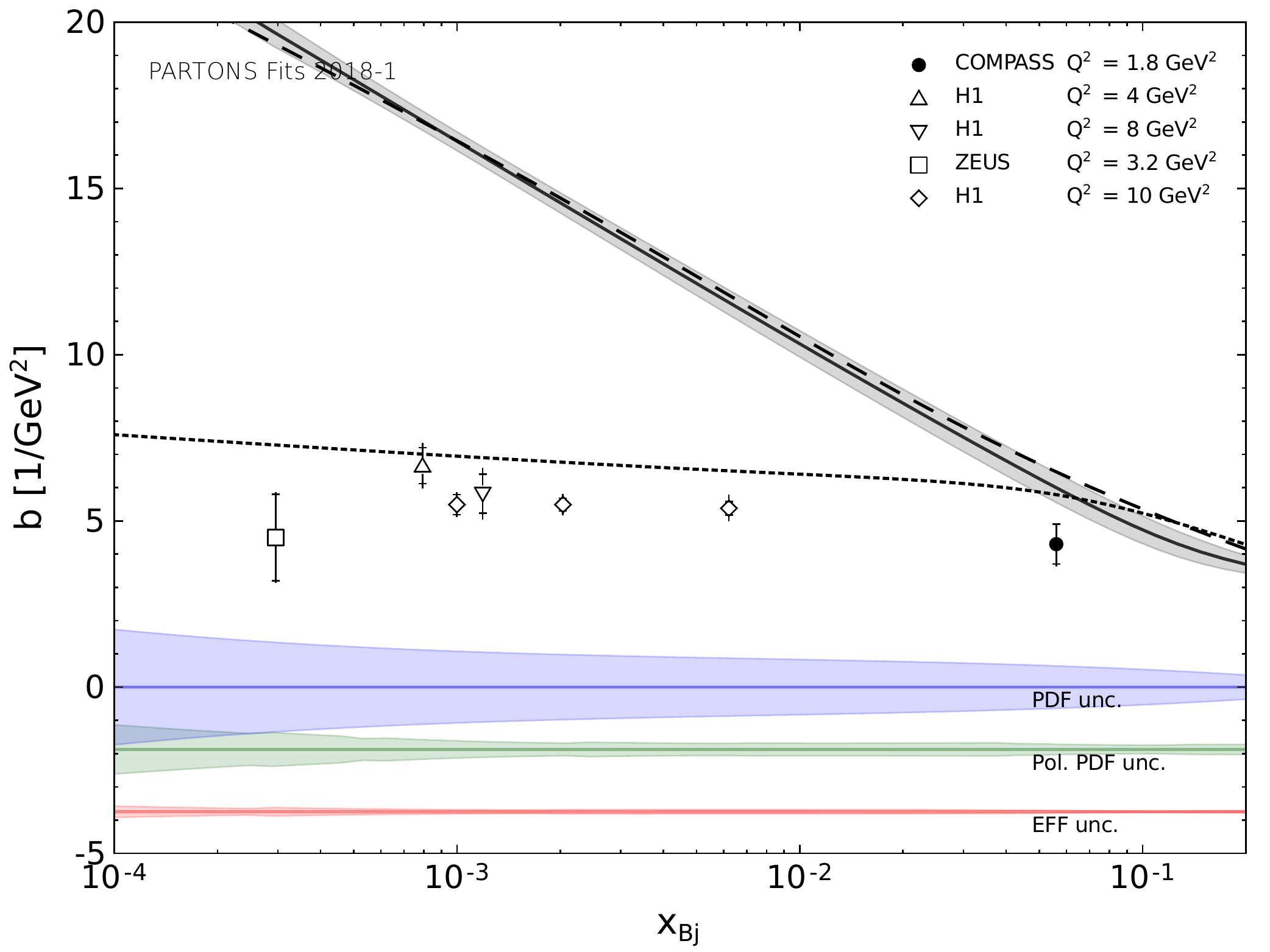}
\caption{Comparison between the results of this analysis, selected GPD models and experimental data published by COMPASS in Ref. \cite{Akhunzyanov:2018nut} for the slope $b$ at $Q^{2} = 1.8~\mathrm{GeV}^2$. Except COMPASS data also those obtained in ZEUS \cite{Chekanov:2008vy} and H1 \cite{Aktas:2005ty, Aaron:2009ac} experiments are shown. Note that the shown data differ by $Q^{2}$ values indicated in the legend. For further description see Fig. \ref{fig:results:clas}.}
\label{fig:results:compass}
\end{center}
\end{figure*}

Figures \ref{fig:results:clas}-\ref{fig:results:compass} provide a straightforward comparison between the results of our analysis and some selected data sets coming from various experiments. These plots can be used in particular to estimate the effect of PDF and EFF uncertainties, which one can not judge from Table \ref{tab:chi2_per_dataset}. 

In Fig. \ref{fig:results:halla} the comparison for both $d^{4}\sigma_{UU}^{-}$ and $\Delta d^{4}\sigma_{LU}^{-}$ data coming from Hall A is shown, however one should keep in mind that only those for $\Delta d^{4}\sigma_{LU}^{-}$ are used in this analysis. In general the agreement between the unpolarized cross section data published by Hall A \cite{Defurne:2015kxq, Defurne:2017paw} and our fit is poor, which can be qualitatively expressed be the reduced value of $\chi^{2}$ function evaluated for those data equal to $5144.4 / 594 \approx 8.66$. The agreement is better for $\phi \simeq 0$, where for $0 < \phi < 45^{\circ}$ and $315^{\circ} < \phi < 360^{\circ}$ one has $232.4 / 132 \approx 1.76$. We note that the KM model also has a problem to describe Hall A unpolarized cross section data at LT and LO accuracy \cite{Kumericki:2015lhb}, and it was suggested \cite{Defurne:2017paw} that including HT and NLO corrections may be needed to improve the quality of the fit. 

Figure \ref{fig:results:compass} demonstrates the comparison between results of our fit and experimental data for the $t$-slope $b$ coming from the measurements by COMPASS \cite{Akhunzyanov:2018nut}, ZEUS \cite{Chekanov:2008vy} and H1 \cite{Aktas:2005ty, Aaron:2009ac} experiments. We remind that for this observable only the COMPASS point is used in this analysis. As one can judge from the figure, the agreement between our fit and experimental data is lost for small $\xBj$, \emph{i.e.} for the range where sea quarks and gluons dominate. We only leave the COMPASS point to have a coverage in the intermediate range of $\xBj$, where valence quarks may still contribute significantly. While it is tempting to improve the agreement for small $\xBj$ by adding extra terms to $f_{H}^{q_{\mathrm{sea}}}(x)$, like those proportional to $(1-x)^n$ with large $n$, we refrain from doing that, treating the disagreement as a possible manifestation of NLO effects. The included comparison with GK \cite{Goloskokov:2005sd, Goloskokov:2007nt, Goloskokov:2009ia} and VGG \cite{Vanderhaeghen:1998uc, Vanderhaeghen:1999xj, Goeke:2001tz, Guidal:2004nd} GPD models shows that GK manages to reproduce the trend dictated by the HERA data reasonably well. This is not a full surprise, as this model was constrained in the low-$\xBj$ region, however by Deeply Virtual Meson Production (DVMP) data. It is worth pointing out, that the lowest possible order of contribution to DVMP starts with $\alpha_{S}^1$, while for DVCS one has $\alpha_{S}^{0}$. It is beyond the scope of this paper, but in the future multichannel analyses of exclusive processes the issue of using the same order of pQCD calculations may become important.

\bgroup 
\setlength{\tabcolsep}{5pt}
\begin{table}[!ht]
\centering
\caption{Values of the $\chi^{2}$ function per data set. For a given data set, \emph{cf}. Table \ref{tab:used_data}, given are: $\chi^{2}$ value, the number of experimental points $n$, and the ratio between these two numbers.}
\label{tab:chi2_per_dataset}
\begin{tabular}{ccccccc}
\toprule
No. & Collab. 	& Year & Ref. & $\chi^{2}$ & $n$ &  $\chi^{2}/n$\\ \midrule
1   & HERMES 	& 2001 & \cite{Airapetian:2001yk}   &  $9.8$  	&   $10$ &  $0.98$  \\
2   &  		& 2006 & \cite{Airapetian:2006zr}   &  $2.9$  	&    $4$ &  $0.72$  \\
3   &  		& 2008 & \cite{Airapetian:2008aa}   &  $24.2$ 	&   $18$ &  $1.35$  \\       
4   &  		& 2009 & \cite{Airapetian:2009aa}   &  $40.1$ 	&   $35$ &  $1.15$  \\
5   &  		& 2010 & \cite{Airapetian:2010ab}   &  $40.3$  	&   $18$ &  $2.24$  \\
6   &  		& 2011 & \cite{Airapetian:2011uq}   &  $14.5$ 	&   $24$ &  $0.60$  \\
7   &  		& 2012 & \cite{Airapetian:2012mq}   &  $25.4$ 	&   $35$ &  $0.73$  \\
8   & CLAS 	& 2001 & \cite{Stepanyan:2001sm}    & ---   	&    $0$ & ---    	\\
9   & 	 	& 2006 & \cite{Chen:2006na} 	    &  $0.9$  	&    $2$ &  $0.47$  \\
10  & 	 	& 2008 & \cite{Girod:2007aa} 	    &  $371.1$ 	&  $283$ &  $1.31$  \\
11  & 	 	& 2009 & \cite{Gavalian:2008aa}     &  $36.4$ 	&   $22$ &  $1.66$  \\
12  & 	 	& 2015 & \cite{Pisano:2015iqa} 	    &  $351.4$ 	&  $311$ &  $1.13$  \\
13  & 		& 2015 & \cite{Jo:2015ema} 	    	&  $937.9$ 	& $1333$ &  $0.70$  \\
14  & Hall A 	& 2015 & \cite{Defurne:2015kxq} &  $220.2$ 	&  $228$ &  $0.97$  \\
15  & 		& 2017 & \cite{Defurne:2017paw}     &  $258.8$	&  $276$ &  $0.94$  \\
16  & COMPASS 	& 2018 & \cite{Akhunzyanov:2018nut} &  $10.7$  	&    $1$ &  $10.67$  \\  \bottomrule
\end{tabular}
\end{table}
\egroup 

\begin{table*}[!ht]
\centering
\caption{Values of the parameters fitted to DVCS data together with estimated uncertainties coming from those data, (un-)polarized PDFs and EFFs. Two last columns indicate the limits in which the minimization routine was allowed to vary the corresponding parameters. In addition, exemplary values of $b_{G}^{q}$ and $c_{G}^{q}$ parameters evaluated at $Q^{2} = 2~\mathrm{GeV}^2$ from Eqs. \eqref{eq:singregul1} and \eqref{eq:singregul2} are given.}
\label{tab:fitted_param}
\begin{tabular}{cccccccc}
\toprule	
\multirow{2}{*}{Parameter} & \multirow{2}{*}{Mean} & \multirow{2}{*}{Data unc.} & \multirow{2}{*}{Unpol. PDF unc.} & \multirow{2}{*}{Pol. PDF unc.} & \multirow{2}{*}{EFF unc.} & \multicolumn{2}{c}{Limit} \\
& & & & & & min & max \\ \midrule
$a_{H}^{q_{\mathrm{val}}}$  & $\phantom{-}0.81$     & $0.04$	& $0.17$	& $0.02$	& $<0.01$ 	& $0.2$	    & $2.0$	\\
$a_{H}^{q_{\mathrm{sea}}}$  & $\phantom{-}0.99$ 	& $0.01$	& $0.02$	& $<0.01$ 	& $<0.01$	& $0.2$	    & $2.0$	\\
$a_{\widetilde{H}}^{q}$	    & $\phantom{-}1.03$     & $0.04$	& $0.30$	& $0.24$ 	& $0.01$ 	& $0.2$	    & $2.0$	\\ 
\midrule[0pt]
$N_{\widetilde{E}}$	    	& 			$-0.46$		& $0.10$	& $0.09$	& $0.06$ 	& $0.01$ 	& $-10$	    & $10$	\\ 
\midrule[0pt]
$A_{H}^{q_{\mathrm{sea}}}$  & $\phantom{-}2.56$	    & $0.23$	& $0.30$	& $0.09$ 	& $0.03$ 	& $0.1$	    & $10$	\\ 
$B_{H}^{q_{\mathrm{sea}}}$  & $-5$	& \multicolumn{4}{c}{at the limit} 		  			& $-5$	    & $20$	\\
$C_{H}^{q_{\mathrm{sea}}}$  & $\phantom{-}34$	    & $27$      & $49$		& $14$ 		& $3$ 		& $-5$	    & $200$	\\ 
\midrule[0pt]
$A_{\widetilde{H}}^{u_{\mathrm{val}}}$	& $\phantom{-}0.77$		& $0.12$    & $0.30$	& $0.23$	& $0.07$    & $0.1$	& $10$	\\
$B_{\widetilde{H}}^{u_{\mathrm{val}}}$	& 				$-0.02$	& $0.26$	& $0.75$	& $0.24$ 	& $0.15$    & $-5$	& $20$	\\
$C_{\widetilde{H}}^{u_{\mathrm{val}}}$  & 				$-0.92$ & $0.07$	& $0.44$	& $0.24$    & $0.04$ 	& $-5$	& $200$	\\ 
\midrule[0pt]
$A_{\widetilde{H}}^{d_{\mathrm{val}}}$	& $\phantom{-}0.64$		& $0.24$	& $0.30$	& $0.28$ 	& $0.05$   	& $0.1$	& $10$	\\
$B_{\widetilde{H}}^{d_{\mathrm{val}}}$	& 				$-1.19$	& $0.45$ 	& $0.91$    & $0.98$ 	& $0.22$    & $-5$	& $20$	\\
$C_{\widetilde{H}}^{d_{\mathrm{val}}}$	& 				$-0.55$	& $0.24$	& $0.26$	& $0.27$ 	& $0.10$    & $-5$	& $200$	\\ 
\midrule[0pt]
$b_{H}^{u_{\mathrm{val}}}$	& $-0.36$	& $0.10$	& $0.15$	& $0.04$ 	& $0.01$   & ---	& ---	\\
$c_{H}^{u_{\mathrm{val}}}$	& $\phantom{-}11.2$		& $3.1$		& $2.7$		& $1.1$ 	& $0.3$   & ---	& ---	\\
$b_{H}^{d_{\mathrm{sea}}}$	& $-0.222$		& $0.062$	& $0.090$	& $0.022$ 	& $0.006$   & ---	& ---	\\
$c_{H}^{d_{\mathrm{sea}}}$	& $\phantom{-}14$	& $4$		& $15$		& $1$ 	& $1$   & ---	& ---	\\
\bottomrule
\end{tabular}
\end{table*}

\subsubsection*{Subtraction constant}

\begin{figure*}[!ht]
\begin{center}
\includegraphics[width=0.49\textwidth]{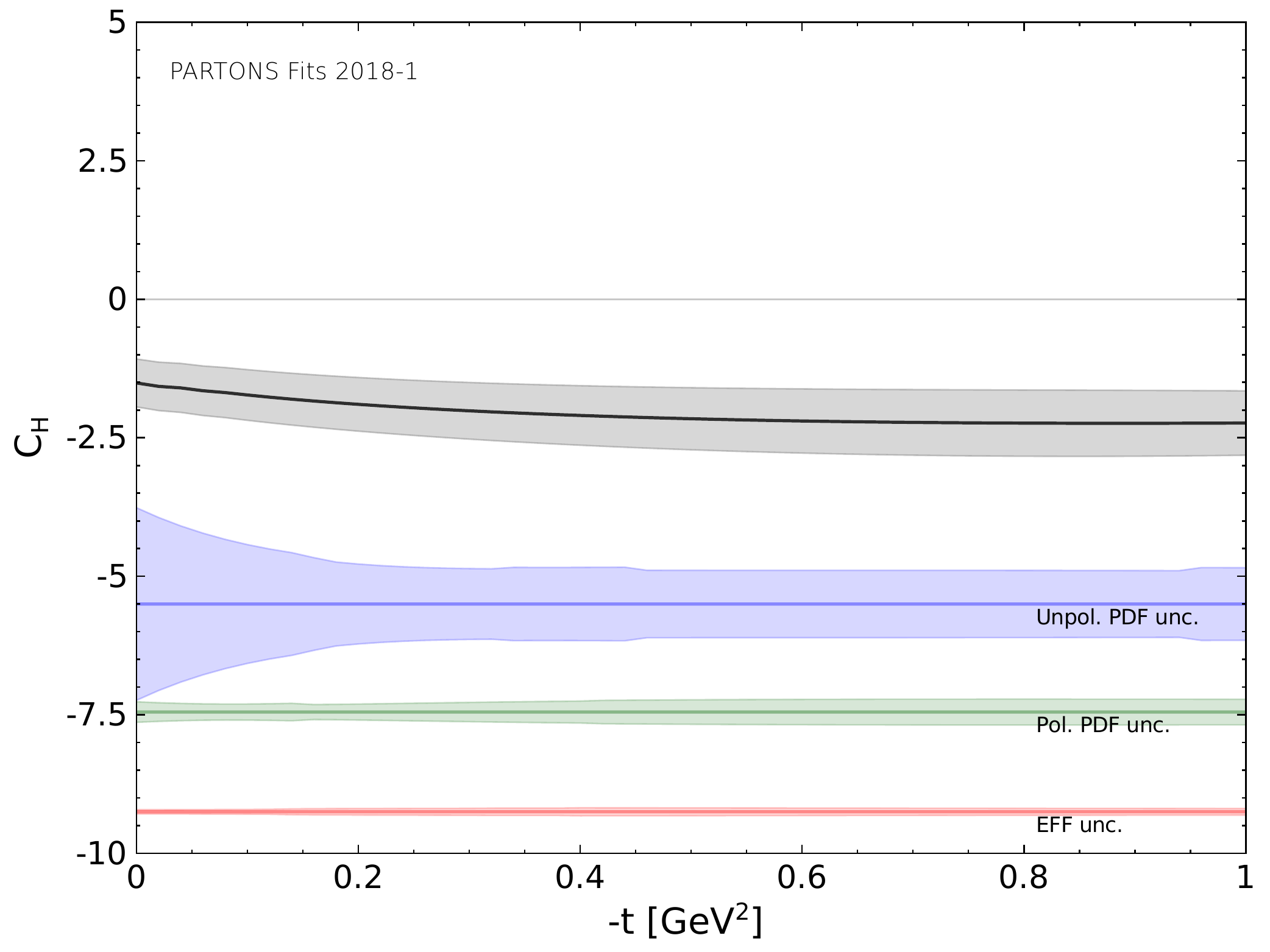}
\includegraphics[width=0.49\textwidth]{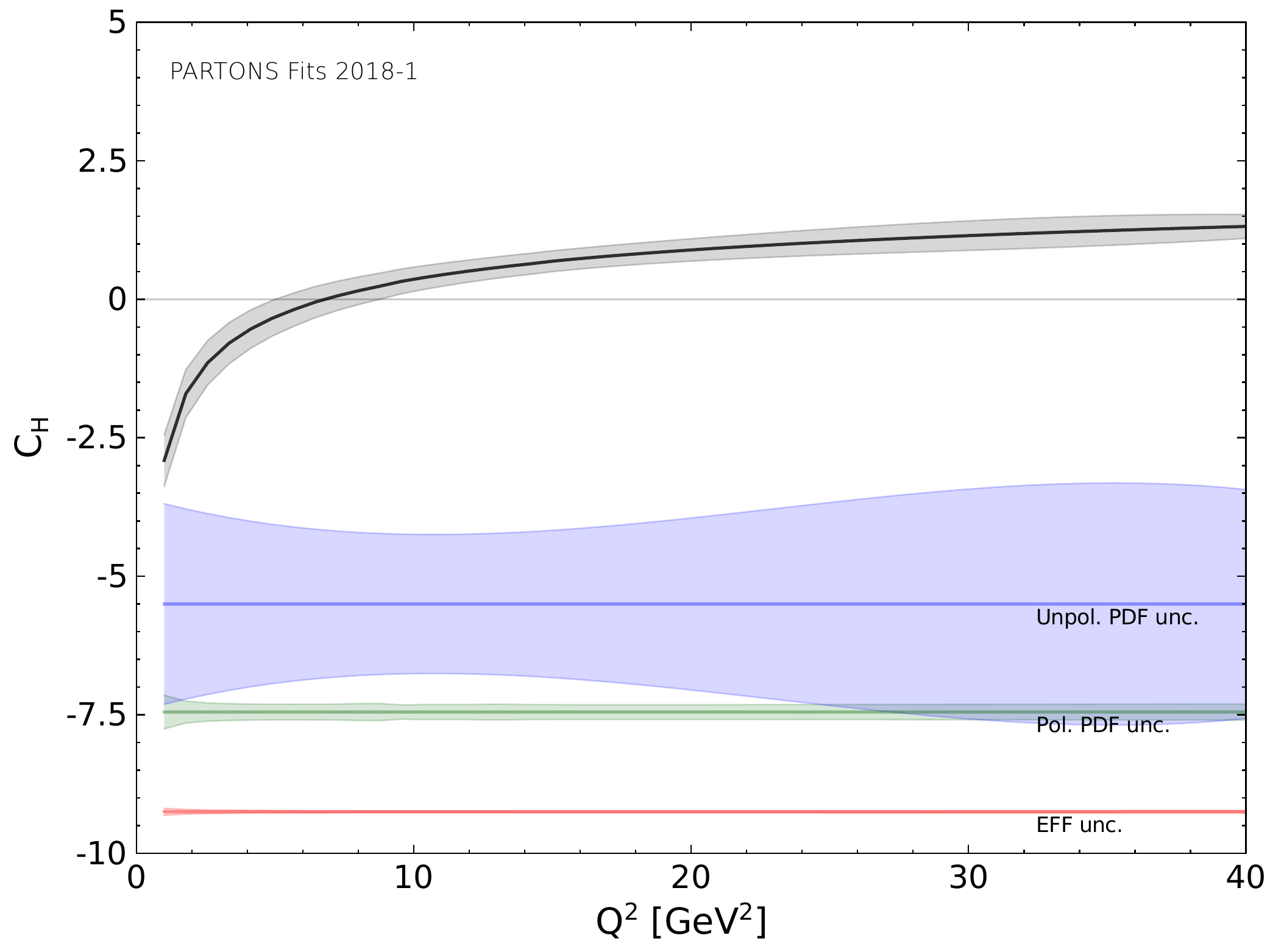}
\caption{Subtraction constant obtained in this work as a function of $t$ at $Q^{2} = 2~\mathrm{GeV}^{2}$ (left) and as a function of $Q^{2}$ at $t = 0$ (right). For further description see Fig. \ref{fig:results:clas}.}
\label{fig:SC}
\end{center}
\end{figure*}

The subtraction constant obtained in this analysis is shown as a function of $t$ at $Q^{2} = 2~\mathrm{GeV}^{2}$ and as a function of $Q^{2}$ at $t = 0$ in Fig. \ref{fig:SC}. One may observe a large uncertainty coming from PDF parameterizations, however in general small values of the subtraction constant are preferred. We also observe a general trend for the saturation of the subtraction constant at large values of $Q^{2}$. 

\subsubsection*{Compton Form Factors}

\begin{figure*}[!ht]
\begin{center}
\includegraphics[width=0.49\textwidth]{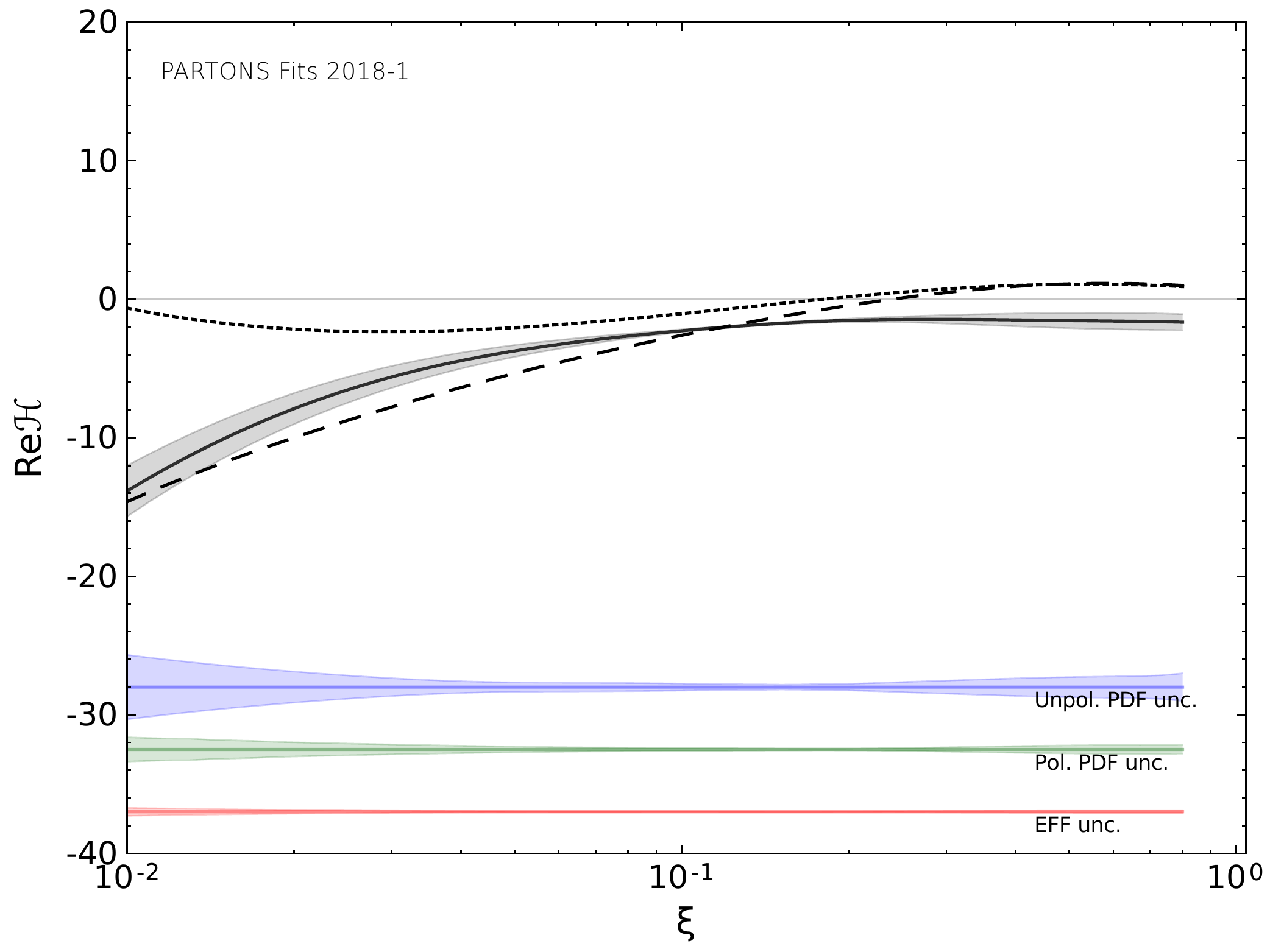}
\includegraphics[width=0.49\textwidth]{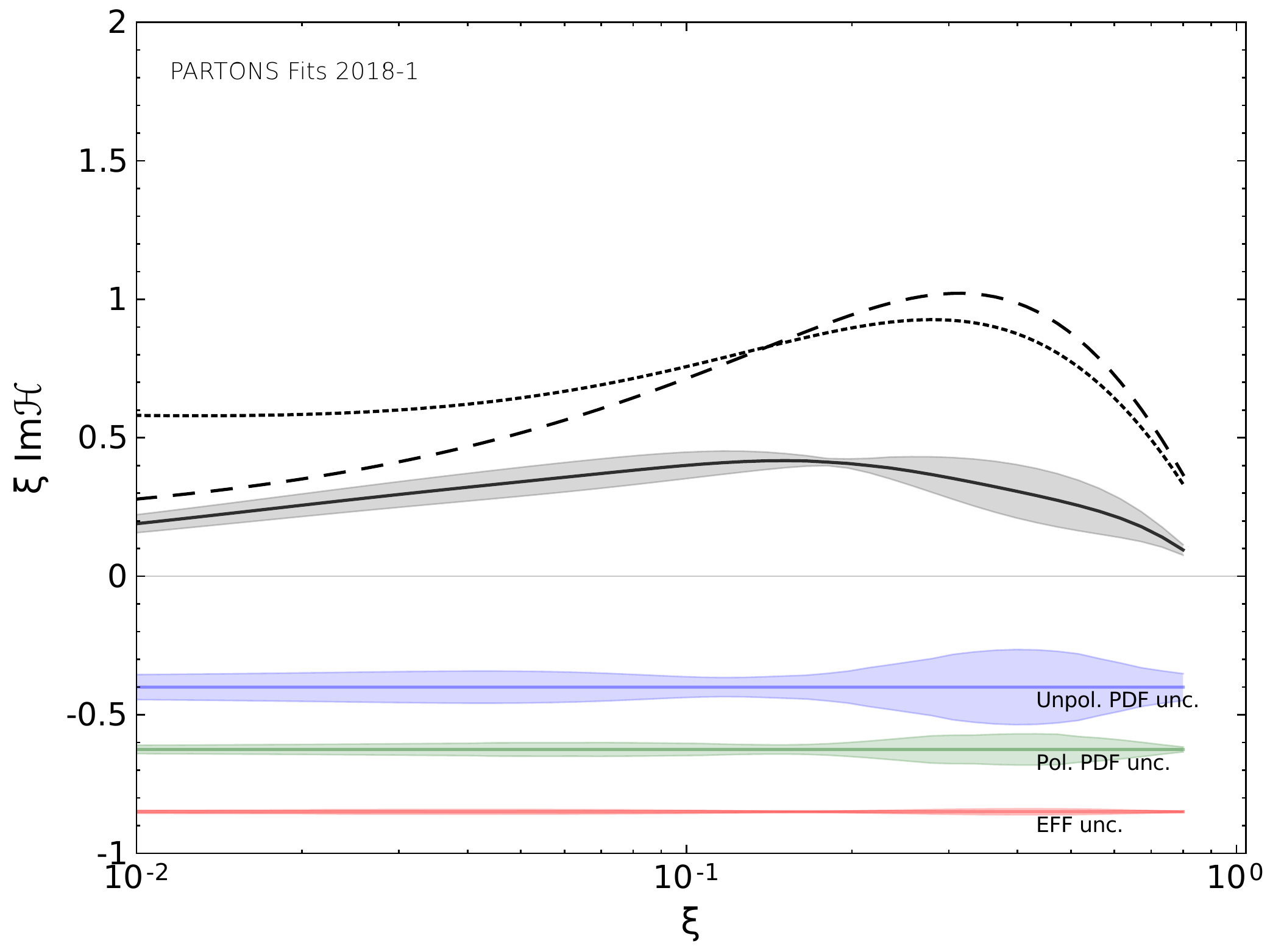}
\caption{Real (left) and imaginary (right) parts of the CFF $\mathcal{H}$ obtained in this work as a function of $\xi$ at $t = -0.3~\mathrm{GeV}^{2}$ and $Q^{2} = 2~\mathrm{GeV}^{2}$. For further description see Fig. \ref{fig:results:clas}.}
\label{fig:CFF:H}
\end{center}
\end{figure*}

\begin{figure*}[!ht]
\begin{center}
\includegraphics[width=0.49\textwidth]{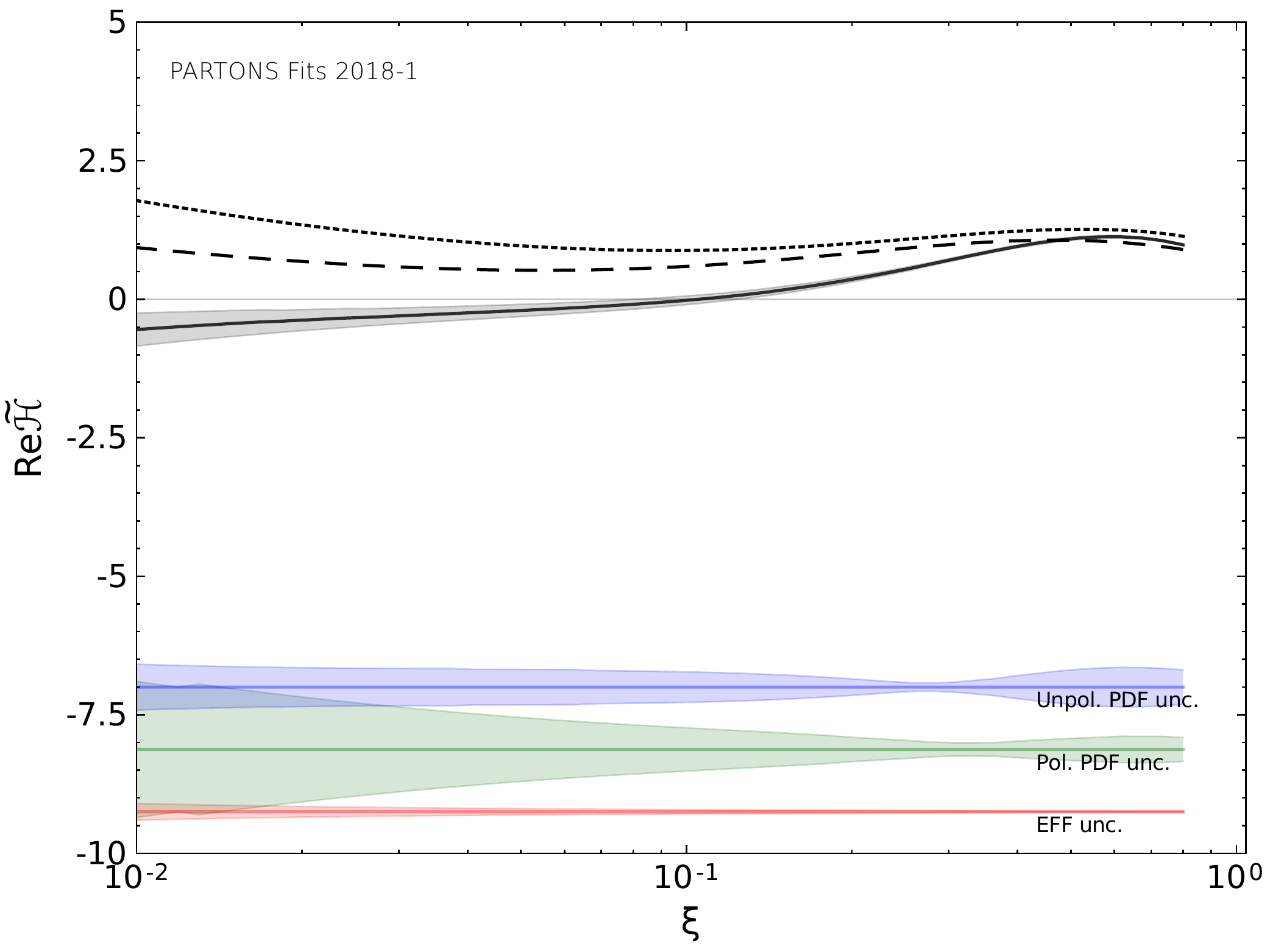}
\includegraphics[width=0.49\textwidth]{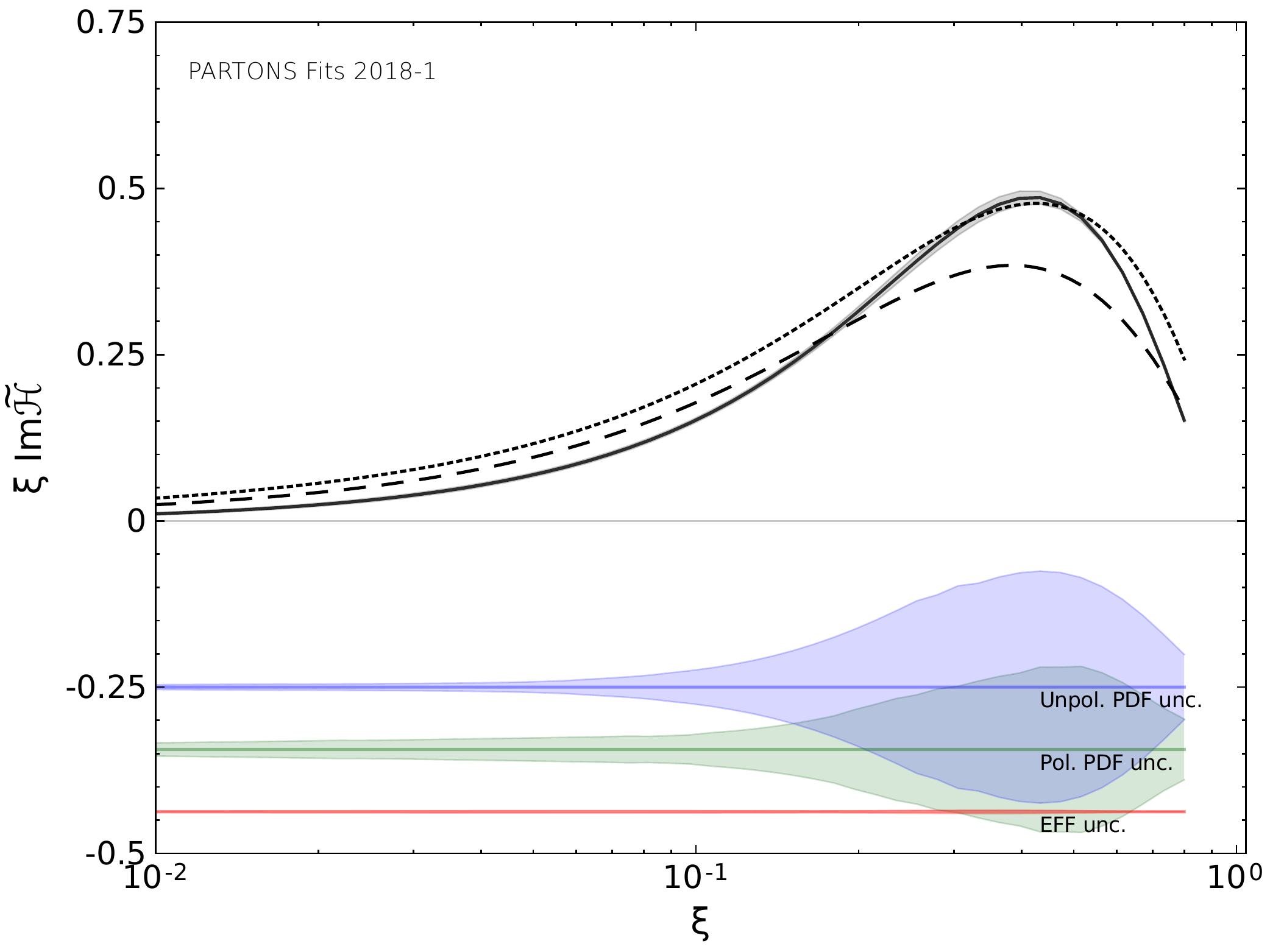}
\caption{Real (left) and imaginary (right) parts of the CFF $\widetilde{\mathcal{H}}$ obtained in this work as a function of $\xi$ at $t = -0.3~\mathrm{GeV}^{2}$ and $Q^{2} = 2~\mathrm{GeV}^{2}$. For further description see Fig. \ref{fig:results:clas}.}
\label{fig:CFF:Ht}
\end{center}
\end{figure*}

\begin{figure*}[!ht]
\begin{center}
\includegraphics[width=0.49\textwidth]{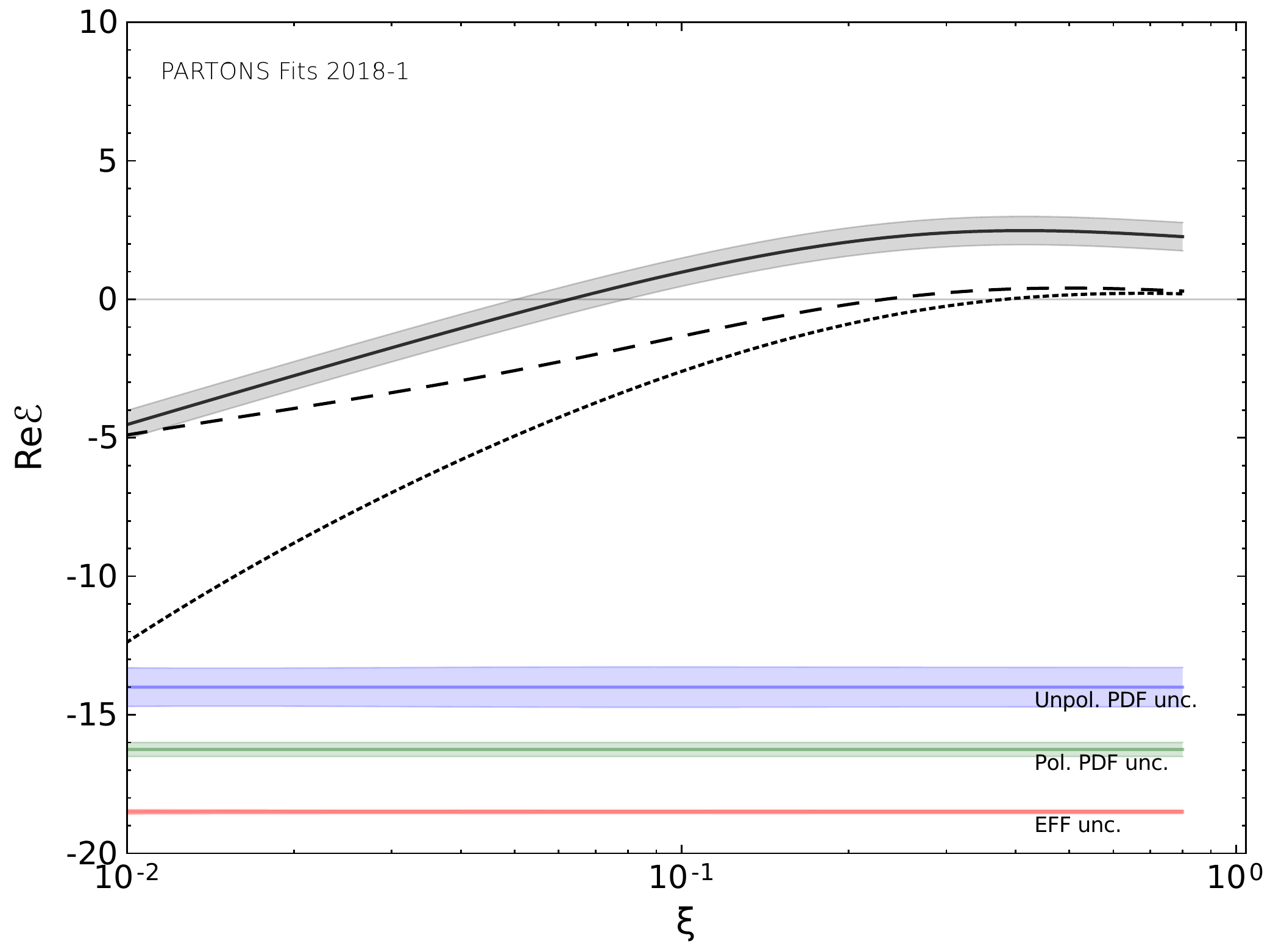}
\includegraphics[width=0.49\textwidth]{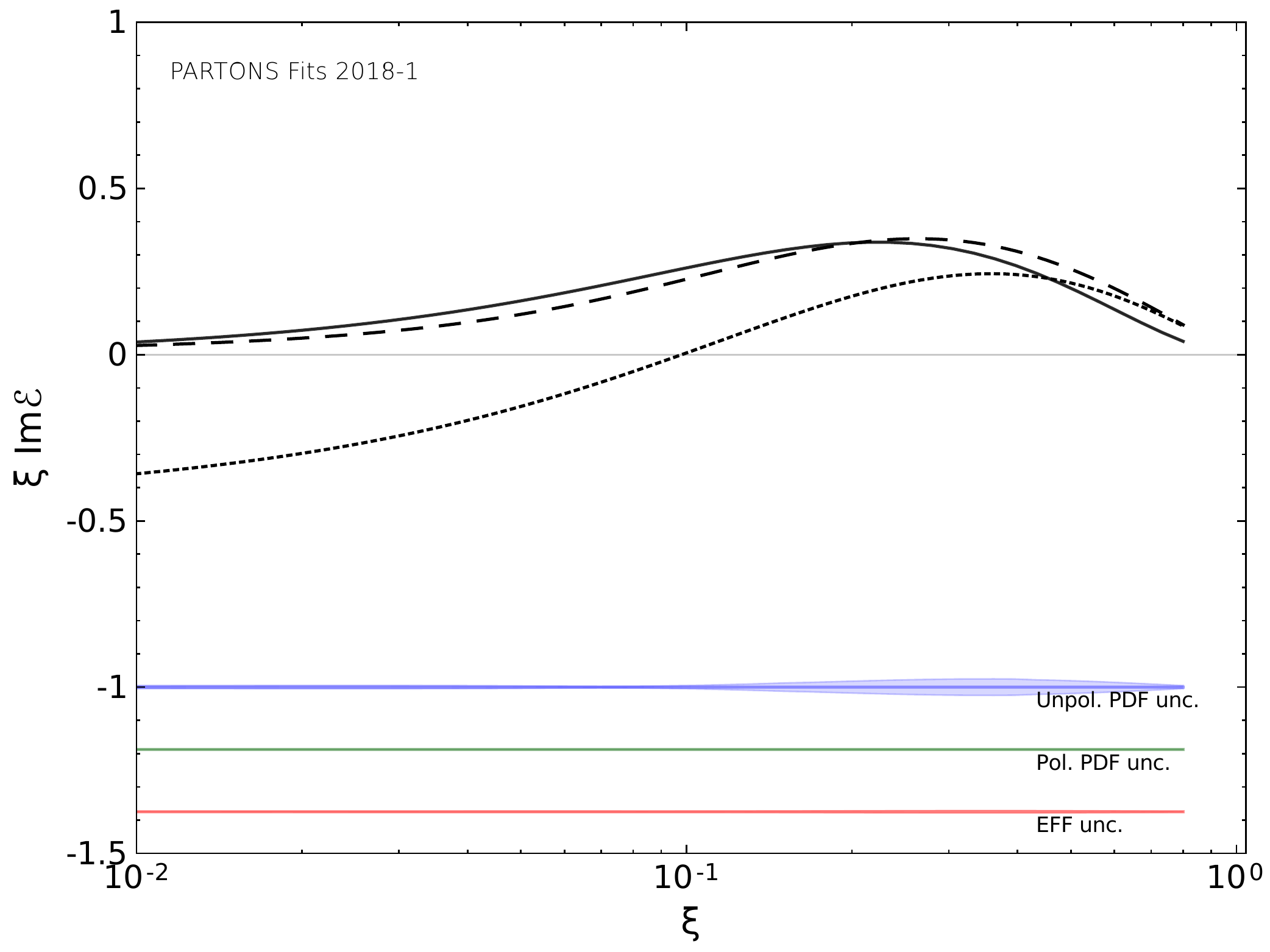}
\caption{Real (left) and imaginary (right) parts of the CFF $\mathcal{E}$ obtained in this work as a function of $\xi$ at $t = -0.3~\mathrm{GeV}^{2}$ and $Q^{2} = 2~\mathrm{GeV}^{2}$. For further description see Fig. \ref{fig:results:clas}.}
\label{fig:CFF:E}
\end{center}
\end{figure*}

\begin{figure*}[!ht]
\begin{center}
\includegraphics[width=0.49\textwidth]{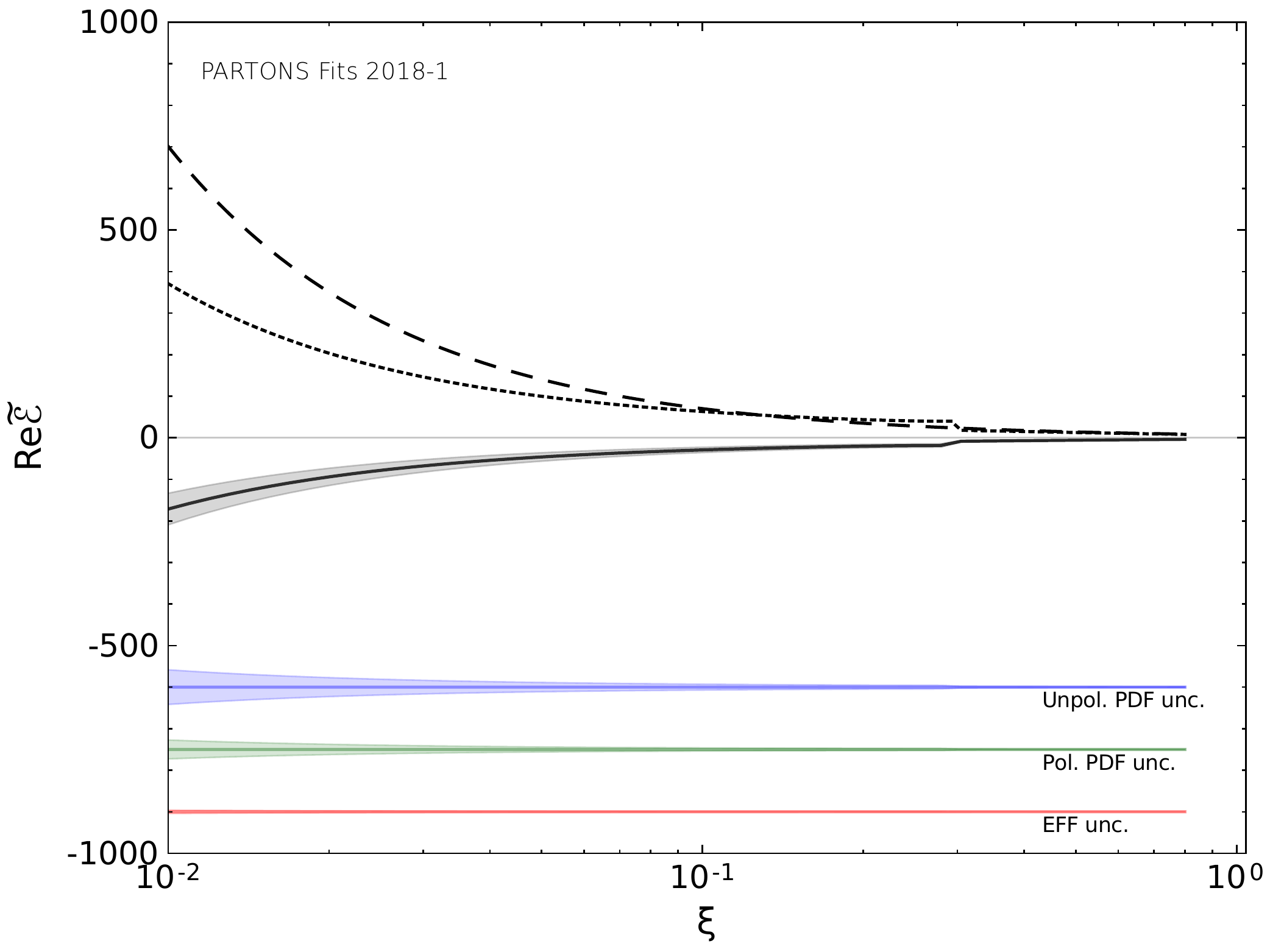}
\includegraphics[width=0.49\textwidth]{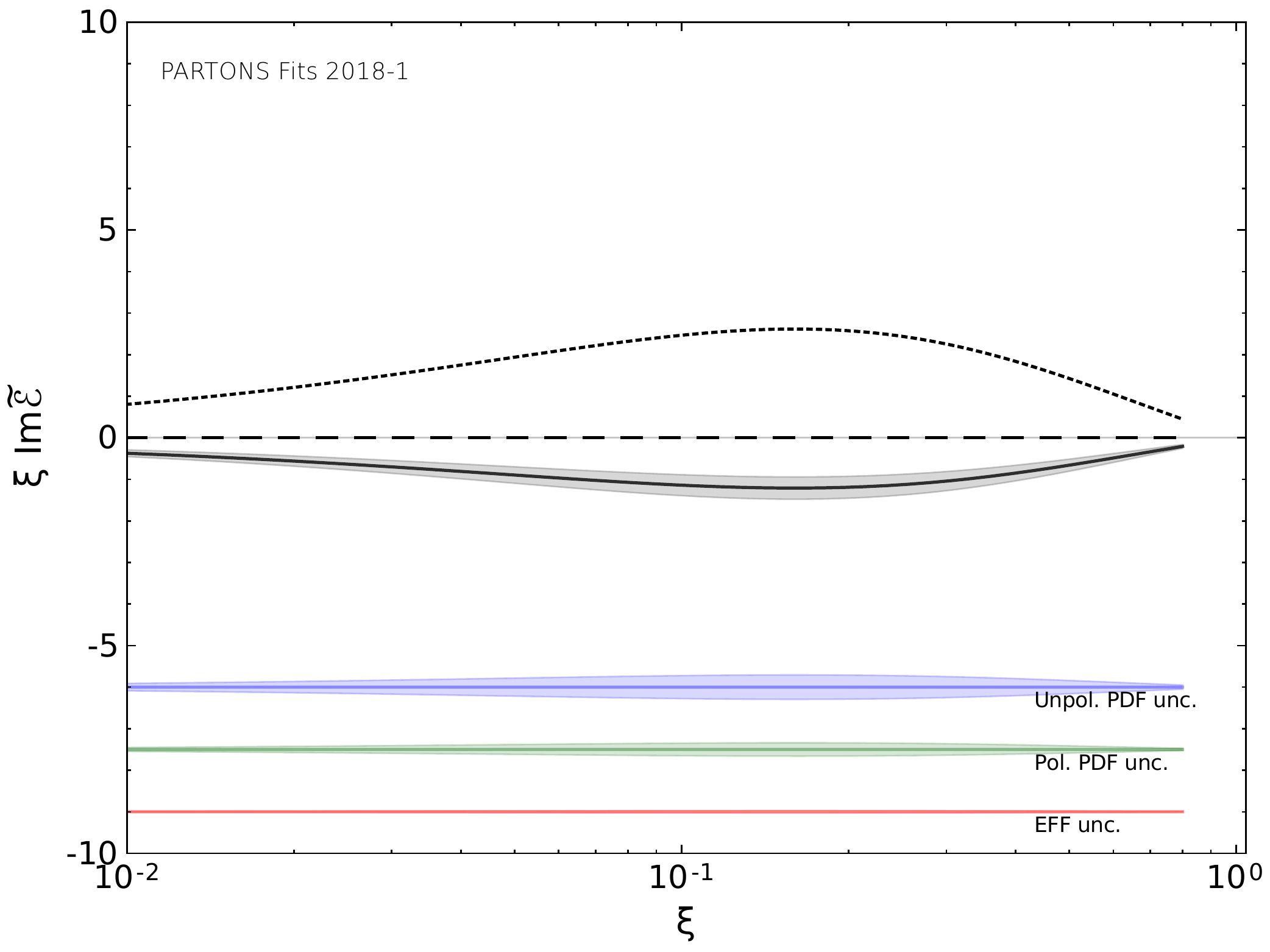}
\caption{Real (left) and imaginary (right) parts of the CFF $\widetilde{\mathcal{E}}$ obtained in this work as a function of $\xi$ at $t = -0.3~\mathrm{GeV}^{2}$ and $Q^{2} = 2~\mathrm{GeV}^{2}$. For further description see Fig. \ref{fig:results:clas}.}
\label{fig:CFF:Et}
\end{center}
\end{figure*}

The CFFs obtained in this analysis are shown as a function of $\xi$ for the exemplary kinematics of $t = -0.3~\mathrm{GeV}^{2}$ and $Q^{2} = 2~\mathrm{GeV}^{2}$ in Figs. \ref{fig:CFF:H}-\ref{fig:CFF:Et}. In complement of our results, we also show the curves evaluated at the same kinematics for the GK \cite{Goloskokov:2005sd, Goloskokov:2007nt, Goloskokov:2009ia} and VGG \cite{Vanderhaeghen:1998uc, Vanderhaeghen:1999xj, Goeke:2001tz, Guidal:2004nd} models. In general, VGG is closer to the results of our fit, which is expected, as this model was primarily designed to reproduce DVCS data in the valence region. We conclude that both models overestimate the imaginary part of the CFF $\mathcal{H}$, which for VGG is compatible with the conclusions of Ref. \cite{Dupre:2016mai}, where local fits to recent JLab DVCS data are reported.

\subsubsection*{Nucleon tomography}

\begin{figure*}[!ht]
\begin{center}
\includegraphics[width=0.95\textwidth]{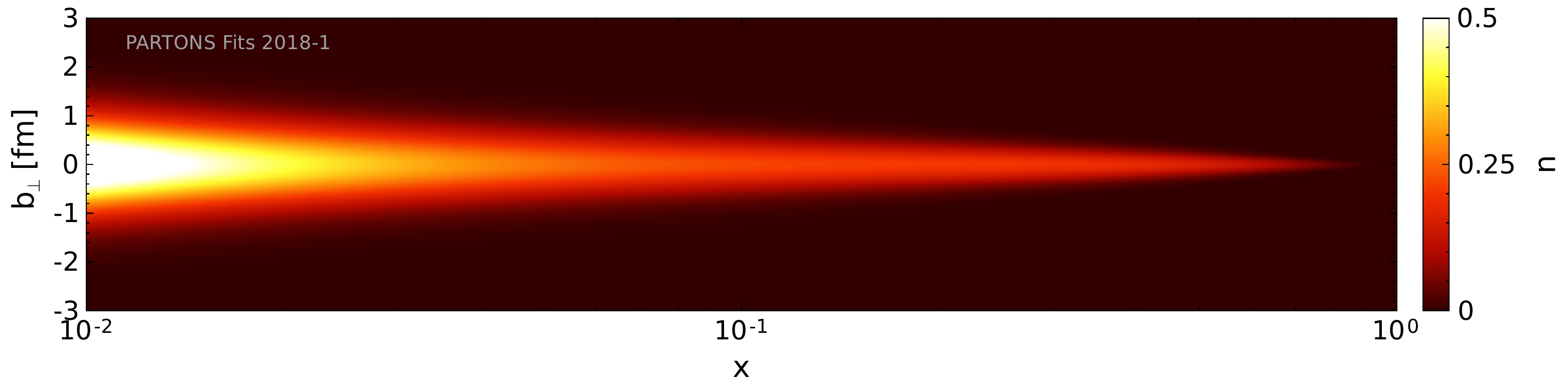}
\includegraphics[width=0.95\textwidth]{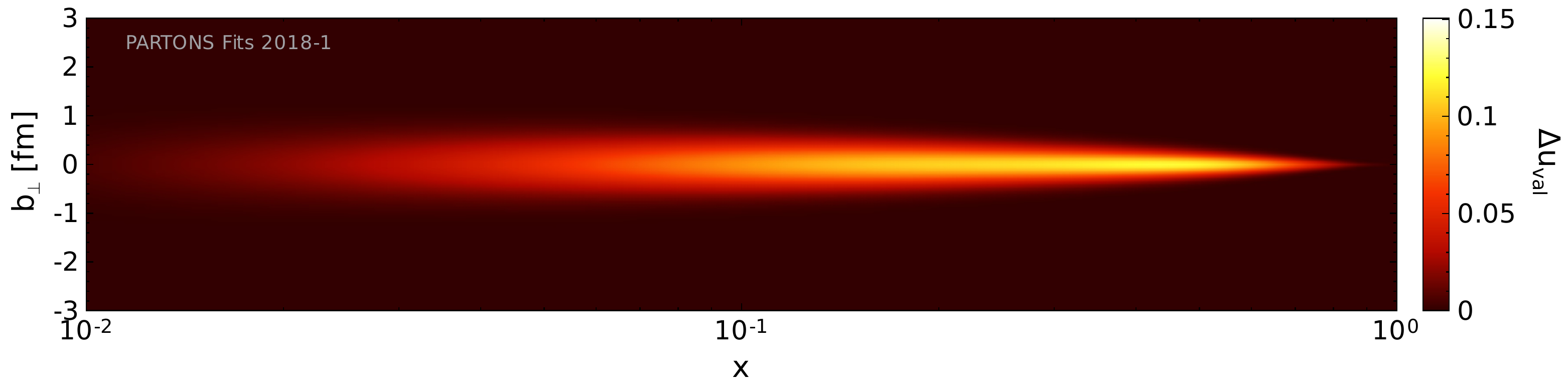}
\caption{Position of up quarks in an unpolarized proton (upper plot) and longitudinal polarization of those quarks in a longitudinally polarized proton (lower plot) as a function of the longitudinal momentum fraction $x$. For the lower plot only the valence contribution is shown.}
\label{fig:NT:2D}
\end{center}
\end{figure*}

Exemplary distributions of $u(x, b_{\perp})$ and $\Delta u_{\mathrm{val}}(x, b_{\perp})$ observables corresponding to Eqs. \eqref{eq:theory:nt_H} and \eqref{eq:theory:nt_Ht}, respectively, are shown in Fig. \ref{fig:NT:2D}. These plots illustrate how parton densities (longitudinal polarization of those partons) are distributed inside the unpolarized (longitudinally polarized) nucleon, however without the possibility of showing the corresponding uncertainties. 

\begin{figure*}[!ht]
\begin{center}
\includegraphics[width=0.49\textwidth]{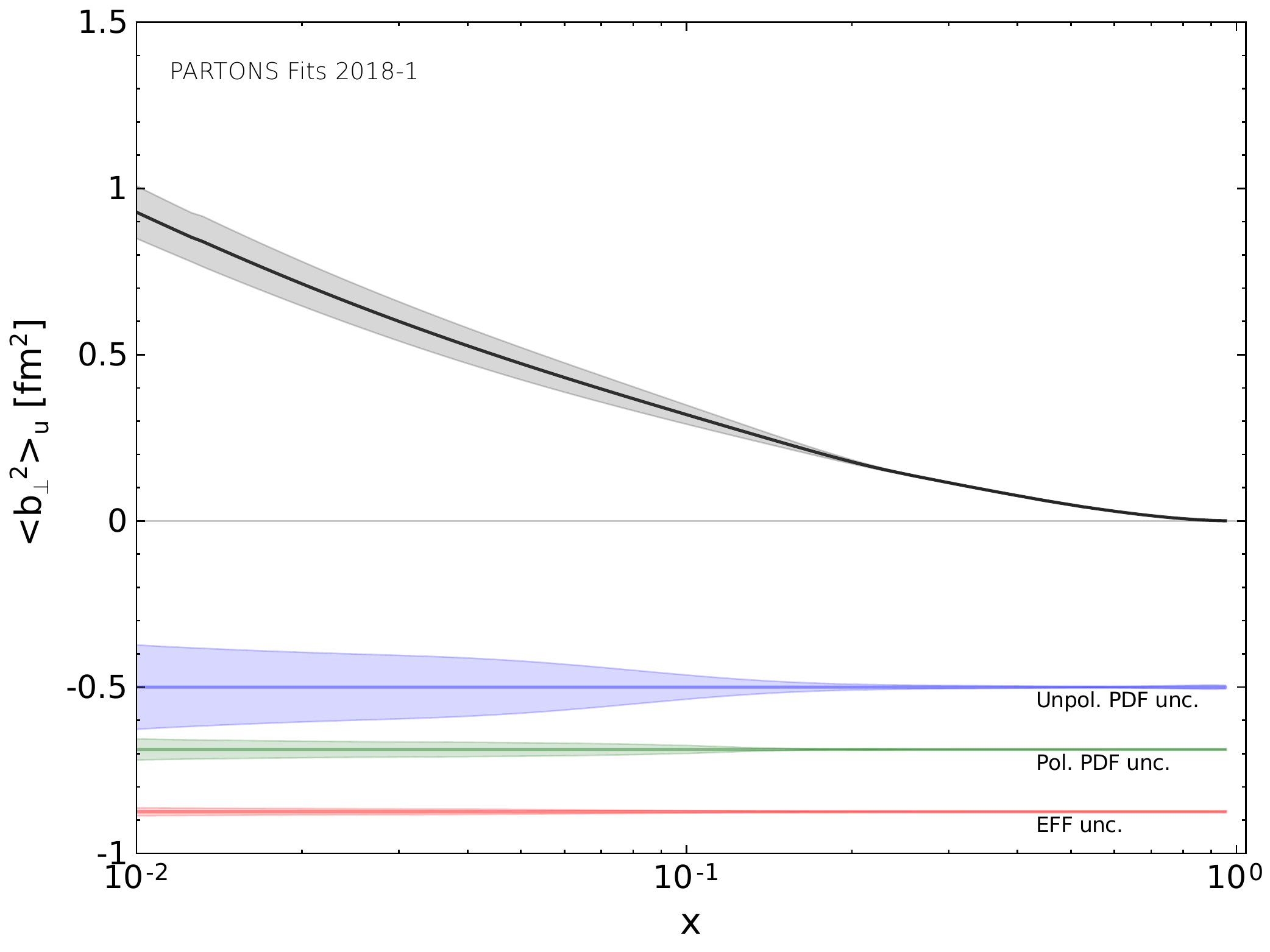}
\includegraphics[width=0.49\textwidth]{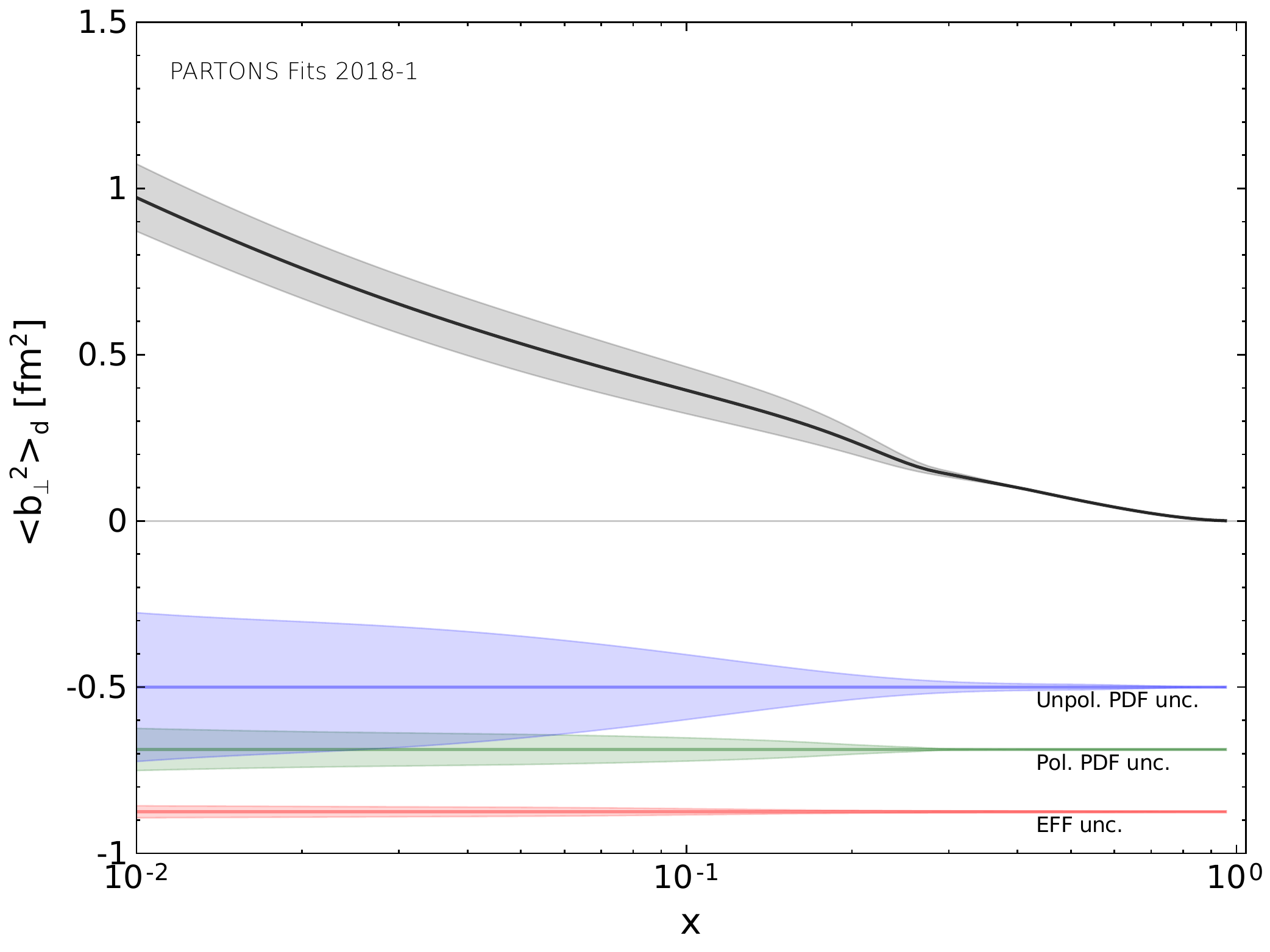}
\caption{Normalized second moment of the $q(x, \mathbf{b}_{\mathbf{\perp}})$ distribution for up (left) and down (right) quarks as a function of the longitudinal momentum fraction $x$. For further description see Fig. \ref{fig:results:clas}.}
\label{fig:NT:r2:H}
\end{center}
\end{figure*}

\begin{figure*}[!ht]
\begin{center}
\includegraphics[width=0.49\textwidth]{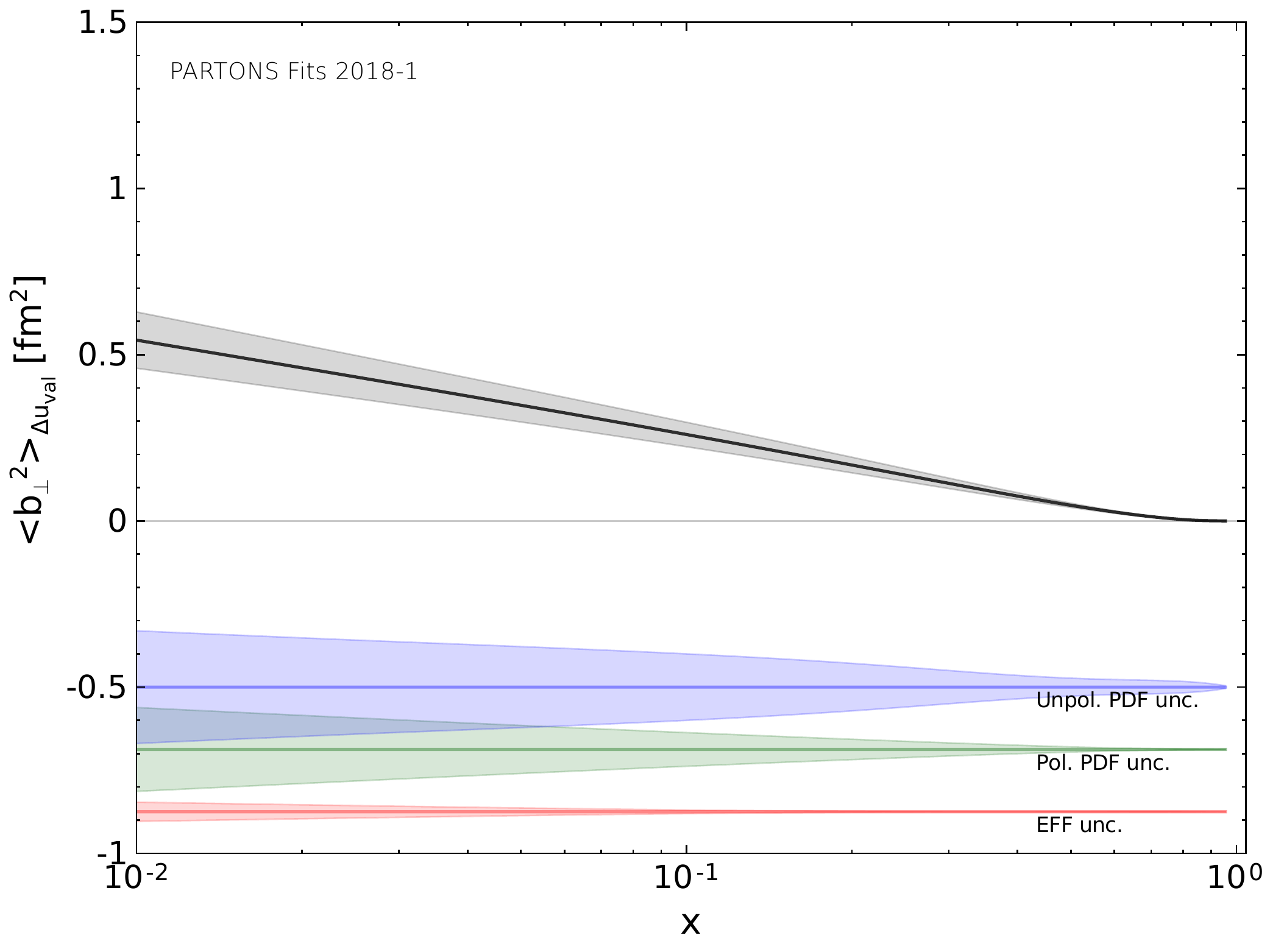}
\includegraphics[width=0.49\textwidth]{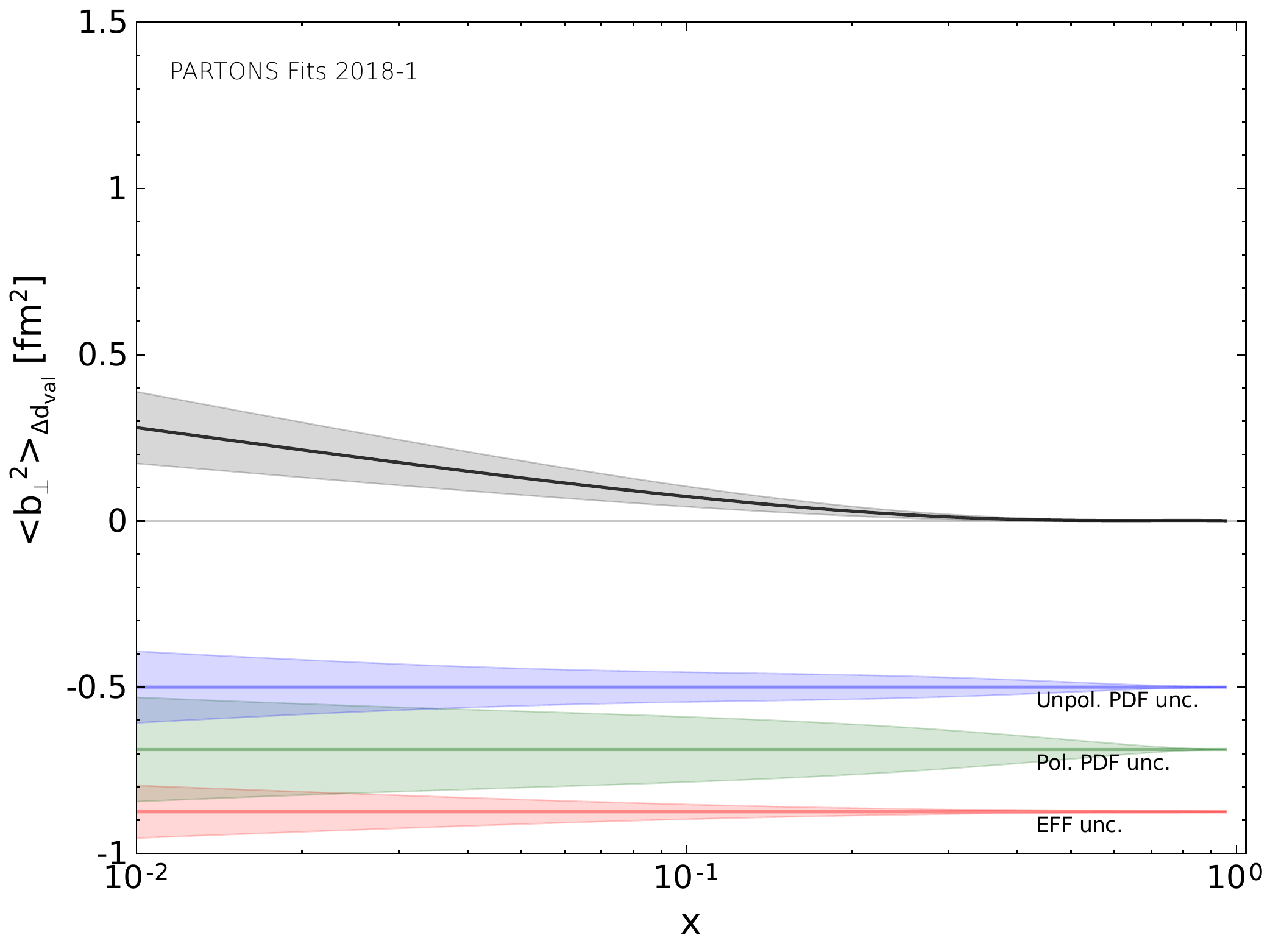}
\caption{Normalized second moment of the $\Delta q_{\mathrm{val}}(x, \mathbf{b}_{\mathbf{\perp}})$ distribution for up (left) and down (right) quarks as a function of the longitudinal momentum fraction $x$. For further description see Fig. \ref{fig:results:clas}.}
\label{fig:NT:r2:Ht}
\end{center}
\end{figure*}

\begin{figure*}[!ht]
\begin{center}
\includegraphics[width=0.49\textwidth]{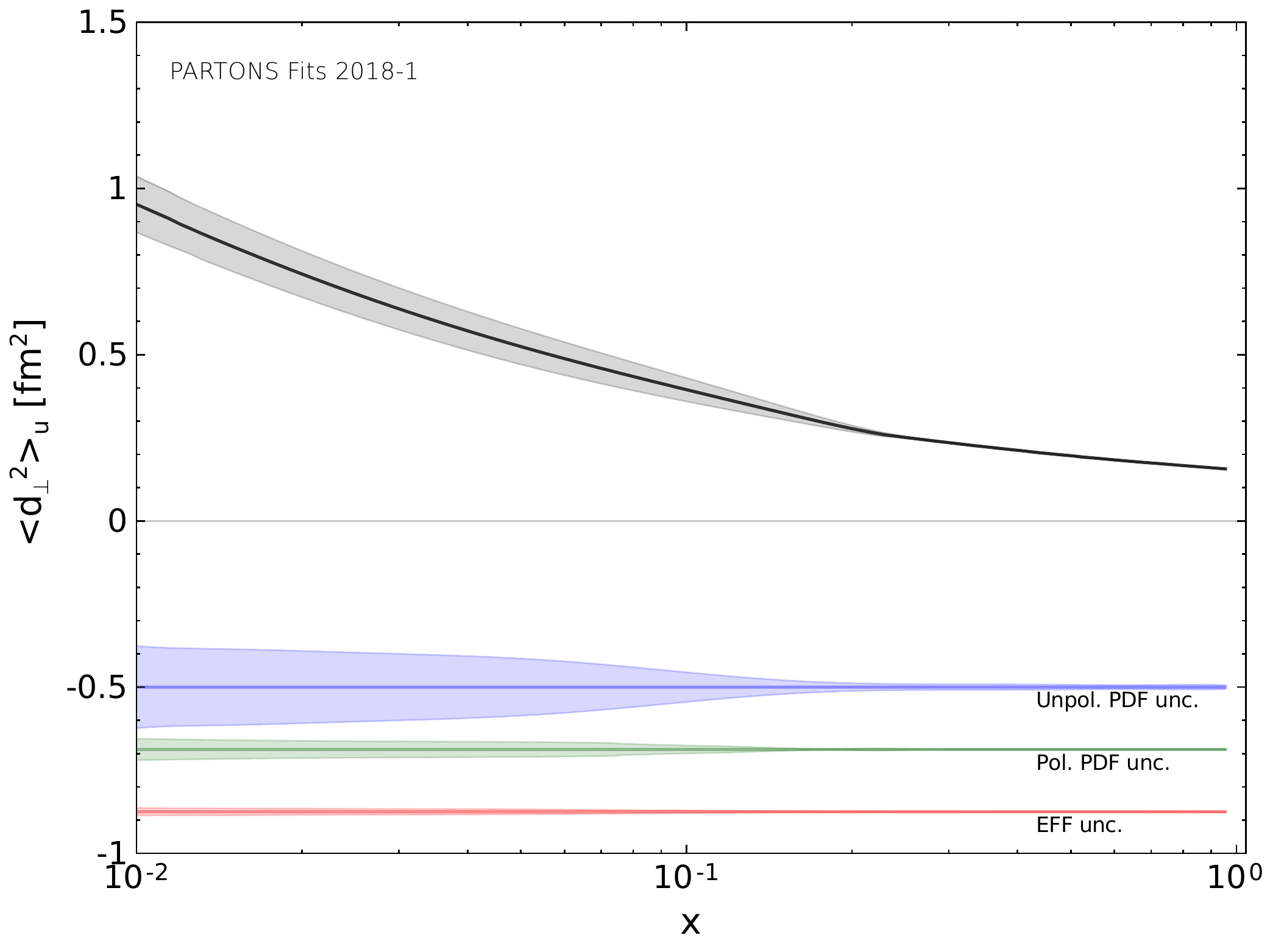}
\includegraphics[width=0.49\textwidth]{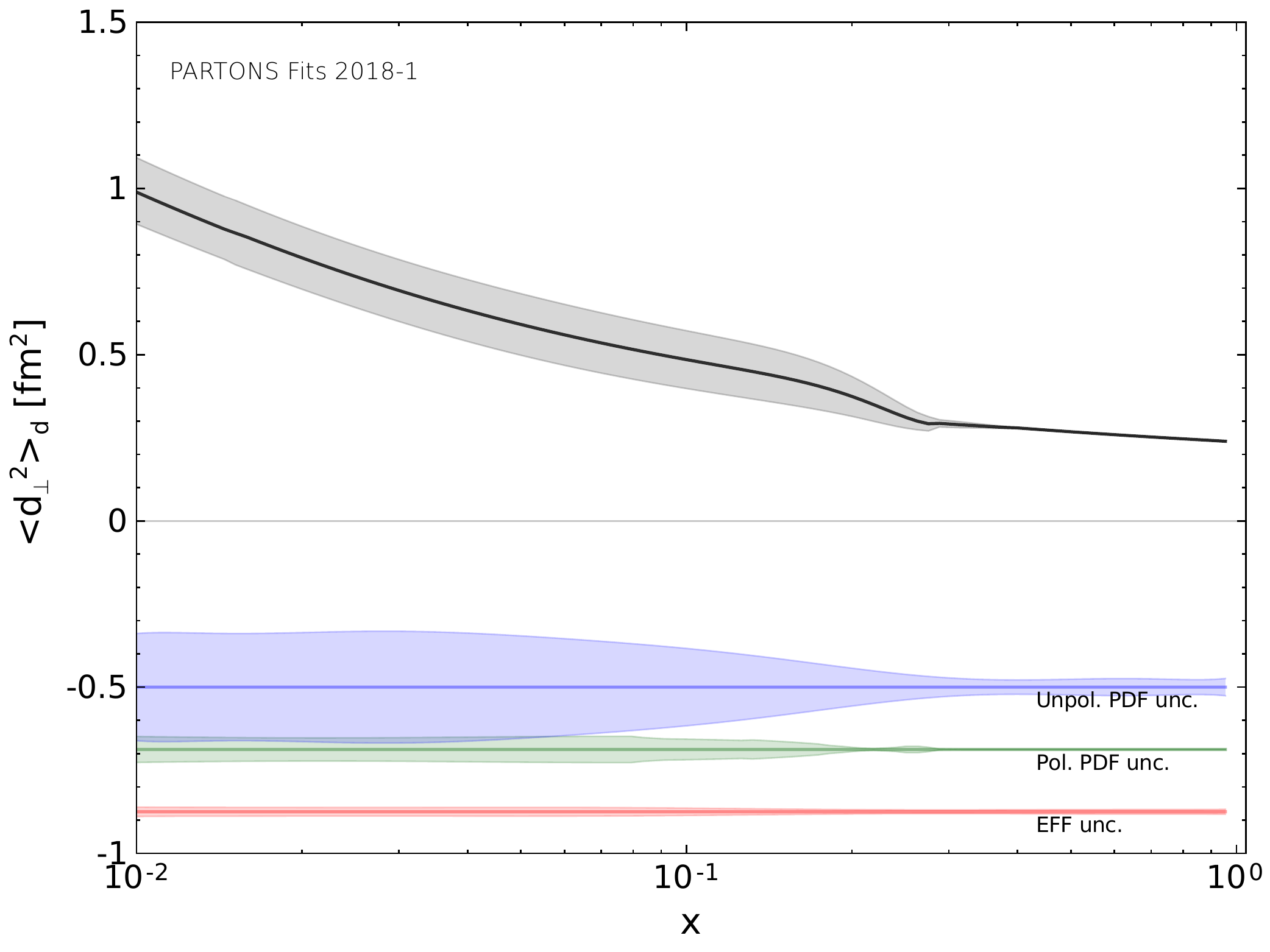}
\caption{Mean distance between the active quark and the spectator system for up (left) and down (right) quarks as a function of the longitudinal momentum fraction $x$. For further description see Fig. \ref{fig:results:clas}.}
\label{fig:NT:d2:H}
\end{center}
\end{figure*}

\begin{figure*}[!ht]
\begin{center}
\includegraphics[width=0.49\textwidth]{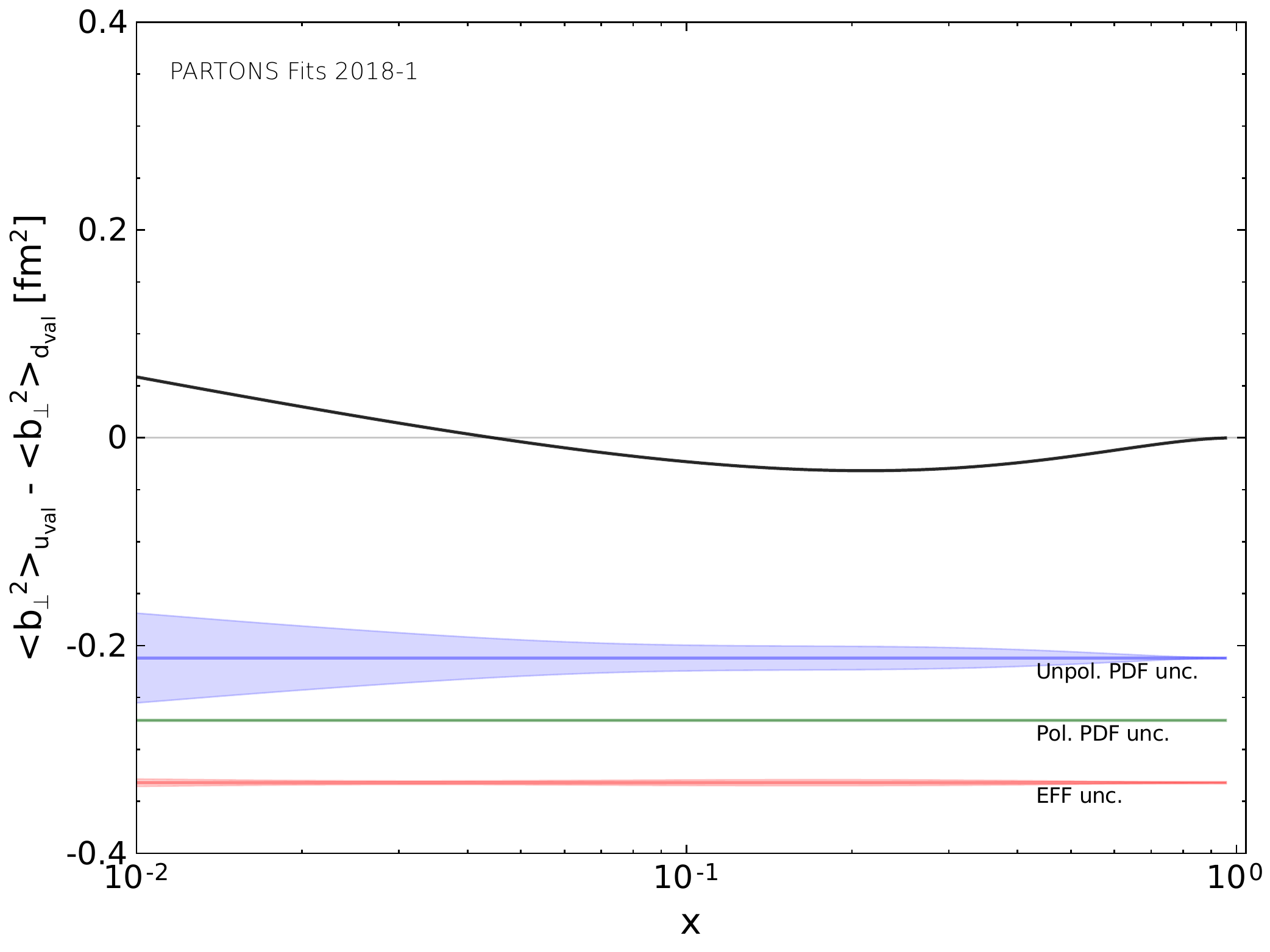}
\caption{Difference between ${\langle b_{\perp}^{2} \rangle}_{u_{\mathrm{val}}}(x)$ and ${\langle b_{\perp}^{2} \rangle}_{d_{\mathrm{val}}}(x)$, \emph{cf}. Fig. \ref{fig:NT:r2:H}. For further description see Fig. \ref{fig:results:clas}.}
\label{fig:NT:r2:diff_H}
\end{center}
\end{figure*}

This difficulty can be overcome by showing the normalized second moments of $q(x, \mathbf{b}_{\mathbf{\perp}})$ and $\Delta q_{\mathrm{val}}(x, \mathbf{b}_{\mathbf{\perp}})$ distributions, see Eqs. \eqref{eq:theory:distance_to_center_unpol} and \eqref{eq:theory:distance_to_center_pol} and Figs. \ref{fig:NT:r2:H} and \ref{fig:NT:r2:Ht}, respectively. One may note that ${\langle b_{\perp}^{2} \rangle}_{q}(x) \to 0$ when $x \to 1$, which is expected, as in this limit the position of the active quark is equivalent to the center of the coordinate system in which the impact parameter is defined. The corresponding distributions illustrating the mean squared distance between the active quark and the spectator system, see Eq. \eqref{eq:theory:distance_to_spectator}, are shown in Fig. \ref{fig:NT:d2:H}. In our studies $\langle d_{\perp}^{2} \rangle_{q}$ is finite when $x \to 1$, which we consider to be a welcome feature imposed by our Ansatz. We also study a possible difference between ${\langle b_{\perp}^{2} \rangle}_{u_{\mathrm{val}}}(x)$ and ${\langle b_{\perp}^{2} \rangle}_{d_{\mathrm{val}}}(x)$, see Fig. \ref{fig:NT:r2:diff_H}. This plot suggests that the distribution of $u_{\mathrm{val}}(x, \mathbf{b}_{\mathbf{\perp}})$ is narrower (broader) than $d_{\mathrm{val}}(x, \mathbf{b}_{\mathbf{\perp}})$ in the region of high (low) $x$. A firm conclusion however is not possible at this moment because of the large uncertainties. 

\section{Summary}
\label{sec:summary}

In this paper we proposed new parameterizations for the border and skewness functions. Together with the assumption about the analyticity properties of the Mellin moments of GPDs, those two ingredients allowed the evaluation of DVCS CFFs with the LO and LT accuracy. The evaluation was done with the dispersion relation technique and it included a determination of the DVCS subtraction constant, which is related to the QCD energy-momentum tensor.

In order to build and constrain our parameterizations we utilized many basic properties of GPDs, like their relation to PDFs and EFFs, the positivity bounds, the power behavior in the limit of $x \to 1$ and even the polynomiality property allowing the evaluation of the subtraction constant by comparing two equivalent ways of CFF computation. Our parameterizations provide a genuine access to GPDs at $(x, 0, t)$ kinematics and therefore they can be used for nucleon tomography. 

We performed the analysis of PDFs and obtained a set of functional parameterizations allowing the reproduction of original values and uncertainties. A small number of free parameters appearing in our approach was constrained by EFF and DVCS data. We considered all proton DVCS data, however not all of them entered the final analysis because of the used kinematic cuts and the initial data exploration. Our work was done within the PARTONS project that provides a modern platform for the study of GPDs and related topics.  

The quality of our fits is quite good. The fit to EFF data returns $\chi^{2} = 129.6$ for $178$ data points and $9$ free parameters, while that to DVCS data returns $\chi^{2} = 2346.3$ for $2600$ data points and $13$ free parameters. The good performance proves that our parameterizations, including the assumed analyticity property, are not contradicted by the used experimental data. We consider the unsurprising discrepancy between our results and the HERA data as a possible manifestation of gluon effects, and the surprising discrepancy with unpolarized cross sections published by Hall A as a possible need for higher twist or kinematic corrections. 

Our analysis also favors small values of the subtraction constant. We extract the distributions of $q(x, \mathbf{b}_{\mathbf{\perp}})$ and $\Delta q_{\mathrm{val}}(x, \mathbf{b}_{\mathbf{\perp}})$ and we study the properties of those distributions.

This analysis is characterized by a careful propagation of uncertainties coming from PDFs parameterizations, EFF and DVCS data, which we achieved by using the replica method. The first successful step towards the reduction of model uncertainties has been done already by selecting PDFs parameterizations based on the neural network technique. We plan to extend the usage of neural networks to other sectors of our future analyses for further reduction of model uncertainties.

\begin{acknowledgements}
The authors would like to thank M. Burkardt, M. Gar\c{c}on, F.-X. Girod, B. Pasquini, M.V. Polyakov, A.V. Radyushkin, K.M. Semenov-Tian-Shansky and F. Yuan for fruitful discussions and valuable inputs. This work was supported in part by the Commissariat à l'Energie Atomique et aux Energies Alternatives and by the Grant No. 2017/26/M/ST2/01074 of the National Science Centre, Poland. The computing resources of {\'S}wierk Computing Centre (CI{\'S}) are greatly acknowledged.
\end{acknowledgements}


\bibliographystyle{spphys}
\bibliography{bibliography.bib,bibliography_eff.bib}

\end{document}